\documentclass[leqno,12pt]{article}
\usepackage[hang]{subfigure}
\usepackage{color}
\usepackage{multirow}
\usepackage{latexsym}
\usepackage{float}

\renewcommand{\epsilon}{\varepsilon}

\newcommand{\Var}{\mbox{Var}}

\newcommand{\Bc}{\mathcal{B}}

\newcommand{\cov}{\textnormal{Cov}}

\newcommand{\E}{\mathbb{E}}
\newcommand{\Z}{{\mathbb Z}}
\newcommand{\T}{{\mathbb T}}
\newcommand{\V}{{\mathbb V}}
\newcommand{\W}{{\mathbb W}}
\newcommand{\D}{{\mathbb D}}
\newcommand{\N}{\mathbb N}
\newcommand{\R}{\mathbb R}

\newcommand{\Prob}{\mathbb P}

\newcommand{\argmax}{\textrm{argmax}}
\newcommand{\weak}{\rightsquigarrow}
\newcommand{\Bb}{\mathbb{B}}
\newcommand{\Gb}{\mathbb{G}}
\newcommand{\Dkonv}{\stackrel{\mathcal{D}}{\rightarrow}}
\newcommand{\Dequal}{\stackrel{\mathcal{D}}{=}}
\newcommand{\nor}{\mathcal{N}}
\newcommand{\Sc}{\mathcal{S}}

\newcommand{\eps}{\varepsilon}

\usepackage{verbatim}
\usepackage{amsmath,stackrel}
\usepackage{amssymb}
\usepackage{ntheorem}
\usepackage[ansinew]{inputenc}
\usepackage{graphicx}
\usepackage{epsfig}
\usepackage{dsfont}
\usepackage{bm}
\usepackage{bbm}

\usepackage[authoryear]{natbib}
\newtheorem{theorem}{Theorem}[section]
\newtheorem{prop}{Proposition}[section]
\newtheorem{example}{Example}[section]
\newtheorem{remark}{Remark}[section]

\newtheorem{lemma}{Lemma}[section]

\DeclareMathOperator*{\esssup}{ess\,sup}

\def\3{\ss}

\newcommand{\bea}{\begin{eqnarray*}}
\newcommand{\eea}{\end{eqnarray*}}
\newcommand{\be}{\begin{eqnarray}}
\newcommand{\ee}{\end{eqnarray}}

\newcommand{\ba}{\begin{array}}
\newcommand{\ea}{\end{array}}
\def\3{\ss}

\newcommand{\Hc}{\mathcal{H}}

\newcommand{\floor}[1]{\lfloor{#1}\rfloor}

\begin{document}

\title{Testing  relevant hypotheses in functional time series via self-normalization}

\author{
{\small Holger Dette, Kevin Kokot} \\
{\small Ruhr-Universit\"at Bochum} \\
{\small Fakult\"at f\"ur Mathematik}\\
{\small Bochum, Germany} \\
{\small e-mail: $\{$holger.dette, kevin.kokot$\}$@rub.de}\\
\and
{\small Stanislav Volgushev } \\
{\small University of Toronto} \\
{\small  Department of Statisical Sciences }\\
{\small  Toronto, Canada}  \\
{\small email: stanislav.volgushev@utoronto.ca}\\
}

  \maketitle

\begin{abstract}
In this paper we develop methodology for testing relevant hypotheses {about functional time series} in a tuning-free way. Instead of testing for exact equality, for example for the equality of two mean functions from two independent time series, we propose to test the null hypothesis of no {\it relevant}  deviation. In the two sample problem this means that an $L^2$-distance between the two mean functions is smaller than a pre-specified threshold. For such hypotheses self-normalization, which was introduced by \cite{shao2010} and \cite{shazha2010} and is commonly used to avoid the estimation of nuisance parameters, is not directly applicable. We develop new self-normalized  procedures  for  testing relevant hypotheses in the one sample, two sample and change point problem and investigate their asymptotic properties. Finite sample properties of the proposed tests are illustrated by means of a simulation study and {data examples}. Our main focus is on functional time series, but extensions to other settings are also briefly discussed.
\end{abstract}


\section{Introduction}  \label{sec1}
\def\theequation{1.\arabic{equation}}
\setcounter{equation}{0}

Statistics for functional data  has found considerable interest in the last twenty years as documented in the various monographs by \cite{RamsaySilverman2005},  \cite{FerratyVieu2010} and    \cite{HorvathKokoskza2012} among others. The available methodology includes explorative tools such as shift and feature registration, warping or principal components, and methods
for statistical inference such as testing of hypotheses and change point analysis. In this context a large portion of the literature attacks the problem of hypotheses testing by considering hypotheses of the form
\be \label{null}
H_0:  d=0  \text{ versus }  H_1:  d  \not =0
\ee
where $d$ is a real valued parameter such as the norm of the  mean function in one sample or the norm  of the difference of two mean functions or two covariance operators from two samples. For example \cite{hallVanKeilegom2007} study the effect of smoothing when converting discrete observations into functional data, \cite{Horkok2009} compare linear operators in two functional regression models, \cite{benHaeKne2009} propose functional principal component analysis (FPCA) for two sample inference while \cite{pankramad2010} and \cite{frehorkokste2012}
consider a test for the equality of covariance operators. More recently \cite{horvathKokoszkaReeder2013} suggest tests for the comparison of two mean functions from  temporally dependent curves under model-free assumptions and \cite{porstagho2016} compare the distributions of two samples by methods which are based on FPCA. Another important research area in functional data analysis is  change point detection and we refer to \cite{bergabhokok2009},  \cite{hoermannkokocka2010}, \cite{astonKirch2012a}, \cite{zhang2011}, \cite{horkokric2013}, \cite{BucchiaWendler2015} among others who investigate change point  problems from various perspectives.

Several authors consider methods for independent data. In this case the  quantiles for corresponding tests can be easily obtained by asymptotic theory as the unknown quantities in the limit distribution of the test statistics can be reliably estimated (for example  the asymptotic variance of a standardized mean). However, for functional samples exhibiting temporal dependence, the asymptotic distribution of many commonly used tests statistics
involves the long-run variance, which makes the statistical inference substantially more difficult. Several authors propose to estimate the long-run variance [see \cite{kokoszka2012} or
 \cite{horvathKokoszkaReeder2013} among others], but the commonly used estimators depend on regularization parameters. As alternative, bootstrap methods can be applied to obtain critical values and we refer to \cite{benHaeKne2009}, \cite{cuefebfra2003}, \cite{zhapenzha2010},  \cite{BucchiaWendler2015} and \cite{papsap2016} among many others. A third method to obtain  (asymptotically) pivotal test statistics is  the concept of self-normalization, which was introduced in the  seminal  papers of  \cite{shao2010} for the construction of confidence intervals and \cite{shazha2010} for change point analysis. More recently it has been developed further  for the specific needs of functional data by \cite{zhang2011} and \cite{zhangshao2015}
 [see also  \cite{shao2015} for a recent review].

This list of references is by no means complete but a  common feature of all of these references is that they usually address hypotheses
 of the form \eqref{null}, which we call  ``classical''  hypotheses  in the following discussion. However, in many applications one might not be
 interested in detecting very small deviations of the parameter $d$ from $0$ (often the researcher  even knows that $d$ is not exactly equal to $0$, before any experiments have been carried out). For  example, in change point detection a modification of the statistical analysis for prediction might not be necessary if the difference between the parameters before and after the change point is rather small. This discussion may be viewed as a particular case of   the common bias variance trade-off in statistics. {Therefore   we argue that  one should carefully think about the size of the difference in which one is interested. In particular we propose to replace the hypotheses \eqref{null}  by the hypotheses of  \emph{relevant differences}, }
that is
\be \label{relevant}
H_0 :  d  \leq \Delta \quad \mbox{versus} \quad H_1:  d > \Delta \, ,
\ee
where  $\Delta$ is a pre-specified constant representing the ``maximal''  value  for the parameter $d$, which
can be accepted  as not scientifically significant.  { If the null hypothesis in \eqref{relevant} holds we speak of a null of
{\it  no relevant difference}.}
This formulation of the  testing problem
requires the specification of the threshold  $\Delta >0 $, which depends on the specific application. ``Classical''
hypotheses tests simply use $\Delta = 0 $, but we argue that from a practical point of view
it might be very reasonable to think about this choice more carefully and to define the  size of the change in which
one is really interested from a scientific viewpoint.

We also note that the  formulation of  the testing  problem   in the form \eqref{relevant} avoids the consistency problem  mentioned in \cite{berkson1938}, that is:
any consistent test will detect any arbitrary small change in the parameters if the sample size is sufficiently large.
Moreover, by interchanging the hypotheses, that is considering the hypotheses of  {\it equivalence}
\be \label{equiv}
H_0 : d  >  \Delta \quad \mbox{versus} \quad H_1: d  \leq \Delta \, ,
\ee
one is able to decide for a  ``small  parameter''  $d$    at a controlled type I error (for example that the norm $d$ of the difference between  the  mean functions
of two samples is smaller than a given threshold). Hypotheses of the form \eqref{relevant} and \eqref{equiv} are called \emph{precise hypotheses} or \emph{relevant hypotheses}
in the literature [see \cite{bergdela1987}] and are frequently used in biostatistics. We refer to \cite{chowliu1992} and \cite{wellek2010} for  more details and applications.

In this paper we discuss the problem of testing relevant hypotheses in the context of functional dependent data.  We are particularly interested in methods based on
 self-normalization in order to avoid estimation of the long-run  variance or resampling methods. {The construction of efficient long-run variance estimates and resampling techniques is more difficult for testing relevant hypotheses, because - in contrast to ''classical''  hypotheses  - the null hypothesis usually corresponds to
 an infinite dimensional set (for example the set of mean function with squared $L^{2}$-norm less or equal than $\Delta$). }

For this purpose we modify the classical approaches to self-normalization based testing proposed by \cite{shao2010} and \cite{shazha2010} in order to make them applicable for testing relevant hypotheses.  { \cite{zhang2011} and \cite{zhangshao2015} also use the concept of self-normalization  to develop statistical methodology for  functional data analysis.
In particular they  construct tests  for  a change in the
  mean function and in the lag-$1$ autocovariance operator and for comparing  the covariance operators and associated  eigenvalues or eigenvectors
from two samples.  The main differences between their approach and the methods presented here are the following.
   First, these references do not consider the problem of testing relevant hypotheses, but deal with ``classical''  hypotheses of the form
   \eqref{null}. Thus the present  paper addresses a different statistical problem, where  currently available  methods are not applicable.
    Second, their approach is based  on a dimension reduction  projecting the functions on a finite dimensional
   vector (for example principal components), which  is  then  used for the subsequent  statistical inference using common self-normalization techniques.
 In contrast to their work   we are able  to develop a self-normalized test for  the problem of testing relevant hypotheses  of the form \eqref{relevant},
 which does not require dimension reduction. For this purpose the common concepts of self-normalization have to be further extended.
 }
This modification is of independent interest besides the field of functional data analysis
and applicable in many other problems.

The remaining part of this paper is organized as follows. Our basic idea is explained in Section \ref{sec2} for the one and two sample case, where it is most transparent. Roughly speaking, we construct an asymptotic confidence interval for the parameter $d$ to obtain tests for hypotheses of the form \eqref{relevant} and \eqref{equiv}.
 In Section \ref{sec3} we  address the problem of relevant change point analysis by the new way of self-normalization; here an additional challenge arises from the fact that the change point location is unknown and needs to be estimated.
{ While the methodology in Section \ref{sec2}  and  \ref{sec3} refers to statistical inference for  mean functions we illustrate in
 Section \ref{sec4}  how those ideas can be extended to inference for  covariance operators.}
 Some finite sample results are presented in Section \ref{sec5}, where we also {illustrate the proposed methodology on two data examples.}
 {Here we also provide brief discussion of  self-normalization  and  estimation of the long-run variance in
 the context of testing relevant hypotheses.}
{Finally, in an online supplement we present additional finite sample results (Section \ref{addsim}), give the  proofs of our results
 (Section \ref{sec6})  and  discuss extensions beyond functional time series  (Section \ref{sec:concalt}). }

\section{Relevant hypotheses and self normalization }\label{sec2}
\def\theequation{2.\arabic{equation}}
\setcounter{equation}{0}

Let $T$ be a compact set in $\R^d$ and let $L^2(T)$ denote the Hilbert space of square
integrable functions on the set $T$  with the usual inner product $\langle \cdot, \cdot \rangle $
and  corresponding norm $\| \cdot \| $.

\subsection{One sample problems}
\label{sec21}  Let $\{X_n\}_{n\in \Z} $ denote a strictly stationary functional time series where the random variables $X_n$ are elements in $L^2(T)$ (with expectation $\mu := \E[X_1] \in L^2(T)$, see Section 2.1 in \cite{buecher2018} for a detailed discussion of expected values in Hilbert spaces). {For the sake of simplicity we will assume that $T = [0,1]$, but all methods proposed in this paper can be generalized to other subsets of $\R^d$. To avoid confusion between the interval $[0,1]$ corresponding to $\lambda$, which defines the sub-sample  $X_1, \ldots , X_{\lfloor n\lambda \rfloor}$ and the interval $T = [0,1]$, we write $T$ for the interval $[0,1]$ belonging to the argument $t$ of $X_n$.} Based on a sample $X_1,...,X_n$ we are interested in relevant hypotheses regarding the parameter $d= \int_T \mu^2(t) dt$, that is
\begin{equation} \label{nullone}
H_0: \int_T \mu^2(t) dt \leq  \Delta  \quad\mbox{ versus  }\quad H_1: \int_T \mu^2(t) dt >  \Delta \, .
\end{equation}
Define the partial sums
\begin{equation} \label{parsum}
S_n(t,\lambda) := \frac{1}{n} \sum_{j=1}^{\lfloor n\lambda \rfloor} X_j(t) \, , \quad \lambda \in [0,1] \, ,
\end{equation}
then, under suitable assumptions, the statistic $\int_T  S_n^2(t,1) dt $ is a consistent estimator of $ \int_T \mu^2(t) dt$. Consequently a test for the hypotheses \eqref{nullone} is obtained by rejecting the {null hypothesis of no relevant difference}  for large values of
\begin{align}\label{tstatone}
\hat \T_n  = \int_T  S_n^2(t,1) dt \, .
\end{align}
It will be shown in the proof of Theorem \ref{thm_relevantfunc} that under some technical assumptions the asymptotic distribution of an appropriately standardized version of $\hat \T_n$ takes the form
\[
\sqrt{n} \Big (\hat \T_n  -\int_T \mu^2(t) dt \Big  ) \Dkonv  \nor (0 , \tau^2)
\]
with long-run variance 
\begin{equation}\label{tau}
\tau^2 = 4\int_T\int_T \mu (s) \mu(t) C(s,t) ds \, dt \, ,
\end{equation}
where
\begin{eqnarray}
 \label{approx4}
C(s,t) &=& \cov (  X_0(s),X_0 (t)) +   \sum_{\ell=1}^\infty \text{Cov} (  X_0(s),X_\ell (t) ) +   \sum_{\ell=1}^\infty \text{Cov} ( X_0(s),X_{-\ell }(t) )
\end{eqnarray}
is the long-run covariance operator of the process $\{X_n\}_{n\in \Z} $.
Here we note that the above weak convergence is also true when $\mu \equiv 0$, in which case the limit is a degenerate normal distribution with a point mass at zero. Unfortunately, the long-run variance $\tau^2$ is difficult to estimate in practice. This motivates us to adopt a self-normalization approach which avoids direct estimation of $\tau^2$. To be more precise let $\nu $ denote a probability measure on the interval $(0,1)$ and define
\begin{equation}  \label{vn}
\hat{\mathbb{V}}_n := \Big( \int_0^1  \Big[ \int_T S_n^2(t,\lambda ) dt - \lambda^2 \int_T S_n^2(t,1) dt\Big]^2 \nu (d\lambda) \Big)^{1/2} \, .
\end{equation}
As we will show later we have
\begin{equation}\label{vntn_joint}
\Big(\sqrt{n}(\hat{\mathbb{T}}_n - d), \sqrt{n} \, \hat{\mathbb{V}}_n\Big) \Dkonv \bigg( \tau \Bb(1), \tau \Big( \int_0^1 \lambda^2 (\Bb(\lambda) - \lambda\Bb(1))^2 \nu (d\lambda ) \Big)^{1/2} \bigg) \, ,
\end{equation}
where $\Bb$ denotes a standard Brownian motion on the interval $[0,1]$. In particular, this implies that, in the case $\tau \neq 0$, the ratio $(\hat{\mathbb{T}}_n - d)/\hat{\mathbb{V}}_n$ converges to a pivotal distribution. This suggests that a test for~\eqref{nullone} can be constructed by rejecting the {null hypothesis of no relevant difference}  in \eqref{nullone}, whenever
\begin{align} \label{testrelone}
\quad \hat \T_n > \Delta + q_{1-\alpha}(\mathbb{W}) {\hat{\mathbb{V}}_n} \, ,
\end{align}
where $q_{1-\alpha}(\mathbb{W})$ denotes the $(1-\alpha)$-quantile of the distribution of the pivotal random variable
\begin{equation}\label{Wdef}
\mathbb{W} := \frac{\Bb(1)}{\big( \int_0^1 \lambda^2 (\Bb(\lambda) - \lambda\Bb(1))^2 \nu (d\lambda ) \big)^{1/2}} \, .
\end{equation}
It is worthwhile to mention that the distribution of $\mathbb{W}$ is not the same as the one in previous work on self-normalization [see for example  \cite{shao2010} or \cite{shao2015}] and quantiles of this distribution need to be simulated first. In Table~\ref{Wquantiles} we display quantiles of this distribution, where  $\nu$ is the discrete uniform distribution supported on the points $\lambda_i=i/5$ $(i=1,\ldots ,4)$, on the points $\lambda_i=i/20$ $(i=1,\ldots ,19)$ and on the points $\lambda_i=i/100$ $(i=1,\ldots ,99)$, respectively.

\vspace{15pt}
\begin{table} \label{Wquantiles}
\begin{center}
\begin{tabular}{c|cccccc}

    & 99\% & 95\% & 90\% \\
\hline \hline
1)  & 18.257 & 10.998 & 7.855  \\
2)  & 16.081 & 10.530 & 7.619  \\
3)  & 16.282 & 10.583 & 7.662  \\

\end{tabular}
\end{center}
\caption{\it Simulated quantiles (based on $1000$ replications) of the distribution of the statistic $\mathbb{W}$ defined by \eqref{Wdef}, where  $\nu$ is the discrete
uniform distribution supported on the points 1) $\lambda_i=i/5$ $(i=1,\ldots ,4)$, 2)
$\lambda_i=i/20$ $(i=1,\ldots ,19)$ and 3) $\lambda_i=i/100$ $(i=1,\ldots ,99)$. }
\end{table}

Next we prove that the decision rule in \eqref{testrelone} indeed provides an asymptotic level $\alpha$ test. For this purpose we make the following assumptions [see also \cite{berhorric2013,horkokric2013}]:

\smallskip

\begin{itemize}
\item [(A1)]   For all $j \in \Z$ we have  $X_j = \mu + \eta_j$, where $( \eta_j)_{j\in \Z} $ is a centered error process which satisfies (A2)--(A4).
\item [(A2)]  $( \eta_j)_{j\in \Z} $ is a sequence of Bernoulli shifts, that is: there exists a measurable space, say $\Sc $ and a function $f:\Sc^\infty \longrightarrow L^2([0,1]) $ such that
$$
\eta_j = f(\varepsilon_j, \varepsilon_{j-1}, \ldots ) \qquad \text{  for all } j \in \Z \, ,
$$
where $( \varepsilon_j)_{j\in \Z} $ is a sequence of i.i.d $\Sc$-valued functions, such that
$\varepsilon_j(t) = \varepsilon_j (t, \omega) $ is jointly measurable $(j\in \Z)$.
\item [(A3)]  $ \E \|  \eta_j \|^{2+\psi} < \infty  $  for some $\psi \in (0,1)$.
\item [(A4)]  The sequence $( \eta_j)_{j\in \Z} $  can be approximated by $\ell$-dependent sequences
$( \eta_{j,\ell} )_{j\in \Z} $ in the sense that for some $\kappa > 2+\psi$
\begin{equation} \label{eq:A4main}
\sum_{\ell=1}^\infty \big ( \E \| \eta_{0} -\eta_{0,\ell} \|^{2+\psi} \big)^{1/\kappa} < \infty \, ,
\end{equation}
where $\eta_{j,\ell}$ is defined by
\begin{align*}
\eta_{j,\ell} &= f(\varepsilon_j, \varepsilon_{j-1}, \ldots
 \varepsilon_{j-\ell + 1},\boldsymbol{\varepsilon}_{j,\ell}^*) \\
 \boldsymbol{\varepsilon}_{j,\ell}^* &= (\varepsilon_{j,\ell,j-\ell }^*,\varepsilon_{j,\ell,j-\ell -1}^*, \ldots ) \, ,
 \end{align*}
and the random variables $\varepsilon_{j,\ell,k}^*$ are i.i.d. copies of $\varepsilon_{0}$, and independent of the sequence $( \varepsilon_j)_{j\in \Z}$.
\end{itemize}

\medskip

\begin{theorem} \label{thm_relevantfunc}
Assume that $\Delta > 0$. Under the assumptions (A1)-(A4) the test decision given in~\eqref{testrelone} satisfies
\begin{align*}
\lim_{n \to \infty}
\mathbb{P} \Big( \hat \T_n > \Delta + q_{1-\alpha}(\mathbb{W}) {\hat{\mathbb{V}}_n} \Big)
&=
\begin{cases}
0 & \text{ if }  \int_T \mu^2(t) dt < \Delta \, ,
\\
\alpha & \text{ if }  \int_T \mu^2(t) dt = \Delta \,   \text{ and }  \tau^{2} >0 ,
\\
1 & \text{ if }  \int_T \mu^2(t) dt > \Delta \, .
\end{cases}
\end{align*}
\end{theorem}

\bigskip

A detailed proof of Theorem \ref{thm_relevantfunc} is given in Section \ref{sec61}. In what follows we provide an informal overview of the main steps in the proof. If $\int \mu^2(t) dt \neq 0$ and assumptions (A1) - (A4) hold, it can be shown that
\begin{align} \label{haupt}
\Big\{\sqrt{n}\Big( \int_T S_n^2(t,\lambda) dt - \lambda^2 \int_T \mu^2(t) dt\Big)\Big\}_{\lambda\in [0,1]} \weak
 \Big\{ \lambda \tau \Bb(\lambda) \Big\}_{\lambda\in [0,1]} \, ,
\end{align}
where the symbol $\weak$ means weak convergence in $\ell^{\infty}([0,1])$
 and $\tau^2$ is defined in~\eqref{tau}. Now an application of the  continuous mapping theorem directly yields the joint weak convergence \eqref{vntn_joint}. This implies the statement of Theorem~\ref{thm_relevantfunc} when $\int \mu^2(t) dt > 0$ after some simple computations.

{If $\int \mu^2(t) dt = 0$ it is possible to prove that $\hat \T_n = o_\Prob(1), \hat{\mathbb{V}}_n = o_\Prob(1)$. This implies
\[
\lim_{n \to \infty}
\mathbb{P} \Big( \hat \T_n > \Delta + q_{1-\alpha}(\mathbb{W}) {\hat{\mathbb{V}}_n}  \Big)
= \lim_{n \to \infty} \mathbb{P} \Big( o_\Prob(1)  > \Delta \Big) = 0 \, ,
\]
where we used that $\Delta > 0$ is fixed.
}

\begin{remark} 
{\rm 
\label{remdelta0}
In general, the rejection rule in~\eqref{testrelone} does \textit{not} lead to an asymptotic level $\alpha$ test when $\Delta = 0$. To see this note that for $\Delta = 0$ the null hypothesis contains only one point $\mu \equiv 0$ and we reject the null when $\hat{\mathbb{T}}_n/\hat {\mathbb{V}}_n > q_{1-\alpha}(\mathbb{W})$. However, a slight extension of the arguments given in the proof of~\eqref{haupt} shows that for $\mu \equiv 0$ we have 
\[
\frac{\hat{\mathbb{T}}_n}{\hat{\mathbb{V}}_n} \stackrel{\mathcal{D}}{\rightarrow} \tilde{\mathbb{W}} := \frac{\int_T \Gamma^2(t,1) dt}{ \Big\{ \int_{0}^1 \Big(\int_T \Gamma^2(t,\lambda) dt - \lambda^2 \int_T \Gamma^2(t,1) dt\Big)^2 \nu(d\lambda) \Big\}^{1/2}}
\]  
where $\Gamma(t,\lambda)$ is a centered Gaussian process with covariance function
\[
{\rm 
Cov} \big(\Gamma(t,\lambda),\Gamma(s,\lambda')\big) = (\lambda \wedge \lambda')C(s,t),
\]
where $C$ is  the long-run covariance operator defined in \eqref{approx4}.
The distribution of $ \tilde{\mathbb{W}} $ does not match that of $\mathbb{W}$ and is not pivotal. Hence a test based on rejecting $H_0: \mu = 0 $ using the decision rule~\eqref{testrelone} will not have asymptotic 
level $\alpha$.
}
\end{remark}

\begin{remark} \label{rem2} 
{\rm  A test for the hypotheses of equivalence
\begin{equation} \label{nullonesim}
H_0: \int_T \mu^2(t) dt > \Delta  \quad\mbox{ versus  }\quad H_1: \int_T \mu^2(t) dt \leq    \Delta
\end{equation}
can be obtained similarly. The {null hypothesis of a relevant difference}
 in \eqref{nullonesim} is rejected, when
\begin{align*} 
\quad \hat \T_n \leq  \Delta + q_{\alpha}(\mathbb{W}) {\hat{\mathbb{V}}_n}
\, ,
\end{align*}
where $\hat \T_n$ and $\hat{\mathbb{V}}_n$ are defined in \eqref{tstatone} and  \eqref{vn}, respectively and $q_{\alpha}(\mathbb{W})$ is the $\alpha$-quantile of the distribution of $\mathbb{W}$ defined in \eqref{Wdef}. Similar arguments as given in the proof of Theorem~\ref{thm_relevantfunc} show that this test is an asymptotic level $\alpha$ and consistent test for the hypotheses \eqref{nullonesim}, that is
\begin{align*}
\lim_{n \to \infty}
\mathbb{P} \Big( \hat \T_n \leq  \Delta + q_{\alpha}(\mathbb{W}) {\hat{\mathbb{V}}_n}  \Big)
&=
\begin{cases}
1 & \text{ if }  \int_T \mu^2(t) dt < \Delta \, ,   \\
\alpha & \text{ if }  \int_T \mu^2(t) dt = \Delta \,   \text{ and }  \tau^{2} >0,   \\
0 & \text{ if }  \int_T \mu^2(t) dt > \Delta \, .
\end{cases}
\end{align*}
The details are omitted for the sake of brevity.
}
\end{remark}

{
\begin{remark} \label{rem3} {\rm
As pointed out by a referee it is of interest to compare  the test \eqref{testrelone} based  on self-normalization
with a corresponding test using an estimate of the long-run variance.  For this purpose note that such a test rejects
the {null hypothesis of no relevant difference} \eqref{nullone}, whenever
\begin{equation}
\label{testlongrun}
\hat \T_n  >  \Delta+  u_{1-\alpha}   \frac{\hat \tau_{n} }{\sqrt{n} } \, ,
\end{equation}
where $ u_{1-\alpha} $ is the $(1-\alpha)$-quantile of the standard normal distribution and $ \hat \tau^{2}_n$ is an appropriate estimator
of the long-run variance  \eqref{tau}.  In the case of one sample this is still relatively easy. For example, one
could use
\begin{equation} \label{lrve}
\hat \tau^2_{n} = 4\int_0^{1}\int_0^{1}  S_{n} (s,1) S_{n} (t,1) \hat C_{n} (s,t) ds \, dt \, ,
\end{equation}
where $S_{n} (t,1) $ is defined in \eqref{parsum} and $\hat C_{n}$ is an appropriate estimator of the long-run covariance operator. {A numerical illustration of this approach in comparison with self-normalization can be found in Section~\ref{sec:longrunvar}.} 
}
\end{remark}
 }

\
\subsection{Two sample problems} \label{sec22}

Throughout this section let $\{X_n\}_{n\in \Z}, \{Y_n\}_{n\in \Z} $ denote two strictly stationary functional time series with values in $L^2(T)$. Assume that we observe finite stretches, say $X_1,...,X_m$ and $Y_1,...,Y_n$ from $\{X_n\}_{n\in \Z}$ and $\{Y_n\}_{n\in \Z} $.
 Denote by $\mu_1 =\E[X_1]$ and $\mu_2 =\E[Y_1]$ the corresponding mean functions, by $D(t) =\mu_1(t) - \mu_2(t) $ their difference and define the partial sum
\[
D_{m,n} (t,\lambda) := \frac{1}{m} \sum_{j=1}^{\lfloor m\lambda \rfloor} X_j(t) - \frac{1}{n} \sum_{j=1}^{\lfloor n\lambda \rfloor} Y_j(t) \, .
\]
From this  definition we see that
\begin{equation}  \label{expappr}
\E [ D_{m.n} (t,\lambda)] = \lambda D(t) + O( (m \wedge  n )  ^{-1}) \, .
\end{equation}
For the sake of brevity we restrict ourselves to the problem of testing the relevant hypotheses
\begin{equation} \label{nulltwo}
H_0: \int_T D ^2(t) dt \leq  \Delta  \quad\mbox{ versus  }\quad H_1: \int_T D^2(t) dt >  \Delta \, ,
\end{equation}
where $\Delta$ is a pre-specified threshold. A corresponding test  for the hypotheses of equivalence can be derived along the lines given in Remark \ref{rem2}. Following the discussion in Section \ref{sec21} we propose to reject the
{null hypothesis of no relevant difference} in
\eqref{nulltwo}, whenever
\begin{align} \label{testreltwo}
\quad \hat \D_{m,n} > \Delta + q_{1-\alpha}(\mathbb{W}) {\hat{\mathbb{V}}_{m,n}}
\, ,
\end{align}
where $q_{1-\alpha}(\mathbb{W})$ is the $(1-\alpha)$-quantile of the distribution of the random variable $\mathbb{W}$ in~\eqref{Wdef}.
The statistics $ \hat \D_{m,n}$ and $ \hat{\mathbb{V}}_{m,n}$ are defined by
\begin{align} \label{dmn}
\hat \D_{m,n}    &= \int_T  D_{m,n}^2(t,1) dt ~, \\
 \hat{\mathbb{V}}_{m,n} &= \Big( \int_0^1  \Big[ \int_T D_{m,n}^2(t,\lambda ) dt - \lambda^2 \int_T D_{m,n}^2(t,1) dt\Big]^2 \nu (d\lambda) \Big)^{1/2}~,
 \label{vmn}
\end{align}
respectively, where $\nu$ is a probability measure on the interval $(0,1)$. 
The asymptotic properties of this test procedure will be established under the following assumptions.

\medskip

\begin{itemize}
\item [(B1)] The sample sizes satisfy: $m \to \infty $ and $n\to \infty $  and $m/(m+n) \to \rho \in (0,1)$.
\item [(B2)] The processes $\{X_n \}_{n\in \Z} $ and $\{Y_n \}_{n\in \Z}$ are independent and satisfy assumptions (A1) - (A4) stated in Section \ref{sec21} with $\E[X_1] = \mu_1, \E[Y_1] = \mu_2$.
\end{itemize}
{We also define the quantity 
$$
\tau^2_{D} =  4 \int_{T}\int_{T} D(s)D(t) \big (\tfrac{1}{\rho} C_X(s,t) + \tfrac{1}{1-\rho} C_Y(s,t) \big ) ds dt,
$$
where $C_X$ and $C_Y$ are the long-run covariance operators corresponding to  the processes  $\{ X_{n}\}_{n\in \Z}$ and $\{ Y_{n}\}_{n\in \Z}$, respectively.
}

\begin{theorem} \label{thm_2s}
Assume that $\Delta > 0$. Under assumptions (B1)-(B2) the test decision given in~\eqref{testreltwo} satisfies
\begin{align*}
\lim_{n \to \infty}
\mathbb{P} \Big( \hat \D_{m,n} > \Delta + q_{1-\alpha}(\mathbb{W}) {\hat{ \mathbb{V}}_{m,n}} \Big)
&=
\begin{cases}
0 & \text{ if }  \int_T D^2(t) dt < \Delta \, ,
\\
\alpha & \text{ if }  \int_T D^2(t) dt = \Delta    \text{ and  }  \tau_{D}^{2}  >0 \, ,
\\
1 & \text{ if }  \int_T D^2(t) dt > \Delta \, .
\end{cases}
\end{align*}
\end{theorem}

We note that similarly to the one-sample case it can be shown that the rejection rule in~\eqref{testreltwo} does not lead to an asymptotic level $\alpha$ test when $\Delta = 0$.

\begin{remark}
\rm \label{knownCP} ~~\\
(a)  The statement in Theorem~\ref{thm_2s} continues to hold if the observations $X_i,Y_i$ are generated according to $X_i = \mu_1 + f_1(\eps_i,\eps_{i-1,...}), i=1,...,n$ and $Y_i = \mu_2 + f_2(\eps_{n+i},\eps_{n+i-1,...}), i=1,...,m$ where $(\eps_j)_{j \in \Z}$ denotes an i.i.d. sequence of $\mathcal{S}$-valued functions with the property that $\eps_j(t,\omega)$ is jointly measurable as in (A2) and $f_1,f_2: \Sc^\infty \to L^2([0,1])$ are functions such that the processes $(f_1(\eps_i,\eps_{i-1,...}))_{i \in \Z}$ and $(f_2(\eps_i,\eps_{i-1,...}))_{i \in \Z}$ satisfy conditions (A3) and (A4). This essentially corresponds to the setting discussed in Section~\ref{sec3} when the change point location is known. 

(b) 
A test based on  estimation  of the  long-run variance of the  statistic
$ \hat \D_{m,n}  $
can be constructed along the lines  given  in Remark \ref{rem3}.  The details are omitted for the sake of brevity.
\end{remark}

{
\begin{remark} \label{rem2b}
{\rm  
As pointed out by the Associate Editor the proposed way of self normalization is not unique and one could also think about alternative constructions. For instance, one could also use the statistics
\begin{align}
\label{alt1}
\hat{\mathbb{V}}_{m,n}^\star &= \nu\text{-}\esssup_{\lambda\in[0,1]} \Big|
\int_T D_{m,n}^2(t,\lambda ) dt
- \lambda^2 \int_T D_{m,n}^2(t,1) dt\Big| \\
\label{alt2}
\hat{\mathbb{V}}_{m,n}^{\star\star} &= \int_0^1  \Big|
\int_T D_{m,n}^2(t,\lambda ) dt
- \lambda^2 \int_T D_{m,n}^2(t,1) dt\Big| \nu (d\lambda)
\end{align}
in  the decision rule \eqref{testrelone} if the quantile $q_{1-\alpha}(\mathbb{W})$ is
replaced by the $(1-\alpha)$-quantile of the random variables
\begin{equation}\label{Wdefalt}
\mathbb{W}^{\star} := \frac{\Bb(1)}{ \nu\text{-}\esssup_{\lambda\in[0,1]} |
\lambda ( \Bb(\lambda)
- \lambda \Bb^{2}(1))  |} ~,~~
\mathbb{W}^{\star\star} := \frac{\Bb(1)}{ \int_0^1 |
\lambda ( \Bb(\lambda)
- \lambda \Bb^{2}(1))  |  \nu (d\lambda)}
\end{equation}
respectively.  The self-normalizing factors  \eqref{alt1} and  \eqref{alt2} might have some advantages for
heavy-tailed data. However, it will be demonstrated in Section \ref{sec52} that the  finite sample  properties
of these two alternative tests  are very similar to those of the test \eqref{testreltwo}.
}
\end{remark}
}

\section{Relevant change points in {the mean function}}
\label{sec3}
\def\theequation{3.\arabic{equation}}
\setcounter{equation}{0}

In this section we consider data that are generated from the following (triangular array) model
\begin{equation}\label{modcp}
X_i =
\begin{cases}
\mu + f_1(\eps_i,\eps_{i-1,...}) & \text{ if } i \leq N\theta_0 \, ,
\\
\mu + \delta + f_2(\eps_i,\eps_{i-1,...}) & \text{ if } i > N\theta_0 \, .
\\
\end{cases}
\end{equation}
Here $\mu, \delta$ denote deterministic but unknown elements in $L^2(T)$ and $\theta_0 \in (0,1)$ is fixed but unknown. Moreover, $(\eps_j)_{j \in \Z}$ denotes an i.i.d. sequence of $\mathcal{S}$-valued functions with the property that $\eps_j(t,\omega)$ is jointly measurable as in (A2) and $f_1, f_2: \Sc^\infty \to L^2(T)$ are functions such that the processes $(f_1(\eps_i,\eps_{i-1,...}))_{i \in \Z}$ and $(f_2(\eps_i,\eps_{i-1,...}))_{i \in \Z}$ satisfy conditions (A3) and (A4). This setting is general enough to allow for the whole distribution of the observed functional data to change together with their mean.

We aim to construct a test for the relevant hypothesis
\begin{equation} \label{nullcp}
H_0: \int_T \delta^2(t) dt \leq  \Delta  \quad\mbox{ versus  }\quad H_1: \int_T \delta^2(t) dt > \Delta
\end{equation}
where $\Delta$ is a pre-specified threshold. Note that for known $\theta_0$ a test for $H_0$ can be constructed in a similar fashion as in Section~\ref{sec22}. In this section, we will prove that replacing the known change point by an estimator also leads to an asymptotic level $\alpha$ test for the hypotheses in \eqref{nullcp}. 
To this end we fix a trimming parameter $\eps \in [0,1/2)$ and define the estimator of the unknown change point $\theta_0$ as
\begin{equation}\label{eq:hattheta}
\hat{\theta} := \frac{1}{N}{\argmax_{\floor{N\eps}+1 \leq k \leq N-\floor{N\eps}}} \hat{f}(k) \, ,
\end{equation}
where $\hat{f}(0)=\hat{f}(N)=0$ and for $k=1,...,N-1$
\begin{align} \label{hol0}
\hat{f}(k) := \frac{k}{N}\Big(1-\frac{k}{N}\Big) \int_T \Big( \frac{1}{k} \sum_{j=1}^{k} X_j(t)
  - \frac{1}{N - k} \sum_{j = k + 1}^{N} X_j(t) \Big)^2 dt \, .
\end{align}
{Our first result shows that the estimator $\hat{\theta} $ is consistent.
\begin{prop}  \label{propcpest}
If the data is generated according to model~\eqref{modcp}, $\int\delta^2(t) dt > 0$, $\theta_0 \in (\eps,1-\eps)$, and the assumptions described right below~\eqref{modcp} are satisfied,
 then
\begin{equation}\label{eq:ratecp}
\hat \theta = \theta_0 + o_{\mathbb{P}}(N^{-1/2}) \, .
\end{equation}
\end{prop}
}
Next we introduce the test statistic. For arbitrary $\theta \in [1/N,1)$ define
\begin{align*}
  D_N^{cp}(t, \lambda, \theta) := \frac{1}{\lfloor N \theta \rfloor}
  \sum_{j=1}^{\lfloor \lambda \lfloor \theta N \rfloor \rfloor} X_{j}(t)
  - \frac{1}{N - \lfloor N \theta \rfloor}
  \sum_{j=\lfloor \theta N \rfloor + 1}^{\lfloor \theta N \rfloor + \lfloor \lambda (N - \lfloor \theta N \rfloor ) \rfloor} X_{j}(t) \, .
\end{align*}
Following the developments in Section~\ref{sec22} let
\begin{align} \label{dmncp}
\hat \D_{N}^{cp} &= \int_T  D_{N}^{cp}(t,1,\hat\theta)^2 dt \, ,
\\
\hat{\mathbb{V}}_{N}^{cp} &= \Big( \int_0^1  \Big[ \int_T D_{N}^{cp}(t,\lambda,\hat\theta)^2 dt - \lambda^2 \int_T D_{N}^{cp}(t,1,\hat\theta)^2 dt\Big]^2 \nu (d\lambda) \Big)^{1/2} \, ,
 \label{vmncp}
\end{align}
respectively, where $\nu$ is a probability measure on the interval $(0,1)$. The test for $H_0$ takes the form
\begin{align} \label{testreltwocp}
\quad \hat \D_{N}^{cp} > \Delta + q_{1-\alpha}(\mathbb{W}) {\hat{\mathbb{V}}_{N}^{cp}}
~,
\end{align}
where $q_{1-\alpha}(\mathbb{W})$ is the $(1-\alpha)$-quantile of the distribution of the random variable $\mathbb{W}$ in~\eqref{Wdef}. This test decision is justified in the following theorem. For its precise statement we define the quantity 
\begin{equation}\label{hcp2}
\tau^2_{\delta, \theta_{0}} = 4 \int_{T}\int_{T} \delta (s) \delta (t) \big (\tfrac{1}{\theta_{0}} K_1(s,t) + \tfrac{1}{1-\theta_{0}} K_2(s,t) \big ) ds dt,
\end{equation}
where for $j=1,2$  
\[
K_j(s,t) = \sum_{h \in \mathbb{Z}} \cov(\eta_0^{(j)}(s),\eta_h^{(j)}(t)) \, , \quad (j=1,2) \, 
\]
is  the long-run covariance kernel of $\{  \eta_{i}^{(j) } \}_{i \in \Z} =
\{ f_j(\eps_i,\eps_{i-1, \ldots }) \}_{i \in \Z}$. 

\begin{theorem} \label{thmcp}
Assume $\Delta > 0$. If the data is generated according to model~\eqref{modcp}, $\theta_0 \in (\eps,1-\eps)$, and the assumptions described right below~\eqref{modcp} hold, then the test decision in~\eqref{testreltwocp} satisfies
\begin{align*}
\lim_{n \to \infty}
\mathbb{P} \Big( \hat \D_{N}^{cp} > \Delta + q_{1-\alpha}(\mathbb{W}) {\hat{\mathbb{V}}_{N}^{cp}}  \Big)
&=
\begin{cases}
0 & \text{ if }  \int_T \delta^2(t) dt < \Delta \, ,
\\
\alpha & \text{ if }  \int_T \delta^2(t) dt = \Delta \,  \text{ and }  \tau^2_{\delta, \theta_{0}} >0 \, ,
\\
1 & \text{ if }  \int_T \delta^2(t) dt > \Delta \, .
\end{cases}
\end{align*}
\end{theorem}

Similarly to the one-sample and two-sample case, the rejection rule in~\eqref{testreltwocp} does not lead to an asymptotic level $\alpha$ test when $\Delta = 0$.

The proof of Proposition \ref{propcpest} and Theorem \ref{thmcp} is technically difficult and deferred to Section~\ref{sec63}, but the main idea is as follows.
A straightforward calculation shows that the processes $\hat{\mathbb{D}}_{N}^{cp}$ and $\hat{\mathbb{V}}_{N}^{cp}$
in \eqref{dmncp}  and  \eqref{vmncp}  are continuous functionals of the process
$$
  \Z_N(\lambda, \hat \theta) = \sqrt{N} \int_T \big( D_N^{cp}(t,\lambda,  \hat \theta)^2    - \lambda^2 \delta(t)^2 \big) dt \, .
$$
Using Proposition \ref{propcpest} it can be shown that
$$
  \sup_{\lambda \in [0,1]} |\Z_N(\lambda,\theta_0) - \Z_N(\lambda,\hat{\theta})|
  = o_{\mathbb{P}}(1) \, ,
$$
where $\theta_0$ is the true change point.  We can then establish the weak convergence
$$
  \big\{ \Z_N(\lambda,\theta_0) \big\}_{\lambda \in [0,1] }
  \weak \big\{ \lambda \tau_{\delta,\theta_0} \Bb(\lambda) \big\}_{\lambda\in [0,1]} \, ,
$$
where $\tau_{\delta,\theta}^2$ defined in~\eqref{hcp2}.
Using the continuous mapping theorem  we then find
$$
\frac{\hat{\mathbb{D}}_{N}^{cp}}{\hat{\mathbb{V}}_{N}^{cp} }  \Dkonv \W \, ,
$$
where the random variable  $\mathbb{W}$ is defined in~\eqref{Wdef}. When $\int_T \delta^2(t) dt  > 0 $, the assertion of Theorem~\ref{thmcp} now follows directly. { In the remaining case  $\int_T \delta^2(t) dt = 0 $ one can show $\hat{\mathbb{D}}_{N}^{cp} = o_{\Prob}(1), \hat{\mathbb{V}}_{N}^{cp}  = o_\Prob(1)$ and the assertion follows with the same arguments as given in Section~\ref{sec21}. }

\begin{remark} \label{rem6}
\rm
In this remark we briefly explain how -  in principle -  a test can be constructed using an appropriate estimate of the long-run variance
of the statistic $\hat \D_{N}^{cp}$ in \eqref{dmncp}. 
A careful inspection of the proof of Theorem \ref{thmcp}, (see in particular the discussion at the end of the proof of~\eqref{Z_N knownCP} in the online supplement) 
shows that 
\begin{equation}\label{hcp1}
\sqrt{N} \Big({\hat{\mathbb{D}}}^{cp}_N - \int_T  \delta^2 (t) dt \Big) \stackrel{\mathcal{D}}{\longrightarrow} \mathcal{N} (0, \tau^2_{\delta,\theta_0}) \, , 
\end{equation}
where the asymptotic variance is  defined in \eqref{hcp2}.
If $\hat \tau^2_N$ denotes an estimator of $ \tau^2_{\delta,\theta_0}$, an asymptotic level $\alpha$ test is obtained by rejecting the null hypothesis in \eqref{nullcp}, whenever
\begin{equation*}
{\hat {\mathbb{D}}}^{cp}_N > \Delta + \frac {\hat \tau_N}{\sqrt{N}} u_{1 - \alpha} \, .
\end{equation*}
One possibility to estimate $\tau_{\delta,\theta_0}^2$ is to replace $\theta_0$ by the estimator $\hat \theta$ defined in~\eqref{eq:hattheta} and to use appropriate plug-in estimators for the covariance kernels $K_j$ and unknown difference $\delta$ based on the sub-samples $\{X_i: i \leq \floor{N\hat \theta}\}$ and $\{X_i: i > \floor{N\hat \theta}\}$. Details are omitted for the sake of brevity. 
\end{remark}

{
\begin{remark} \label{rem7}
{\rm
The extension of the methodology to the analysis of multiple change point problems  is of great practical interest and briefly indicated
here.
 Let $0 < \theta_1 < \theta_2 < \ldots < \theta_K  < 1$ denote the unknown change points and assume  that
\begin{equation} \label{modmult}
X_i = \mu + \delta_{j-1} + f(\varepsilon_i, \varepsilon_{i-1}, \ldots) \quad \text{if} \quad \lfloor N\theta_{j-1}\rfloor +1 \leq i \leq \lfloor N \theta_j\rfloor; \quad 1 \leq j \leq K+1
\end{equation}
where $\delta_0 = 0; \theta_0=0, \theta_{K+1}=1$. For the sake of simplicity we consider the same function $f$ as filter for the error process on $j$'th segment (in contrast to model \eqref{modcp}) and further consider $K$ as known. Defining the vector of integrated squared differences 
\[
\Psi := \Big(\int_T \delta_1^2(t) dt, \int_T \{\delta_2(t) - \delta_{1}(t)\}^2 dt,...,\int_T \{\delta_K(t) - \delta_{K-1}(t)\}^2 dt \Big)^\top \, ,
\]
there are several possibilities to formulate relevant hypotheses in this setting. Here we will focus on 
\begin{align} \label{L2}
H_0^{L^2}: \sum_{j=1}^K \Psi_j \leq \Delta
\end{align}
which corresponds to the null that the sum of all integrated squared changes does not exceed a threshold $\Delta$ and
\begin{align} \label{max}
H_0^{\infty}: \max_{j=1}^K \Psi_j \leq \Delta
\end{align}
meaning that no single integrated squared change exceeds $\Delta$. Note that $H_0^{\infty}$ can be equivalently formulated as the intersection of the following hypotheses
\begin{equation}\label{remh1}
 H^{(j)}_0 :  \Psi_j \leq \Delta \qquad j = 1,\ldots,K \, .
\end{equation}
For testing either of the hypotheses, we propose to first estimate the multiple change points by adapting one of the commonly used methods such as binary segmentation or wild binary segmentation [see for example \cite{vostrikova1981,harchlevy2010,fryzlewicz2014,zhanglav2018} among many others] to dependent functional data. The resulting estimator is denoted  by $\hat \theta=(\hat \theta_1, \ldots, \hat \theta_K)$ and we put $\hat \theta_{K+1}=1, \hat \theta_0=0$. 
Next we consider generalizations of the statistics $\hat {\mathbb{D}}^{cp}_N$ and $\hat {\mathbb{V}}^{cp}_N$ in \eqref{dmncp} and \eqref{vmncp} on each segment
\[
S_j := \{ X_i \mid \lfloor N \hat  \theta_{j-1}  \rfloor +1 \leq i \leq \lfloor N  \hat \theta_{j+1} \rfloor \}  \quad (j=1,\ldots,K) \, .
\]
More precisely, define $\hat N_j := \lfloor N \hat \theta_{j}\rfloor - \lfloor N \hat \theta_{j-1}\rfloor, j = 1,...,K+1$ as the sample size between the $(j-1)$'st and $j$'th estimated change point and let 
\begin{align*}
 \hat D^{cp}_{N,j} (t, \lambda, \hat \theta) = \frac {1}{\hat N_{j}}
  \sum^{\lfloor \lambda \hat N_{j} \rfloor}_{i= 1}  X_{\lfloor N \hat \theta_{j-1} \rfloor + i }  (t) -
 \frac {1}{\hat N_{j+1}}
  \sum_{i= 1}^{ \lfloor \lambda \hat N_{j+1}\rfloor}  X_{\lfloor N \hat \theta_{j} \rfloor + i }  (t) , \quad j = 1,...,K \, .
\end{align*}
Further, define
\begin{equation} \label{remDj}
\hat{\mathbb{D}}^{cp}_{N,j}(\lambda,\hat\theta)  := \int_T ( \hat D^{cp}_{N,j} (t,\lambda, \hat \theta))^2dt \qquad (j = 1,...,K) \, .
\end{equation}
With this preparation, we first discuss a test for the hypothesis $H_0^{L^2}$ 
in \eqref{L2}. Define
\begin{align} \label{remD}
\hat{\mathbb{D}}^{L^2}_{N}(\lambda,\hat\theta) &:= \sum_{j=1}^K \hat{\mathbb{D}}^{cp}_{N,j}(\lambda,\hat\theta)
\\ \label{remV}
\hat{\mathbb{V}}^{L^2}_{N}(\hat \theta) &:= \Big\{ \int_0^1 \Big[\hat{\mathbb{D}}^{L^2}_{N}(\lambda,\hat\theta) - \lambda^2\hat{\mathbb{D}}^{L^2}_{N}(1,\hat\theta) \Big]^2\nu(d\lambda) \Big\}^{1/2}.
\end{align}
We propose to reject $H_0^{L^2}$ whenever
\[
\hat{\mathbb{D}}^{L^2}_{N}(1,\hat\theta) > \Delta + q_{1-\alpha}(\mathbb{W}) \hat{\mathbb{V}}^{L^2}_{N}(\hat\theta) \, ,
\] 
where $q_{1-\alpha}(\mathbb{W})$ denotes the $(1-\alpha)$ quantile of the random variable $\mathbb{W}$ defined in~\eqref{Wdef}. In Section~\ref{proofrem} of the Online Supplement, we will show that this provides a consistent and asymptotic level $\alpha$ test for $H_0^{L^2}$ under the following conditions: 
\begin{enumerate} 
\item[(m1)] The data are generated from model~\eqref{modmult} with $f, \{\eps_i\}_{i \in \Z}$ satisfying (A2)--(A4).
\item[(m2)] The true number of change points is $K$ (i.e. all entries of $\Psi$ are non-zero and there are no other change points) and $\hat \theta_j = \theta_j + o_P(N^{-1/2})$, $(j=1,...,K)$.
\end{enumerate}
Those assumptions are made for the sake of a simpler presentation. It is possible to generalize (m1) to the case where the filter $f$ changes in each segment as in model~\eqref{modcp}. Similarly, (m2) can be weakened to i.e. $\Psi_j = 0$ for some values $j$. In this case the requirement $\hat \theta_j = \theta_j + o_P(N^{-1/2})$ for $j$ with $\Phi_j = 0$, is not realistic and has to be replaced by a different condition. Details are omitted for the sake of brevity.\\
Finally, we briefly comment on testing the hypothesis \eqref{max}  using self-normalization which is  a substantially  more challenging problem. Although it is possible to construct self-normalized test statistics for each of the hypotheses $H_0^{(j)}$ in~\eqref{remh1} separately, note that neighbouring segments $S_j, S_{j+1}$ overlap so that in the limit those self-normalized statistics become dependent. It can further be shown that, although the marginal distributions are pivotal, the resulting joint distribution is not pivotal any more. Therefore one option to construct a  test for  the hypothesis \eqref{max}    is to apply a multiple testing correction after testing each $H_0^{(j)}$ separately based on self-normalization. {Problems of this type have been discussed more intensively the context of testing simultaneously  several hypotheses of equivalence  of the form \eqref{equiv}
[see \cite{munk1999,wang1999}]. These references  indicate the  general difficulty  to construct tests for multiple precise hypotheses using the joint distribution of a vector of test statistics, and we leave a detailed investigation of this problem for future research.  }
} 
\end{remark}
}


\section{ Inference for covariance operators}\label{sec4}
\def\theequation{4.\arabic{equation}}
\setcounter{equation}{0}

In this section we briefly discuss extensions of the methodology in Sections~\ref{sec2} and~\ref{sec3} to test similar hypotheses 
regarding the covariance operators of functional time series. For the sake of brevity we omit the one sample case and focus on the two sample case and the  change point setting. 
The problem of testing the classical hypothesis  has found considerable attention in the literature and we refer
to \cite{pankramad2010}, \cite{frehorkokste2012},   \cite{Guo2016} and  \cite{papsap2016}, who developed methodology for comparing covariance operators 
for  independent functional data.  \cite{Pilavakis2019} proposed a test for the equality  of the  lag-$0$ auto-covariance operators of $K$ functional time series,
and \cite{sharwend2019} considered  bootstrap-based statistical inference for covariance operators of functional time series.   Change point analysis for   covariance 
operators  was developed by  \cite{Aston2012} and \cite{stoehr2019}, among others, while 
\cite{zhangshao2015} and \cite{AueRiceSonmez2018}  discussed statistical inference tools for the eigen-system of covariance operators.


We will repeatedly make use of the following strengthening of Assumptions (A3)-(A4) on a sequence of $L^2(T)$-valued random elements $\{\eta_j\}_{j \in\Z}$ that satisfy (A2):
\begin{itemize}
\item [(A3')]  $ \E \|  \eta_j \|^{4+\psi} < \infty  $ for some $\psi \in (0,1)$.
\item [(A4')] The sequence $( \eta_j)_{j\in \Z} $ can be approximated by $\ell$-dependent sequences $( \eta_{j,\ell} )_{j\in \Z}$ in the sense that for some $\kappa > 4+\psi$
\[ 
\sum_{\ell=1}^\infty \big ( \E \| \eta_{0} -\eta_{0,\ell} \|^{4+\psi} \big)^{1/\kappa} < \infty \, ,
\]
where $\eta_{j,\ell}$ is defined by
\begin{align*}
\eta_{j,\ell} &= f(\varepsilon_j, \varepsilon_{j-1}, \ldots
 \varepsilon_{j-\ell + 1},\boldsymbol{\varepsilon}_{j,\ell}^*) \\
 \boldsymbol{\varepsilon}_{j,\ell}^* &= (\varepsilon_{j,\ell,j-\ell }^*,\varepsilon_{j,\ell,j-\ell -1}^*, \ldots ) \, ,
 \end{align*}
and the random variables $\varepsilon_{j,\ell,k}^*$ are i.i.d. copies of $\varepsilon_{0}$, and independent of the sequence $( \varepsilon_j)_{j\in \Z}$.
\end{itemize}

\subsection{Two sample problem} \label{sec:2sCov}
{ Given two samples $X_1,\dots,X_m$ and $Y_1,\dots,Y_n$ from independent, strictly stationary functional time series $\{X_t\}_{t \in \Z}, \{Y_t\}_{t \in \Z}$, we are interested in testing the null hypothesis of no relevant difference in the covariance operators $C^X, C^Y$ where 
$$C^X(s,t) := {\E\big[(X_0(s)-\E[X_0(s)])(X_0(t)-\E[X_0(t)])\big]}$$ 
and $C^Y$ is defined similarly. Thus we investigate the hypotheses
\begin{align} \label{nullCov}
H_0^C:   d_{C} =  \int_T \int_T D_C^2(s,t) ds dt \leq  \Delta
\quad\mbox{ versus  }\quad
H_1^C:  d_{C}   >  \Delta \, ,
\end{align}
where $D_C(s,t) = C^X(s,t) - C^Y(s,t)$  denotes the difference of the covariance operators at the points $s,t \in T$.  
We reject the null hypothesis in \eqref{nullCov}, whenever
\begin{align} \label{eq:cov2s}
\hat \D_{m,n}^C > \Delta + q_{1-\alpha} (\mathbb{W}) \hat{\mathbb{V}}_{m,n} \, ,
\end{align}
where $q_{1-\alpha}(\mathbb{W})$ denotes the $(1-\alpha)$-quantile of the random variable $\mathbb{W}$ defined in~\eqref{Wdef}, 
\begin{align*}
  \hat \D_{m,n}^C    &= \int_T \int_T D_{m,n}^2(s,t,1) dsdt ~, \\
  \hat{\mathbb{V}}_{m,n}^C &=
  \Big( \int_0^1  \Big[ \int_T\int_T D_{m,n}^2(s,t,\lambda ) dsdt
  - \lambda^2 \int_T\int_T D_{m,n}^2(s,t,1) dsdt\Big]^2 \nu (d\lambda) \Big)^{1/2}
\end{align*}
and $D_{m,n}$ is the partial sum defined by
\begin{align*}
  D_{m,n}(s,t,\lambda) =& \frac{1}{m-1} \sum_{j=1}^{\lfloor m \lambda \rfloor}
  \Big (X_j(s)-\frac{1}{\lfloor m \lambda \rfloor\vee 1}\sum_{i=1}^{\lfloor m \lambda \rfloor} X_i(s)\Big)
  \Big(X_j(t)-\frac{1}{\lfloor m \lambda \rfloor\vee 1}\sum_{i=1}^{\lfloor m \lambda \rfloor} X_i(t)\Big) \\
  &- \frac{1}{n-1} \sum_{j=1}^{\lfloor n \lambda \rfloor}
  \Big(Y_j(s)-\frac{1}{\lfloor n \lambda \rfloor\vee 1}\sum_{i=1}^{\lfloor n \lambda \rfloor} Y_i(s)\Big)
  \Big(Y_j(t)-\frac{1}{\lfloor n \lambda \rfloor\vee 1}\sum_{i=1}^{\lfloor n \lambda \rfloor} Y_i(t)\Big) ~.
\end{align*}
The following result shows that  the decision rule in~\eqref{eq:cov2s}  defines a consistent asymptotic level $\alpha$-test
for the hypotheses $H_0^C$ versus $H_1^C$ provided that $\Delta >0$.
For  a precise statement, define
\begin{align*}
  \tau_{D_C}^2 =  4 \int_{T}\int_{T}\int_{T}\int_{T} D_C(s,t)D_C(s',t') \big (\tfrac{1}{\rho} C_X((s,t),(s',t')) + \tfrac{1}{1-\rho} C_Y((s,t),(s',t')) \big ) ds dt ds' dt' 
\end{align*}
where $\rho$ comes from assumption~(B1) and  $C_X$ and $C_Y$ denote 
the long-run covariance kernels corresponding to the time series 
$\{X_t\otimes X_t\}_{t\in\Z}$ and $\{Y_t\otimes Y_t\}_{t\in\Z}$, respectively.

\begin{theorem} \label{th:2scov}
Assume that $\Delta > 0$. Let assumption (B1) from Section~\ref{sec22} hold and assume that the functional time series $\{X_t\}_{t \in \Z}, \{Y_t\}_{t \in \Z}$ satisfy (A1) with means $\mu_X, \mu_Y$ and errors $\eta^X_j, \eta^Y_j$, (A2), and that $\eta^X_j, \eta^Y_j$ satisfy (A3'), (A4'). Finally, assume that $\nu$ puts no mass in a neighbourhood of zero. Then 
\begin{align*}
\lim_{n \to \infty}
\mathbb{P} \Big( \hat \D_{m,n}^C > \Delta + q_{1-\alpha}(\mathbb{W}) {\mathbb{V}_{n,m}^C} \Big)
&=
\begin{cases}
0 & \text{ if  ~} d_{C}  < \Delta \, ,
\\
\alpha & \text{ if  ~}d_{C}  = \Delta  \text{ and  } \tau_{D_C}^2 > 0 \, ,
\\
1 & \text{ if }  d_{C}   > \Delta \, .
\end{cases}
\end{align*}
\end{theorem}
}

Similarly to the setting of testing for a relevant difference in the means it can be shown that the rejection rule in~\eqref{eq:cov2s} does not lead to an asymptotic level $\alpha$ test when $\Delta = 0$.

\subsection{Change point problem} \label{sec:cpCov}
For an extension of the methodology to testing for relevant changes in the covariance structure of a time series, we will assume that data are generated from the model
\begin{equation}\label{modcp_cov}
X_i =
\begin{cases}
\mu + f_1(\eps_i,\eps_{i-1,...}) & \text{ if } i \leq N\theta_0 \, ,
\\
\mu + f_2(\eps_i,\eps_{i-1,...}) & \text{ if } i > N\theta_0 \, .
\\
\end{cases}
\end{equation}
Define $\eta_i^{(k)} := f_k(\eps_i,\eps_{i-1,...}), k=1,2$ and denote by $C_1, C_2$ the covariance operators of the process $X_i$ before and after the structural break. We are now interested in testing
the hypotheses 
\begin{equation} \label{nullcpCov}
H_0^{C}: d_{C}^{cp} = 
 \int_T\int_T \{C_1(s,t) - C_2(s,t)\}^2 dsdt \leq  \Delta  \quad\mbox{ versus  }\quad H_1^{C}: d_{C}^{cp}  > \Delta \, .
\end{equation}
Similarly to Section~\ref{sec3} we first construct an estimator for the unknown change point location $\theta_0$. To this end we define 
\[
\bar X_{\ell:k}(t) := \frac{1}{k-\ell{+1}} \sum_{i={\ell}}^k X_i(t)
\]
and consider the covariance estimators 
\begin{align*}
\hat C_{1:k}(s,t) &:= \frac{1}{k-1} \sum_{j=1}^{k}\big (X_j(s)-\bar X_{1:k}(s)\big )
  \big (X_j(t)-\bar X_{1:k}(t)\big ) \, ,
\\
\hat C_{k+1:N}(s,t) &:= \frac{1}{N-k-1} \sum_{j=k+1}^{N}\big (X_j(s)-\bar X_{k+1:N}(s)\big )
  \big (X_j(t)-\bar X_{k+1:N}(t)\big )
\end{align*} 
{for $k = 2,\dots,N-2$.}
Next fix a trimming parameter $\eps \in [0,1/2)$ and define the estimator 
\begin{equation*}
\hat{\theta}^{Cov} := \frac{1}{N}{\argmax_{\floor{N\eps}+1 \leq k \leq N-\floor{N\eps}}} \hat{f}^{Cov}(k) \, ,
\end{equation*}
where $\hat{f}^{Cov}(0)=\hat{f}^{Cov}(N)=0$ and for $k=1,...,N-1$
\begin{align} \label{hol0Cov}
\hat{f}^{Cov}(k) := \frac{k}{N}\Big(1-\frac{k}{N}\Big) \int_T\int_T \big( \hat C_{1:k}(s,t) - \hat C_{k+1:N}(s,t) \big)^2 dsdt \, .
\end{align}   
Following the approach in Section~\ref{sec3}, for arbitrary $\theta \in [2/N,1-1/N)$ define
\begin{multline*}
D_{N}^{cp,Cov}(s,t,\lambda, \theta) = \frac{1}{\lfloor N \theta \rfloor-1} \sum_{j=1}^{\lfloor \lfloor N \theta \rfloor \lambda \rfloor}
\Big(X_j(s)-\bar X_{1:\lfloor \lfloor N \theta \rfloor \lambda \rfloor}(s)\Big)
\Big(X_j(t)-\bar X_{1:\lfloor \lfloor N \theta \rfloor \lambda \rfloor}(t)\Big) 
\\
- \frac{1}{N - \lfloor N \theta \rfloor-1} \sum_{j=\lfloor \theta N \rfloor + 1}^{\lfloor \theta N \rfloor + \lfloor \lambda (N - \lfloor \theta N \rfloor ) \rfloor} 
\Big(X_j(s)-\bar X_{\lfloor N \theta \rfloor+1:(\lfloor  N \theta \rfloor + \lfloor \lambda (N - \lfloor  N \theta \rfloor)}(s)\Big) 
\\
\times
\Big(X_j(t)-\bar X_{\lfloor N \theta \rfloor+1:(\lfloor  N \theta \rfloor + \lfloor \lambda (N - \lfloor  N \theta \rfloor)}(t)\Big)  ~,
\end{multline*}
and consider the statistics 
\begin{align*}
\hat \D_{N}^{cp,Cov} &= \int_T\int_T  \big ( D_{N}^{cp,Cov}(s,t,1,\hat\theta)  \big)^{2}dsdt \, ,
\\
\hat{\mathbb{V}}_{N}^{cp,Cov} &= \Big( \int_0^1  \Big[ \int_T\int_T  
\big ( D_{N}^{cp,Cov}(s,t,\lambda,\hat\theta)\big )^2 dt - \lambda^2 \int_T\int_T  \big ( D_{N}^{cp,Cov}(s,t,1,\hat\theta)^2 dsdt\big ) \Big]^2 \nu (d\lambda) \Big)^{1/2} \, ,
\end{align*}
where $\nu$ is a probability measure on the interval $(0,1)$. The test for the hypotheses \eqref{nullcpCov}  rejects  $H^C_0$, whenever 
\begin{align} \label{testreltwocpCov}
\quad \hat \D_{N}^{cp,Cov} > \Delta + q_{1-\alpha}(\mathbb{W}) {\hat{\mathbb{V}}_{N}^{cp,Cov}}
~,
\end{align}
where $q_{1-\alpha}(\mathbb{W})$ is the $(1-\alpha)$-quantile of the distribution of the random variable $\mathbb{W}$ in~\eqref{Wdef}. 

For a precise statement of the next theorem, we define  
$$
  \tau^2_{\delta_C, \theta_{0}} = 4 \int_{T}\int_{T}\int_{T}\int_{T} \delta_C (s,t) \delta_C (s',t') \big (\tfrac{1}{\theta_{0}} K_1((s,t),(s',t')) + \tfrac{1}{1-\theta_{0}} K_2((s,t),(s',t')) \big ) ds dt ds'dt',
$$
  where $\delta_C(s,t) = C_2(s,t) - C_1(s,t)$ and $K_j $ is  the long-run covariance kernel of $\{  \eta_{i}^{(j)}\otimes \eta_{i}^{(j)} \}_{i \in \Z}$.

\begin{theorem}
Assume that the data is generated according to model~\eqref{modcp_cov}, $\theta_0 \in (\eps,1-\eps)$, and that $(\eps_j)_{j \in \Z}$ together with $f_1,f_2$ satisfy (A2), (A3'), (A4'). Further assume that $\nu$ puts no mass in a neighbourhood of zero. Then the decision rule  in~\eqref{testreltwocpCov} satisfies
\begin{align*}
\lim_{n \to \infty}
\mathbb{P} \Big( \hat \D_{N}^{cp,Cov} > \Delta + q_{1-\alpha}(\mathbb{W}) {\hat{\mathbb{V}}_{N}^{cp,Cov}}  \Big)
&=
\begin{cases}
0 & \text{ if  ~}   d_{C}^{cp} < \Delta \, ,
\\
\alpha & \text{ if ~} d_{C}^{cp} = \Delta  \text{ and  } 
\tau^2_{\delta_C, \theta_{0}} > 0 \, ,
\\
1 & \text{ if ~}  d_{C}^{cp} > \Delta \, .
\end{cases}
\end{align*}
\end{theorem}

Similarly to the setting of testing for a relevant change-point in the mean it can be shown that the rejection rule in~\eqref{testreltwocpCov} does not lead to an asymptotic level $\alpha$ test when $\Delta = 0$.


\section{Finite sample properties}
\label{sec5}
\def\theequation{5.\arabic{equation}}
\setcounter{equation}{0}

In this section we illustrate the finite sample properties of the new procedures by means of a simulation study. Note that one has to specify the measure $\nu$ used in the definition of the
normalizer \eqref{vn}, \eqref{vmn} and \eqref{vmncp} and we use
$
\nu = \frac{1}{19}\sum_{i=1}^{19} \delta_{i/20}
$
throughout this section if not mentioned otherwise; here $\delta_\lambda$ denotes the Dirac measure at the point $\lambda\in [0,1]$. For example, for this choice  the quantity $\hat{\mathbb{V}}_n$ defined in \eqref{vn} is given by
\begin{align*}
\hat{\mathbb{V}}_n&= \Big[\frac{1}{19} \sum_{i=1}^{19} \Big(\int_T S_n^2 \big (t,\tfrac{i}{20} \big )dt -
\big (\tfrac{i}{20} \big )^2 \int_T S_n^2(t,1)dt\Big)^2 \Big]^{1/2}
 \end{align*}
and the other expressions are obtained similarly. In the following sections we discuss the one sample case, the two sample case and change point detection separately. All results are based on $1000$ simulation runs.

\subsection{One sample problems} \label{sec51}

We consider a process \(\{ X_n\}_{n\in \N} \) with expectation function
\begin{align} \label{err0}
\mu(t) = \sqrt{2 \delta}\sin(2\pi t)
\end{align}
 and  different error processes, where we investigate a similar scenario as
\cite{aueDubartNorinhoHormann2015} (see Sections 6.3 and 6.4 in the latter paper).
More precisely,
let $X_n = \mu  + \varepsilon_n$ and consider $B$-spline basis functions \(b_1,\dots , b_D\) (\(D \in \mathbb{N} \)). Define the linear space $\mathbb{H}=$ span$\{ b_1,\dots , b_D \}  \subset  L^2([0,1]) $ and independent processes  \(\eta_1, \dots, \eta_n \in \mathbb{H} \) by
\begin{align} \label{eq:errors}
  \eta_j = \sum_{i=1}^D N_{i,j} b_i ~~~~ (j=1,\dots , n) \, ,
\end{align}
 where \(N_{1,1},N_{2,1}, \dots, N_{D,n} \) are independent $N(0,\sigma_i^2)$ ($i=1,\ldots , D$; $j=1,\ldots , n$) distributed random variables.
Our first example considers independent error processes of the form
\begin{align} \label{err1}
 \varepsilon_j=  \eta_j ~~~(j\in \Z) \, ,
\end{align}
while the second example investigates a functional moving average \(fMA(1)\) process
 given  by
 \begin{align} \label{err2}
  \varepsilon_j =  \eta_j + \Theta  \eta_{j-1} ~~~(j \in \Z) \, .
  \end{align}
   Here
the operator $\Theta: \mathbb{H} \to \mathbb{H} $ (acting on finite dimensional spaces) is   defined by the matrix
 \(\Theta =  ( \Theta_{ij}  )_{i,j=1}^D \in \mathbb{R}^{D\times D} \), where the entries $ \Theta_{ij}$
are normally distributed with mean zero and standard deviation \( \kappa \sigma_i \sigma_j \) and $\kappa$ is a scaling factor such that the resulting matrix \(\Theta \) has (induced) spectral norm equal to \(0.7\). The operator $\Theta$ is newly generated in every simulation run (see Sections 6.3 and 6.4 in \cite{aueDubartNorinhoHormann2015} for a similar approach) and
we use $D=21$. The third error structure under consideration are independent Brownian Bridges.

\begin{figure}[H]
{  \centering
\includegraphics[width=5.65cm,height=5.25cm]{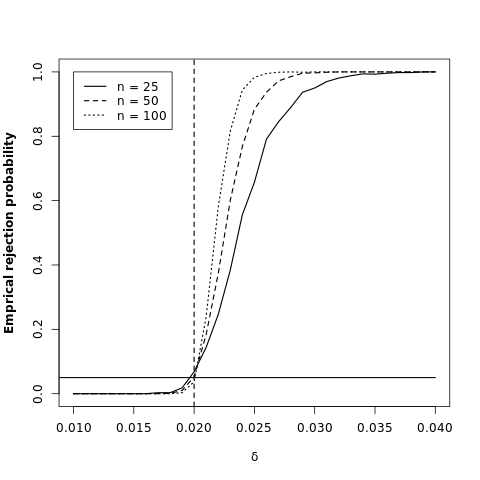}
 \includegraphics[width=5.65cm,height=5.25cm]{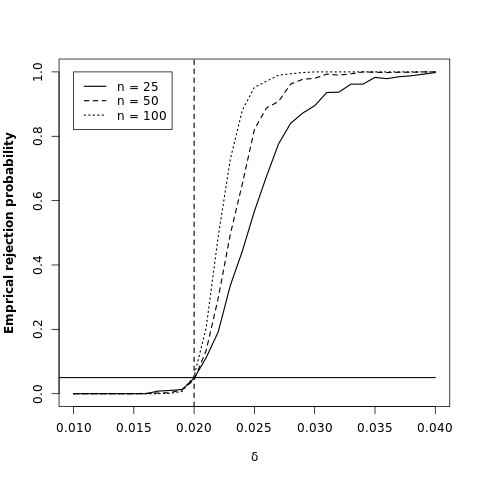}
 \includegraphics[width=5.65cm,height=5.25cm]{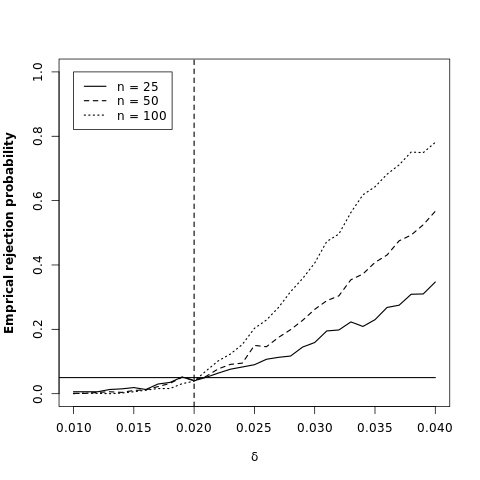}
\caption{\it Simulated rejection probabilities of the test \eqref{testrelone} for the relevant hypotheses \eqref{nullone}
with  $\Delta =0.02$. The mean function
is given by \eqref{err0} and different error processes are considered.
Left  panel:  independent  error processes defined by \eqref{err1}. Middle panel: \(fMA(1)\) processes defined by \eqref{err2}.
Right panel: independent Brownian Bridges.\label{fig1}}
}
\end{figure}

In Figure \ref{fig1} we display the simulated rejection probabilities of the test \eqref{testrelone} for the hypotheses \eqref{nullone},
where $\Delta =0.02$, which corresponds to the value $\delta=0.02$ in model \eqref{err0}. These results show a pattern which is in line with the theoretical findings in Theorem~\ref{thm_relevantfunc}. For example at the boundary of the null hypotheses, i.e. for $\delta=\Delta=0.02$, the simulated level is close to the nominal level. In the interior of the null hypothesis ($\delta < \Delta$) the simulated rejection probabilities are strictly smaller than $\alpha=0.05$, while they are strictly larger than $0.05$ in the interior of the alternative, i.e. $\delta  >  \Delta$.

 \subsubsection{Estimating the long-run variance} \label{sec:longrunvar}

It is of interest to  compare the procedure based on self-normalization with the test   \eqref{testlongrun} defined in Remark~\ref{rem3},
which uses an estimate  of the long-run variance. For this comparison, we also implement the (practically infeasible) test which rejects the null hypothesis of no relevant difference whenever
\begin{equation}
\label{testtruelongrun}
\hat \T_n  >  \Delta+  u_{1-\alpha}   \frac{ \tau }{\sqrt{n} } \, ,
\end{equation}
that is we use the true asymptotic standard deviation $\tau$ instead of its estimate $\hat \tau_n$. Throughout this section we consider fAR(1) error processes defined by  
\begin{align} \label{eq:fAR1}
  \varepsilon_j = \eta_j + \kappa \, \varepsilon_{j-1}  ~~~(j \in \Z)
\end{align}
for some $\kappa \in (0,1)$ and expectation function $\mu$ as in \eqref{err0}. The random functions $\eta_j$, for $j = 1,\dots,n$, are defined as in~\eqref{eq:errors} (again with $D=21$) but in this section, we use the Fourier functions defined by $b_1 \equiv 1$ and
  \begin{align} \label{eq:fourier_basis}
  b_j(t) =
  \begin{cases}
  \sqrt{2} \, \sin(j \pi t), \quad &j \text{ is even} \\
  \sqrt{2} \, \cos((j-1) \pi t), \quad &j>1 \text{ is odd}
  \end{cases}
  \end{align}
as (orthonormal) basis functions such that an explicit calculation of the long-run variance becomes easier. Thus we have  $\cov (\varepsilon_0 (s), \varepsilon_\ell(t)) = \cov(\eta_0(s),\eta_0(t)) \kappa^\ell/(1-\kappa)^2$
which yields
\begin{align*}
  \tau^2 = 4\int_T\int_T \mu (s) \mu(t) C(s,t) ds \, dt
  = \frac{4}{(1-\kappa)^2} \sum_{i=1}^D \frac{1}{i^2} 
  \bigg(\int_0^1 \mu(t) b_i(t) dt \bigg)^2 \, .
\end{align*}
In order to obtain an estimate of the long-run covariance function $C$, we use the ``\texttt{opt\_bandwidth}'' function from the R package ``\texttt{fChange}'' with the Bartlett kernel (both as ``\texttt{kern\_type}'' and as `\texttt{`kern\_type\_ini}'').


\begin{figure}[H]
  {  \centering
    \includegraphics[width=4cm,height=3.75cm]{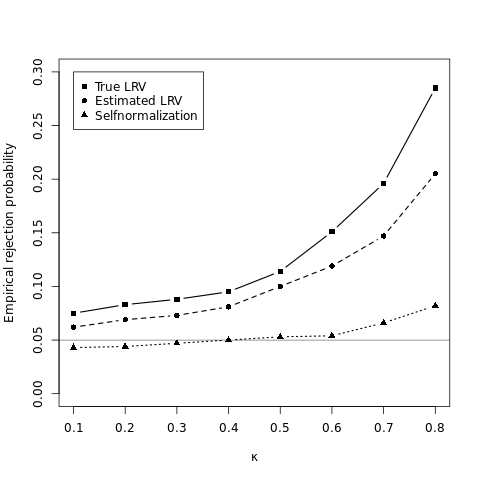}
    \includegraphics[width=4cm,height=3.75cm]{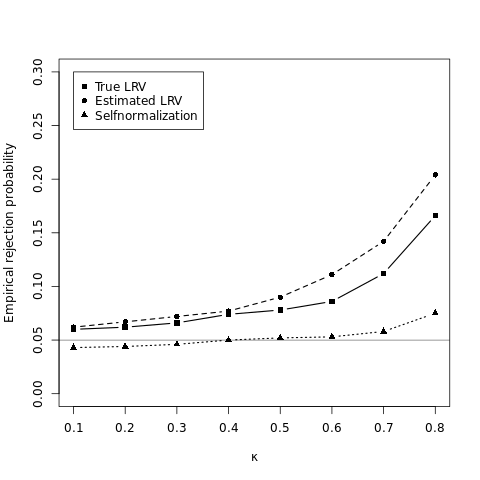}
    \includegraphics[width=4cm,height=3.75cm]{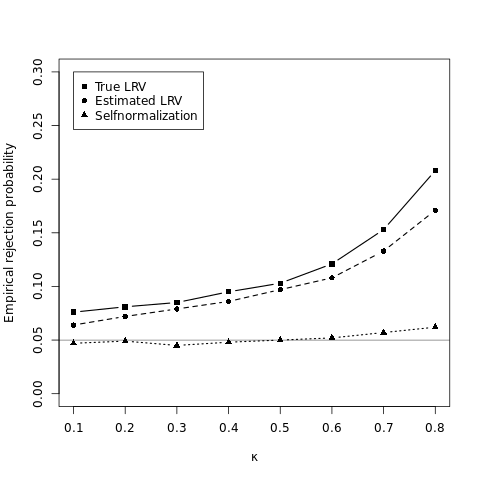}
    \includegraphics[width=4cm,height=3.75cm]{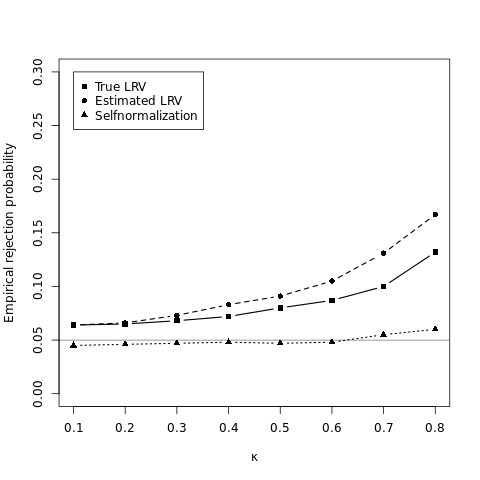}
    \caption{\it Approximation of the test level for different values of 
    $\kappa$. Errors are fAR(1) processes defined by \eqref{eq:fAR1}. Left to right: 
    $\Delta = 0.5, n = 100$;
    $\Delta = 1.5, n = 100$;
    $\Delta = 0.5, n = 200$;
    $\Delta = 1.5, n = 200$.
    }
    \label{fig:LRVlevelComparison}}
\end{figure}

In Figure~\ref{fig:LRVlevelComparison}, we compare the approximation of the nominal  level of the three tests for different values of $n, \kappa, \Delta$ at  the boundary of the null hypothesis, that is for $\int_T \mu^2(t) dt = \Delta$. We observe that the self-normalized test performs well across all settings considered with only a slight inflation of level for the most difficult case $\kappa = 0.8, n=100$. In contrast, even for a large sample size $n=200$, the tests based on the estimated and true (asymptotic) {long-run variance exceed their nominal level for all values of $\kappa$ considered with especially large over-rejections for $\kappa > 0.5$. Interestingly, the test based on the estimated long-run variance performs slightly better compared to the test with the true asymptotic long-run variance when $n=100$. A similar pattern can be observed for data that are more heavy-tailed. Due to space considerations additional details are deferred to Section~\ref{sec:addcompsn} in the online supplement. }

\subsection{Two sample problem} \label{sec52}

We begin considering the case of two independent (stationary) samples, \(X_1,\dots, X_m\) and \(Y_1,\dots, Y_n\), with
\(\E [X_i] = \mu_1\) and \(\E[Y_j] = \mu_2\), where the mean functions are given by
\begin{equation} \label{hd1}
\mu_1 \equiv 0~,~~
 \mu_2(t) = a t(1-t)
\end{equation}
[see Section 4 in \cite{horvathKokoszkaReeder2013} for a similar approach], such that
$ \int_0^1 \mu_2^2(t) dt = a^2 / 30 $.  We are interested in testing the  hypotheses
 \eqref{nulltwo}, that is
$$
H_0: \int_T D ^2(t) dt \leq  \Delta  \quad\mbox{ versus  }\quad H_1: \int_T D^2(t) dt >  \Delta \, ,
$$
where  $D=\mu_1 - \mu_2 $ is the (unknown) difference  of the two mean functions and  the threshold is given by
 \(\Delta  =0.2^2/30\) (note that this corresponds to the choice $a=0.2$). 
 We consider  independent samples,  \(fMA(1)\) processes (generated as described in Section \ref{sec51}) and independent Brownian Bridges as error processes.

\begin{figure}[H]
{  \centering
\includegraphics[width=5.65cm,height=5.25cm]{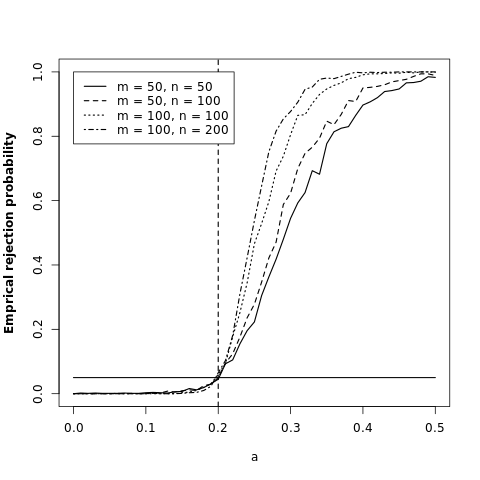}
 \includegraphics[width=5.65cm,height=5.25cm]{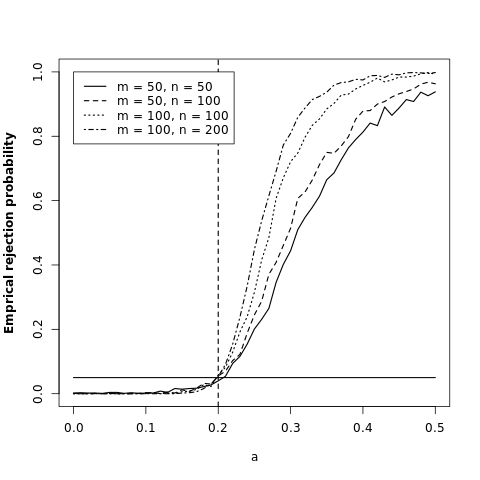}
 \includegraphics[width=5.65cm,height=5.25cm]{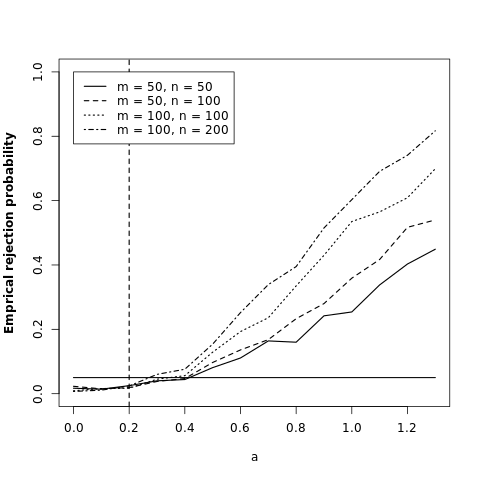}
\vspace{ -.5cm}
\caption{\it Simulated rejection probabilities of the test \eqref{testreltwo} for the relevant hypotheses
\eqref{nulltwo} with  $\Delta = 0.2^2 / 30$. The mean functions
are given by \eqref{hd1} and different independent error  processes are considered.
First panel:  independent  error processes defined by \eqref{err1}. Second panel: \(fMA(1)\) processes defined by \eqref{err2}.
Third panel: Brownian Bridges\label{fig2}.}
}
\end{figure}
 
In Figure  \ref{fig2} we display the rejection
 probabilities of the test \eqref{testreltwo}  as a function of the parameter $a$ for different sample sizes $m$ and $n$.
 We observe that the test yields a good approximation of the nominal level at the boundary  $a=0.2^2/30$ and  
 detects the alternatives with reasonable power. Further results for dependent samples are presented in Section \ref{sec62a} of the online supplement and show a similar picture.

{
We conclude  this section investigating  the effect of more heavy-tailed data  and compare
the test  \eqref{testreltwo} with the two tests obtained by the alternative self-normalizations
in \eqref{alt1} and \eqref{alt2}.
To be  precise these tests reject the null hypothesis of no relevant difference, whenever
\begin{align} \label{testreltwoa}
& \hat \D_{m,n} > \Delta + q_{1-\alpha}(\mathbb{W}^{\star}) \hat{\mathbb{V}}^{\star}_{m,n} \, ,\\
& \hat \D_{m,n} > \Delta + q_{1-\alpha}(\mathbb{W}^{\star\star}) \hat{\mathbb{V}}^{\star\star}_{m,n} \, ,
 \label{testreltwob}
\end{align}
where $q_{1-\alpha} (\mathbb{W}^{\star})$ and  $q_{1-\alpha} (\mathbb{W}^{\star\star})$ are
the $(1-\alpha)$-quantiles of $\mathbb{W}^{\star}$ and $\mathbb{W}^{\star\star}$
in  \eqref{Wdefalt}.}
\begin{figure}[h]
{  \centering
\includegraphics[width=5.65cm,height=5.25cm]{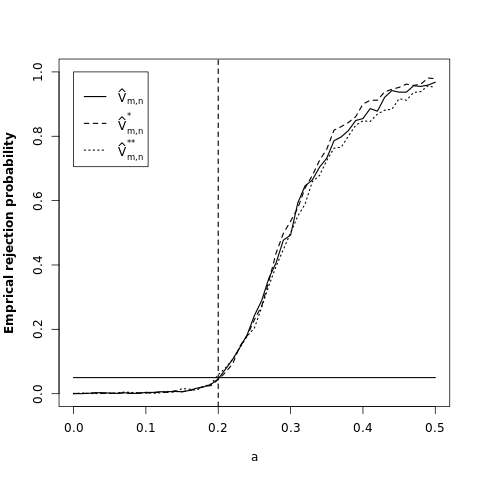}
    ~~~    ~~~     ~~~
\includegraphics[width=5.65cm,height=5.25cm]{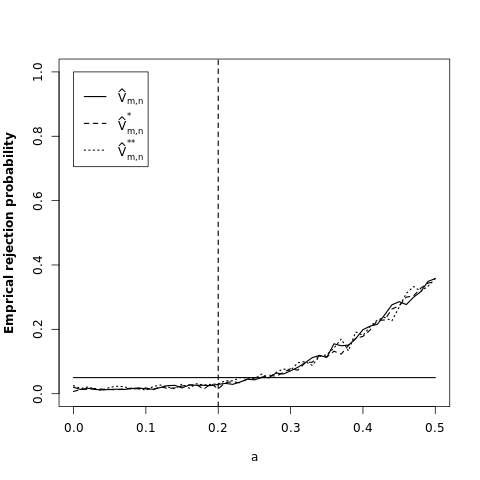}
\vspace{ -.5cm}
\caption{\label{fig4a}
\it Simulated rejection probabilities of the tests \eqref{testreltwo}, \eqref{testreltwoa} and \eqref{testreltwob} using  different self-normalizing factors. The mean functions are given by \eqref{hd1} and different independent error processes are considered with sample sizes $m = 50$, $n = 100$. The threshold is defined as $\Delta = 0.2^2 / 30$ and the errors are \(fMA(1)\) processes given by \eqref{err2} (left panel) and \eqref{eq:heavy-tailed2} (right panel).
 }
  }
\end{figure}

In the left panel of Figure \ref{fig4a}, we display the rejection probabilities of the tests \eqref{testreltwo}, \eqref{testreltwoa} and \eqref{testreltwob} in the situation  considered in Figure \ref{fig2}.
More precisely, the sample sizes are $m=50$, $n=100$, the mean functions are given by \eqref{hd1} and the error process is an \(fMA(1)\) process defined by \eqref{err2}. We observe a very similar behaviour of all three tests under consideration. \\
Next we investigate a similar situation for more heavy-tailed data and  consider similar error processes as used in \cite{Kraus2012}, that is
\begin{align} \label{eq:heavy-tailed2}
\begin{split}
  \eta_{i}(t) &= \frac{1}{\sqrt{10}} \sum_{k=1}^{10} 
  \{ k^{-3/2} \sqrt{2} \sin(2 \pi k t) V_{i,k}
  + 3^{-k/2} \sqrt{2} \cos(2 \pi k t) W_{i,k} \} \\
  \tilde{\eta}_{j}(t) &= \frac{1}{\sqrt{10}} \sum_{k=1}^{10} 
  \{ k^{-3/2} \sqrt{2} \sin(2 \pi k t) \tilde{V}_{j,k}
  + 3^{-k/2} \sqrt{2} \cos(2 \pi k t) \tilde{W}_{j,k} \}
\end{split}
\end{align}
($i=1, \ldots , m, \, j=1, \ldots , n$). Here  the random variables $V_{i,k}, W_{i,k}, \tilde{V}_{j,k}, \tilde{W}_{j,k}$ are independent $t_5$-distributed random variables scaled to have unit variance. The right panel of Figure~\ref{fig4a} shows the empirical rejection probabilities. We observe a very similar behaviour of the three considered tests. Additional results with b-spline basis functions show a similar picture and details are deferred to Section~\ref{sec:addheavy} in the online supplement.
\begin{figure}[H]
{ \centering
  \includegraphics[width=5.25cm,height=5.25cm]{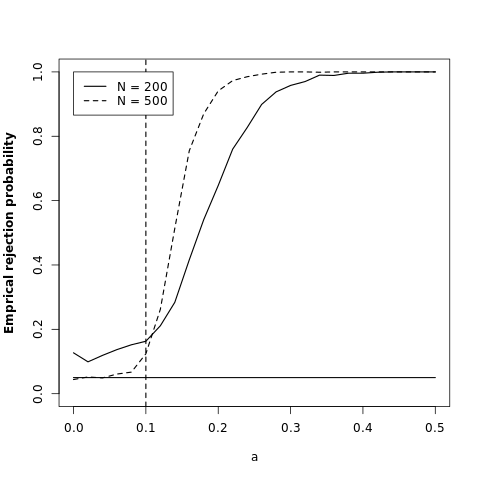} ~
  \includegraphics[width=5.25cm,height=5.25cm]{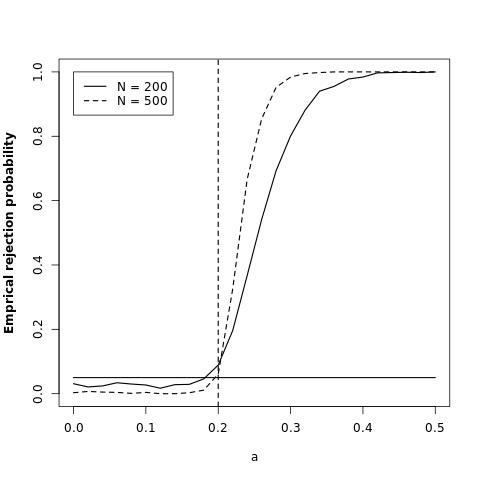} ~
  \includegraphics[width=5.25cm,height=5.25cm]{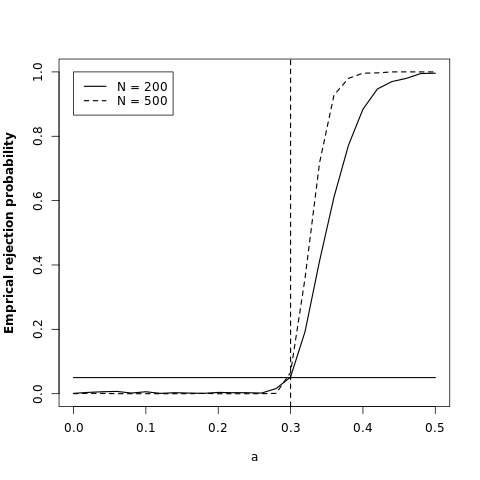} ~
\vspace{ -.75cm}
\caption{\it Simulated rejection probabilities of the test \eqref{testreltwocp} for
the relevant hypotheses \eqref{nullcp} with  $\Delta = 0.1^2/30$ (left), $\Delta = 0.2^2/30$ (middle), $\Delta = 0.3^2/30$ (right).
Data is generated according to model~\eqref{modcp} with $\theta_0 = 0.5, \mu = 0, \delta(t) = at(1-t)$, for $a = 0, 0.02,\dots, 0.5$, and the errors are i.i.d. defined by~\eqref{err1}. The tuning parameter is set to $\eps =0.05$.
\label{fig_CPPfIID}}
}
\end{figure}

\subsection{Change point problem} \label{sec53}

We begin  considering the model~\eqref{modcp} with $\theta_0 = 0.5, \mu = 0, \delta(t) = at(1-t)$, the errors are i.i.d. from~\eqref{err1}. The trimming parameter $\eps$ for estimating the change point location is set to $0.05$. Data are generated with \(a = 0, 0.02,\dots, 0.5\) and then empirical rejection probabilities are calculated using \(\Delta = 0.1^2/30, 0.2^2/30, 0.3^2/30\). These probabilities are shown in Figure \ref{fig_CPPfIID}.

\begin{figure}[t]
{ \centering
  \includegraphics[width=5.45cm,height=5.25cm]{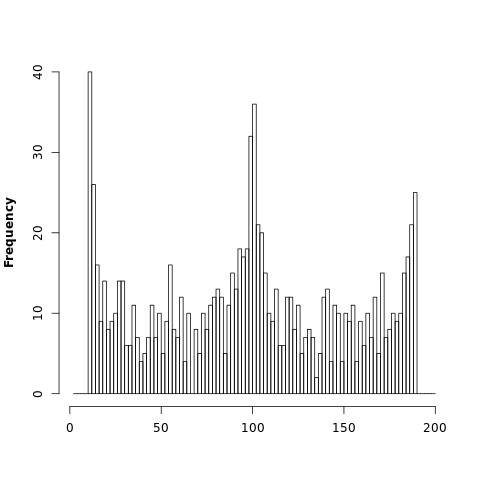}~
  \includegraphics[width=5.45cm,height=5.25cm]{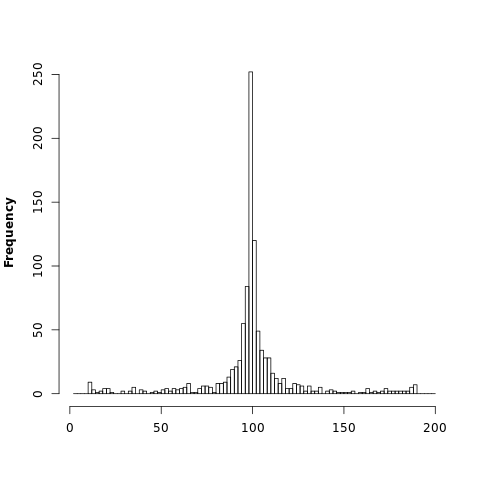}~
  \includegraphics[width=5.45cm,height=5.25cm]{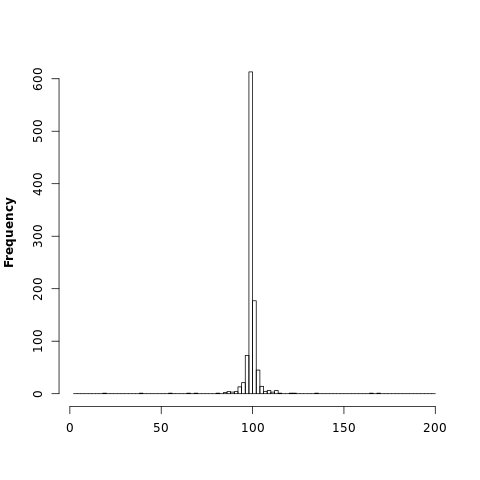}~
\caption{\it Histogram of the change point estimator $\hat \theta$ defined in \eqref{eq:hattheta}. Size $N=200$, {data are generated according to model~\eqref{modcp} with $\theta_0 = 0.5, \mu = 0, \delta(t) = at(1-t)$, for $a = 0.1, 0.2, 0.3$, and the errors are i.i.d. defined by~\eqref{err1}. The tuning parameter is set to $\eps =0.05$.}
\label{fig_CPestimatorfIID}}
}
\end{figure}

From Theorem \ref{thmcp}, we expect that the probability of rejection should be close to $\alpha$ at the boundary of the hypotheses
($\int_0^1 D^2(t) dt = \Delta$), strictly smaller than $\alpha$  in the interior of the null hypothesis ($\int_0^1 D^2(t) dt < \Delta$) and larger than $\alpha$
in the interior of the alternative ($\int_0^1 D^2(t) dt > \Delta$). This pattern is clearly observed for relevant  hypotheses  with threshold $\Delta \geq 0.2^2/30$.
However the proposed test is oversized if relevant hypotheses with
 $\Delta = 0.1^2/30$ are tested (see the left  panel in Figure \ref{fig_CPPfIID}). This is because change point tests for relevant hypotheses require a precise estimate of the change point
For small values of $a$ it is extremely difficult to estimate the true change point location, and an imprecise estimation of the change point results in a less accurate approximation of the nominal level. The difficulty of estimating the true change point location for small
values of $a$ is further illustrated in Figure \ref{fig_CPestimatorfIID} where we show the histogram of the corresponding estimator of the change point for \(a = 0.1, 0.2, 0.3\) with sample size $N = 200$.

\begin{figure}[H]
{ \centering
  \includegraphics[width=5.65cm,height=5.25cm]{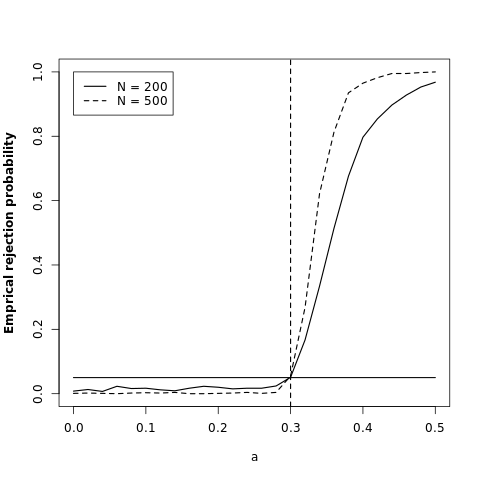}
  \includegraphics[width=5.65cm,height=5.25cm]{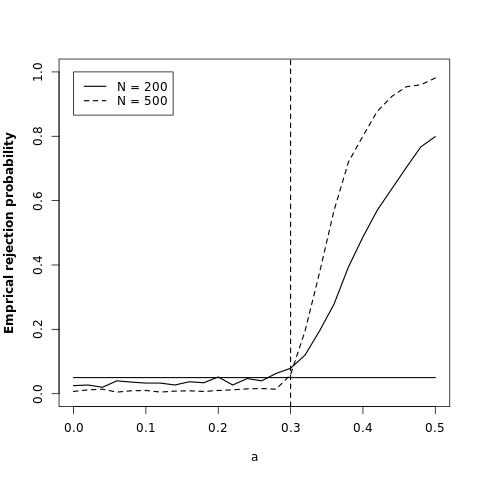}
\caption{\it Simulated rejection probabilities of the test \eqref{testreltwocp} for the relevant hypotheses \eqref{nullcp} with  $\Delta =0.3^2 / 30$
in the case of $fMA(1)$ samples. The mean function after the change point
is given by \eqref{hd1} and the mean function before the change point is the zero function.
Left panel:  error processes defined by  \eqref{CPPfMA1_c1}.   Right panel:
error processes defined by  \eqref{CPPfMA1_c3}.
\label{fig_CPPfMA1}}
}
\end{figure}

Next, we investigate the properties of our test with dependent error processes, i.e we generate a \(fMA(1)\) process $\{ \eta_i \}_{i\in \Z} $ as described in Section \ref{sec51} and define
\begin{eqnarray} \label{CPPfMA1_c1}
X_i = \mu + \eta_{i} ~,~~i=1,\ldots , \lfloor \theta_0 N \rfloor ~;
~~ X_i = \mu + \delta + \eta_{i} ~,~~i = \lfloor \theta_0 N \rfloor+1,\ldots , N \, .
\\
 \label{CPPfMA1_c3}
X_i = \mu + \eta_{i} ~,~~i=1,\ldots , \lfloor \theta_0 N \rfloor ~;
~~ X_i = \mu + \delta + \sqrt{3}~\eta_{i} ~,~~i = \lfloor \theta_0 N \rfloor+1,\ldots , N \, .
\end{eqnarray}
as the first and second scenario. The functions $\mu, \delta$ are as described in the beginning of this section. The corresponding rejection probabilities of the test  \eqref{testreltwocp}  are depicted in Figure \ref{fig_CPPfMA1} where we restrict our attention to the case $\Delta = 0.3^2/30$ for the sake of brevity. We find that for both error settings the test performs reasonably well.

\subsection{Results for Covariance operators} 

In this section we investigate the finite sample properties of the tests for  precise hypotheses regarding the covariance operators as introduced in Section \ref{sec4}. 

\subsubsection{Two sample problem} \label{sec:2scov}
{
For the sake of brevity we only display results for fMA(1) processes which are defined by 
\begin{align} \label{eq:fMA1}
X_j = \eta_j + \kappa \, \eta_{j-1} \, , 
\quad Y_i = \tilde{\eta}_i + \kappa \, \tilde{\eta}_{i-1}
\qquad j = 1,\dots m; \, i = 1,\dots, n 
\end{align}
with $\kappa = 0.7$.
The  error processes are given by 
\begin{align*}
\eta_j = \sum_{l=1}^D N_{l,j} b_l \, \quad
\tilde{\eta}_i = a \, \sum_{l=1}^D N^\prime_{l,j} b_l \, ,
\end{align*}
for $j=1,\dots ,m; i=1,\dots,n$, where the coefficients $N_{i,j} ,N^\prime_{i,j^\prime}$ are independent $N(0,\sigma_i^2)$ ($i=1,\dots,D=21$,   $j = 1,\dots,m$, $j^\prime = 1,\dots,n$) distributed random variables and the (orthonormal) basis functions $b_1,\dots,b_D$ are defined  in \eqref{eq:fourier_basis} ($b_1 \equiv 1$).
Similar as in Section~6.3 in \cite{aueDubartNorinhoHormann2015}, we consider two scenarios for the variance structure of the random coefficients, namely, for any $j=1,\dots,m$, $j^\prime = 1,\dots,n$, 
\begin{equation} \label{eq:varScenario}
\begin{aligned}
  &(A) \quad \sigma_i^2 = \Var(N_{i,j}) = \Var(N^\prime_{i,j^\prime}) = 1/i^2 
  \quad & &(i = 1,\dots, D) \\
  &(B) \quad \sigma_i^2 = 1.2^{-2i} \quad & &(i = 1,\dots, D) \, .
\end{aligned}
\end{equation}
Note that  in this case  $X_{i}$ is a multiple of  $Y_{j}$ in distribution  and the distance between the  covariance operators is given by
\begin{align*}
\int_0^1 \int_0^1 D^2(s,t) ds dt
=  (1-a^2)^2 \, \sum_{i=1}^{D} \sigma_i^4 \, (1 + \kappa^2)^2 ~
\end{align*}
[see~\cite{papsap2016} for a similar approach].
The empirical rejection probabilities  of the test \eqref{eq:cov2s}  for different values of $a$ are displayed in Figure~\ref{fig:TSPcov-fMA1},
where the case $a=1.5$ corresponds to the boundary of the hypotheses. We observe a similar pattern as for the comparison of the mean functions, where the test \eqref{eq:cov2s} is slightly more conservative in the variance scenario (A). Additional results with independent data and heavy-tailed errors  show a  similar picture  and can be found in Section~\ref{sec:add2scov} of the online supplement.

\subsubsection{Change point problem}
\label{sec:cpcov}

In this section we investigate   the test for a relevant change in the  covariance operator, which was  developed in Section~\ref{sec:cpCov}. 
For this purpose   we consider an  fMA(1) process $X^\prime_1,\dots,X^\prime_N$ defined by \eqref{eq:fMA1} (with $\kappa = 0.7$),
where the  basis functions and variances  $\sigma_1^{2},\dots,\sigma_D^{2}$ are given by 
\eqref{eq:fourier_basis} and  \eqref{eq:varScenario}, respectively.
The data $X_1,\dots,X_N$ is  defined by 
\begin{align*}
  X_j = 
  \begin{cases}
  X^\prime_j \, , \quad &j \leq \lfloor N \theta_0 \rfloor \\
  aX^\prime_j \, , \quad &j > \lfloor N \theta_0 \rfloor
  \end{cases}
\end{align*}
for $j = 1,\dots,N$, where  the change point is given by $\theta_0 = 0.5$. In Figure~\ref{fig:CPPcov-fMA1} we show the rejection probabilities of the test
\eqref{testreltwocpCov}  for the hypotheses \eqref{nullcpCov}, where the threshold is given by  $\Delta = (1-1.5^2)^2 \, \sum_{i=1}^{D} \sigma_i^4 \, (1 + 0.7^2)^2$.
  Overall, we observe a similar behaviour as for  the test for a relevant change point in the   mean functions. Further simulations with independent and heavy-tailed data show similar patterns, see Section~\ref{sec:addcpcov} of the online supplement.    

\begin{figure}[H]
  { \centering
    \includegraphics[width=5.65cm,height=5.25cm]{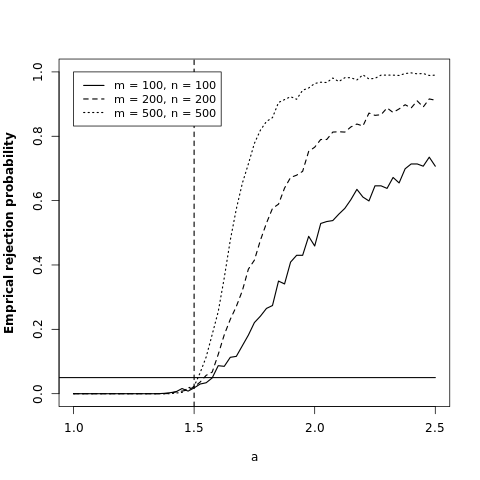}
    \includegraphics[width=5.65cm,height=5.25cm]{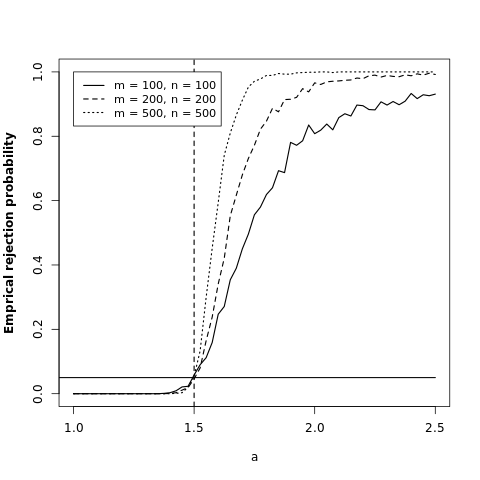}
    \caption{\it Simulated rejection probabilities of the test \eqref{eq:cov2s} 
      for the hypotheses \eqref{nullCov} of a relevant difference between  the covariance operators of two  
      fMA(1) processes ($\Delta = (1-1.5^2)^2 \, \sum_{i=1}^{D} \sigma_i^4 \, (1 + 0.7^2)^2 $).  
      Left panel:  variance scenario (A).   
      Right panel:  variance scenario (B).
      \label{fig:TSPcov-fMA1}}
  }
\end{figure}

\begin{figure}[H]
{ \centering
\includegraphics[width=5.65cm,height=5.25cm]{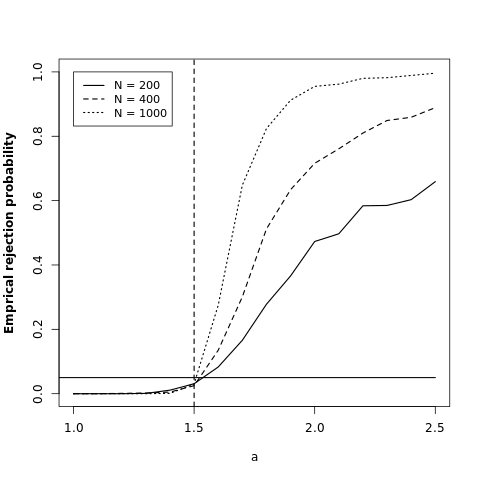}
\includegraphics[width=5.65cm,height=5.25cm]{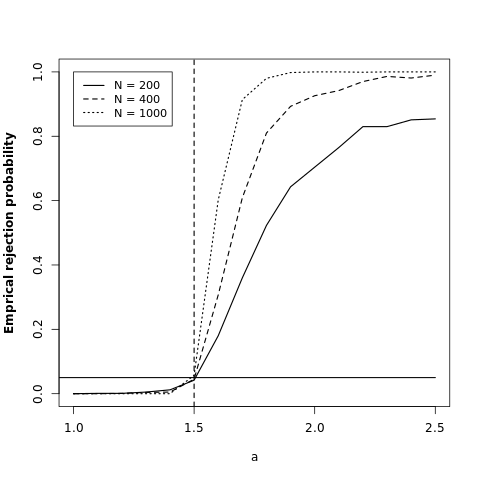}
\caption{\it Simulated rejection probabilities of the test~\eqref{testreltwocpCov} for the  hypotheses \eqref{nullcpCov} 
of a relevant change point  in the covariance operator  of an fMA(1) process ($\Delta = (1-1.5^2)^2 \, \sum_{i=1}^{D} \sigma_i^4 \, (1 + 0.7^2)^2 $).
Left panel:   variance scenario (A).  
Right panel:  variance scenario (B). \label{fig:CPPcov-fMA1}}
}
\end{figure}



\subsection{Data illustrations}  \label{sec:dataExample}

\subsubsection{Two sample test}

In this section we consider an application of the methodology developed in Section~\ref{sec22} to Australian temperature data. The data consists of daily minimum temperatures collected at different meteorological stations in Australia. Following \cite{fremdt2014} we project the daily values of each year on a Fourier basis consisting of 49 basis functions resulting in annual temperature curves for each location under consideration. These authors investigate the temperature data  to illustrate  methodology designed to choose the dimension of the projection space obtained with fPCA and in \cite{aueVandelft2017} the data is considered in the context of stationarity tests for functional time series. 

We investigate annual data curves obtained from the meteorological stations in Cape Otway~(1865-2011) and Sydney~(1859-2011). Cape Otway is a location in the south of Australia and Sydney is a city on the eastern coast of Australia. There is a distance of approximately 1000~km between the two locations such that differences in the temperature profiles are expected and the task of the relevant two sample test is now to specify how big the difference might be. The samples consist of $m=147$ and $n=153$ temperature curves, respectively.


In order to calculate the test decision in \eqref{testreltwo} for
the hypotheses defined in \eqref{nulltwo}, we computed the
statistic in \eqref{dmn} and the normalizer in \eqref{vmn}.
We obtained $\hat \D_{m,n} = 14.115$, $\hat \V_{m,n} = 0.315$
and in Table~\ref{tab:ts-temp}, the test decisions are displayed
for several choices of the level $\alpha$ and the threshold parameter
$\Delta$. In the left panel of Figure \ref{fig:means}
we display the two estimated mean functions.

The results in Table~\ref{tab:ts-temp} provide no evidence for
an integrated squared mean difference larger than $\Delta = 11.8$ but on
the other extreme there is strong evidence that it exceeds
$\Delta = 9$. Choosing $\Delta$ between $9.1$ and $10.7$ led to
rejecting the null at level $\alpha \geq 5\%$ and for
$\Delta \in [10.8,11.7]$, the test rejected the null only at level
$\alpha \geq 10\%$, which means weaker support of the
alternative.

\begin{table}[H]
\vspace{.5cm}
\begin{center}
\begin{tabular}{c@{\qquad}llll}
\hline
$\Delta$ & 99\% & 95\% & 90\%  \\
\hline
9.0  & TRUE  & TRUE  & TRUE   \\
9.1  & FALSE & TRUE  & TRUE   \\
10.7 & FALSE & TRUE  & TRUE   \\
10.8 & FALSE & FALSE & TRUE   \\
11.7   & FALSE & FALSE & TRUE   \\
11.8 & FALSE & FALSE & FALSE  \\
\hline
\end{tabular}
\end{center}
\caption{\it Summary of the two sample test for relevant hypotheses
with varying $\Delta$ for the annual temperature curves. The label
TRUE refers to a rejection of the null, the label FALSE to a failure
to reject the null.}
\label{tab:ts-temp}
\end{table}

\begin{table}[H]
  \vspace{.5cm}
  \begin{center}
    \begin{tabular}{c@{\qquad}llll}
      \hline
      $\Delta$ & 99\% & 95\% & 90\%  \\
      \hline
      0.72  & FALSE & FALSE & TRUE   \\
      0.73  & FALSE & FALSE & FALSE  \\
      \hline
    \end{tabular}
  \end{center}
  \caption{\it Summary of the change point test for relevant hypotheses
    with varying $\Delta$ for the annual river flow curves. The labels
    TRUE, FALSE refers to a rejection of the null and failure
    to reject the null, repectively}
  \label{tab:cpp-riverflow}
\end{table}

\vspace{-.6cm}
\begin{figure}[H]
{  \centering
\includegraphics[width=5.65cm,height=5.2cm]{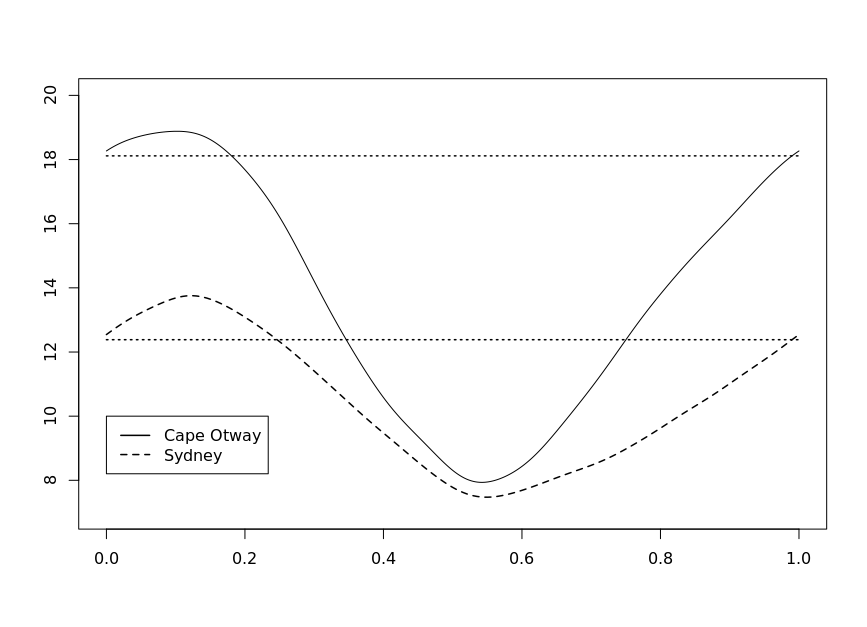}
~~
\includegraphics[width=5.65cm,height=5.2cm]{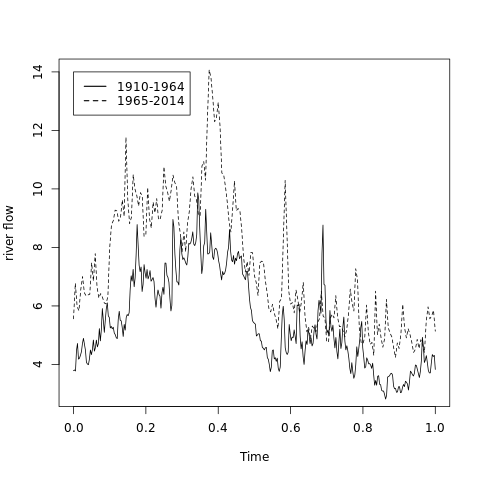}
\caption{\it Left panel: Mean functions of the Cape Otway and Sydney series for the two sample case. Right panel: Mean curves of the river flow (in $m^3/sec$) for the periods 1910-1964 and 1965-2014, respectively.
}
\label{fig:means}}
\end{figure}

\newpage
{\subsubsection{Change point test}}

In this section we consider daily flows (in $m^3/sec$) of the river Chemnitz at G\"oritzhain (located in the East of Germany), where data was recorded for the years $1909$ -  $2014$. One hydrological year (different definitions are possible but we consider the same as \cite{Sharipov2014}) starts at the first of November and ends at the 31-th of October which means that we consider the hydrological years 1910-2014. Note that \cite{Sharipov2014} considered the years 1910-2012. We regard data from each year as one flow curve resulting in a sample of size $N = 105$.

In the definition of the change point estimator, we use the trimming parameter $\varepsilon = 0.1$ and obtain the year 1964 as a possible change point. This is the same year which was identified by \cite{Sharipov2014}. In the right panel of Figure \ref{fig:means}, we display the mean of the curves before and after the estimated change point. Applying the test defined by \eqref{testreltwocp} for two different values of $\Delta$ leads to the test decisions in Table~\ref{tab:cpp-riverflow}. For $\Delta = 0.72$, we reject the null hypothesis of no relevant change at level $\alpha = 0.1$ and we do not reject the null at level $\alpha = 0.05$. For $\Delta \geq 0.73$, we do not reject the null at the test levels under consideration.

\medskip

{\bf Acknowledgements }
This work has been supported in part by the
Collaborative Research Center ``Statistical modeling of nonlinear
dynamic processes'' (SFB 823, Teilprojekt A1 and C1) of the German Research Foundation
(DFG). Parts of this paper were written  while H. Dette was visiting the Department of Statistics, University of Toronto,
and this author would like to thank the institute for its hospitality. The authors would like to thank Tim Kutta for helpful discussions regarding some of the material on covariance operators.
The authors are also grateful to two anonymous referees and the associate editor whose constructive comments lead to a substantial improvement of an earlier version of this manuscript.

\medskip

\setlength{\bibsep}{1pt}
\begin{small}
\bibliography{2020-02-19_SnRelevant}
\end{small}

\newpage

\appendix

\section*{Online supplementary material}

This section contains addition finite sample results (Section \ref{addsim}), proofs of all results in the main part of the paper 
(Section \ref{sec6})
and  an extension of the methodology  to other testing problems for relevant  hypotheses (Section \ref{sec:concalt}).

\section{Additional simulation results} \label{addsim}
\def\theequation{A.\arabic{equation}}
\setcounter{equation}{0}

\subsection{Heavy tailed data in the two sample problem}
\label{sec:addheavy}

Here, we display results for the mean functions in~\eqref{hd1} but as error processes we use 
\begin{align} \label{eq:heavy-tailed1}
\eta_j = 
\sum_{i=1}^D \sqrt{3/(5 \, i^2)} \, t_{i,j} \, b_i ~~~~ (j=1,\dots , m),
\quad \tilde{\eta}_j = \sum_{i=1}^D \sqrt{3/(5 \, i^2)} \, \tilde{t}_{i,j} \, b_i ~~~~ (j=1,\dots , n) \, ,
\end{align}
where \(t_{1,1},t_{2,1}, \dots, t_{D,m}, \tilde{t}_{1,1},\tilde{t}_{2,1}, \dots, \tilde{t}_{D,n} \) are independent $t_5$-distributed random variables and $b_1,\dots,b_D$ are B-spline basis functions. Note that the coefficients of the B-splines are defined such 
that the expectations are zero and the variance of the $i$-th coefficient 
is equal to $\sigma_i^2 = 1/i^2$ ($i=1,\ldots , D$). 

In Figure~\ref{fig4a_add} we show empirical rejection probabilities of the three tests \eqref{testreltwo}, \eqref{testreltwoa} and \eqref{testreltwob} for different values of $a$. We observe again that all three tests yield very similar rejection probabilities.

\begin{figure}[h]
{\centering
\includegraphics[width=5.65cm,height=5.25cm]{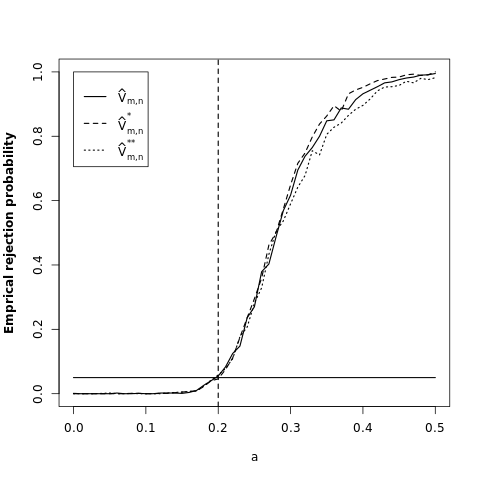}
\caption{\label{fig4a_add}
\it Simulated rejection probabilities of the tests \eqref{testreltwo}, \eqref{testreltwoa} and \eqref{testreltwob} using different self-normalizing factors. The mean functions are given by \eqref{hd1} and two independent samples are considered with sample sizes $m = 50$, $n = 100$. The error process is given  by \eqref{eq:heavy-tailed1}.
}}
\end{figure}

\subsection{Two dependent samples} 
\label{sec62a}

In this section we investigate the important case of dependent samples in the two sample problem. We generate a $fMA(1)$ process $\{ \eta_i \}_{i\in \Z}$ as described in Section \ref{sec51} and define 
\begin{eqnarray} \label{first}
X_i =\mu_1 +  \eta_{i} ~,~~i=1,\ldots , m~;~~ Y_i =\mu_2 +  \eta_{m+i} ~,~~i=1,\ldots , n~
\end{eqnarray}
in the first and
 \begin{eqnarray} \label{second}
X_i =\mu_1 +  \eta_{i} ~,~~i=1,\ldots , m~;~~ Y_i =\mu_2 +  \sqrt{3}~\eta_{m+i} ~,~~i=1,\ldots , n \, .
\end{eqnarray}
in the second scenario; in both cases $\Delta = 0.3^2 / 30$ and $\mu_1,\mu_2$ are given in~\eqref{hd1}. The corresponding rejection probabilities of the test  \eqref{testreltwo}  are depicted in Figure~\ref{fig3}. Overall the test performs well in all settings considered.

\begin{figure}[H]
{  \centering
\includegraphics[width=5.65cm,height=5.25cm]{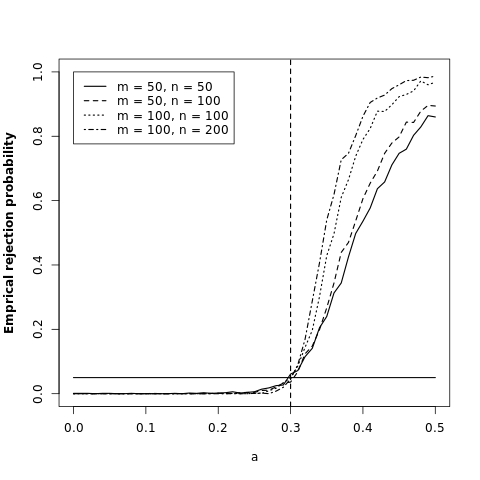}~~~~~
 \includegraphics[width=5.65cm,height=5.25cm]{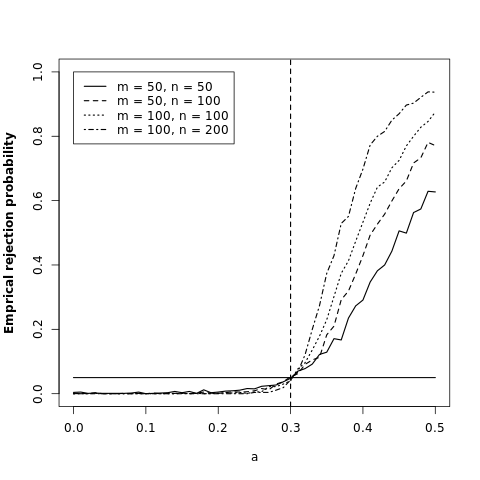}
\caption{\it Simulated rejection probabilities of the test \eqref{testreltwo} for the relevant hypotheses \eqref{nulltwo} with  $\Delta =0.3^2 / 30$
in the case of dependent samples. The mean functions
are given by \eqref{hd1}.
Left panel:  error processes defined by  \eqref{first}.   Right panel:
error processes defined by  \eqref{second}.}
\label{fig3}}
\end{figure}

\subsection{The two sample problem for covariance operators}\label{sec:add2scov}

Recall the setting introduced at the beginning of Section~\ref{sec:2scov}. First, we consider independent data, that is we define 
\begin{align} \label{eq:fIID}
  X_j = \eta_j \, , \quad Y_i = \tilde{\eta}_i 
  \qquad j = 1,\dots m; \, i = 1,\dots, n \, .
\end{align}
In this scenario, we have
\begin{align*}
\int_0^1 \int_0^1 D^2(s,t) ds dt
= (1-a^2)^2 \, \sum_{i=1}^{D} \sigma_i^4 
\end{align*}
and empirical rejection probabilities for different values of $a$ can be seen in  Figure~\ref{fig:TSPcov-fIID},
while Table~\ref{tab:TSPcov-fIID} shows the simulated level at the boundary of the hypotheses.
}

\begin{figure}[H]
  { \centering
    \includegraphics[width=5.65cm,height=5.25cm]{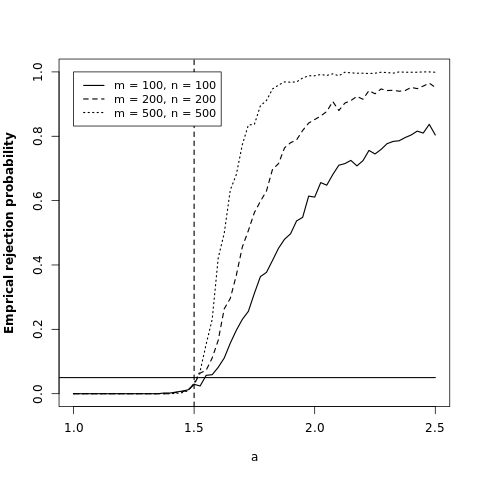}
    \includegraphics[width=5.65cm,height=5.25cm]{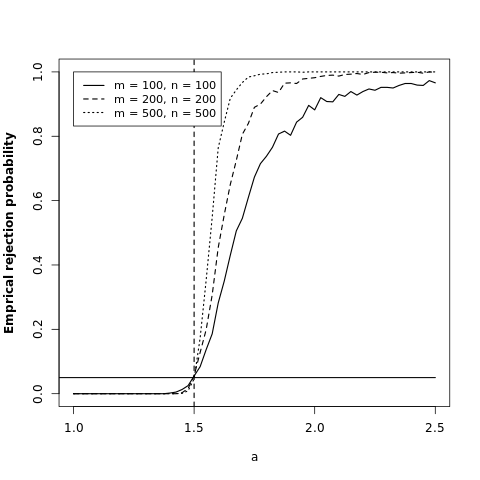}
    \caption{\it Simulated rejection probabilities of the test \eqref{eq:cov2s} 
      for the relevant hypotheses \eqref{nullCov} with  
      $\Delta = (1-1.5^2)^2 \, \sum_{i=1}^{D} \sigma_i^4$.   The model is given by  \eqref{eq:fIID}.
      Left panel: variance scenario (A).   
      Right panel:  variance scenario (B).
      \label{fig:TSPcov-fIID}}
  }
\end{figure}

 \begin{table}[h]
  {  \centering
    \begin{tabular}{c|ccc||ccc}
      & \multicolumn{3}{|c||} {(A)} & \multicolumn{3}{|c} {(B)} \\
      \hline
      & 1\% & 5\% & 10\% & 1\% & 5\% & 10\% \\
      \hline \hline
      \(m=n=100\)  & 0.4 & 2.9 & 6.7 & 0.8 & 5.5 & 14.5 \\
      \(m=n=200\)  & 0.3 & 2.9 & 7.4 & 1.2 & 5.6 & 12   \\
      \(m=n=500\)  & 0.2 & 3.6 & 8.8 & 1.3 & 5.6 & 11.7 
    \end{tabular}
    \caption{\it\it Approximation of the level  at the boundary of the hypotheses in  model \eqref{eq:fIID}.
      \label{tab:TSPcov-fIID}}
  }
\end{table}

\begin{figure}[H]
  { \centering
    \includegraphics[width=5.65cm,height=5.25cm]{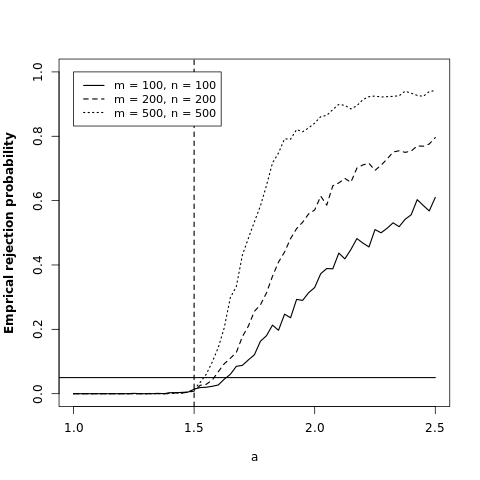}
    \caption{\it Simulated rejection probabilities of the test \eqref{eq:cov2s} 
      for the relevant hypotheses \eqref{nullCov} with  
      $\Delta = 1/10^2 \sum_{k=1}^{10} (1/k^6+1/3^{2k})(1-a^2)^2 $. 
      The errors are defined by \eqref{eq:heavy-tailed2}.
      Left panel:  variance scenario (A).   
      Right panel: variance scenario (B).
      \label{fig:TSPcov-nonGaussian}}
  }
\end{figure}

In Figure~\ref{fig:TSPcov-nonGaussian} we display empirical rejection 
probabilities for heavy-tailed error processes defined by 
\eqref{eq:heavy-tailed2}, which are similar to those used in \cite{Kraus2012} and in \cite{papsap2016}.  

\subsection{Change point problems for covariance operators}
\label{sec:addcpcov}

\begin{figure}[H]
{ \centering
\includegraphics[width=5.65cm,height=5.25cm]{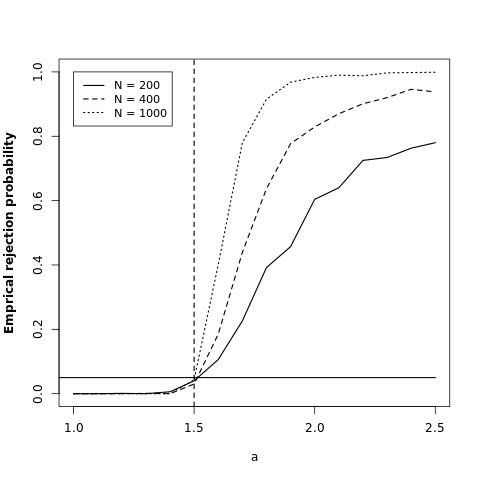}
\includegraphics[width=5.65cm,height=5.25cm]{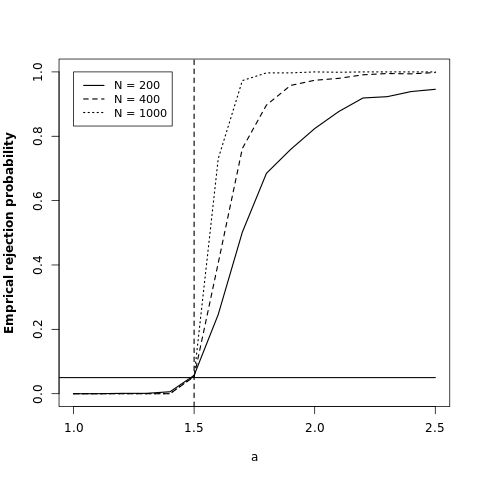}
\caption{\it  Simulated rejection probabilities of the test \eqref{testreltwocpCov} for the relevant hypotheses \eqref{nullcpCov} with $\Delta = (1-1.5^2)^2 \, \sum_{i=1}^{D} \sigma_i^4$.
The model is given by   \eqref{eq:fIID}.  Left panel:   variance scenario (A). Right panel: variance scenario (B). \label{fig:CPPcov-fIID}}
}
\end{figure}

\begin{figure}[H]
{ \centering
\includegraphics[width=5.65cm,height=5.25cm]{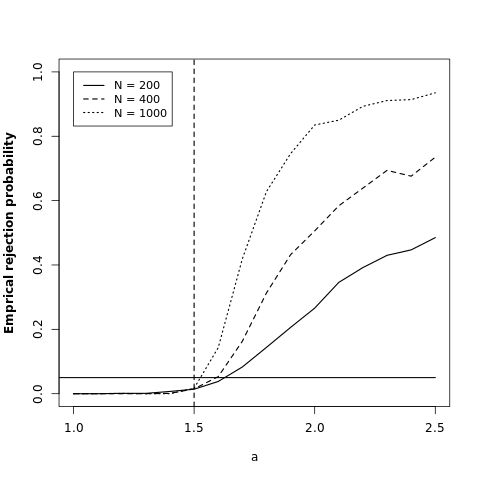}
\caption{\it  Simulated rejection probabilities of the test \eqref{testreltwocpCov} for the relevant hypotheses \eqref{nullcpCov} with $\Delta = (1-1.5^2)^2 \, \sum_{i=1}^{D} \sigma_i^4$. 
The errors are given by \eqref{eq:heavy-tailed2}.
\label{fig:CPPcov-nonGaussian}}}
\end{figure}

The setting considered here is the same as in Section~\ref{sec:cpcov}. In Figure~\ref{fig:CPPcov-fIID}, we consider   independent data as in \eqref{eq:fIID} and display the  
empirical rejection probabilities for different values of $a$ and fixed threshold  $\Delta = (1-1.5^2)^2 \, \sum_{i=1}^{D} \sigma_i^4$. In Figure~\ref{fig:CPPcov-nonGaussian}, we display empirical rejection probabilities for data with heavy-tailed errors as in \eqref{eq:heavy-tailed2}. In this scenario the threshold parameter $\Delta$ is fixed to $\Delta = 1/10^2 \sum_{k=1}^{10} (1/k^6+1/3^{2k})(1-a^2)^2 $.

\subsection{Comparison of self-normalization with long-run variance estimation for heavy-tailed data} \label{sec:addcompsn}
\def\theequation{A.\arabic{equation}}

\begin{figure}[H]
  {  \centering
    \includegraphics[width=5.65cm,height=5.25cm]{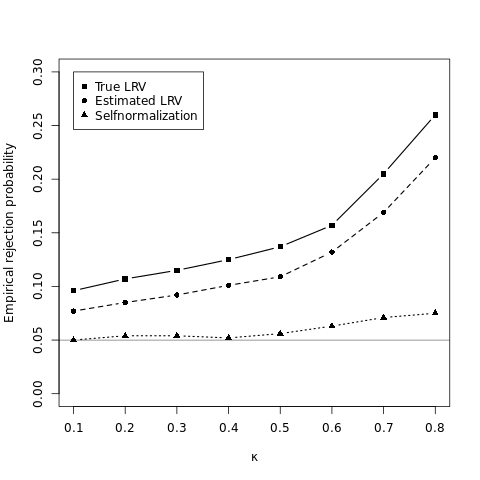}
    \includegraphics[width=5.65cm,height=5.25cm]{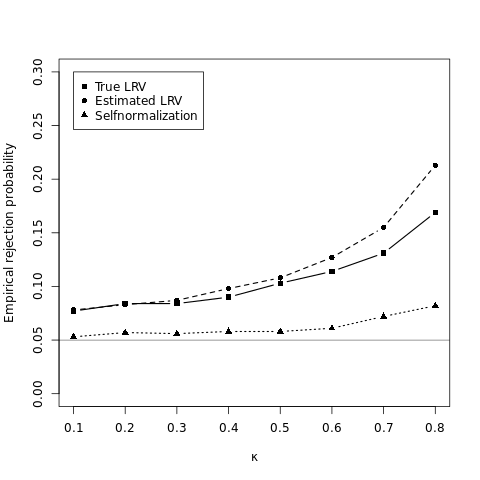} ~~~~
    \includegraphics[width=5.65cm,height=5.25cm]{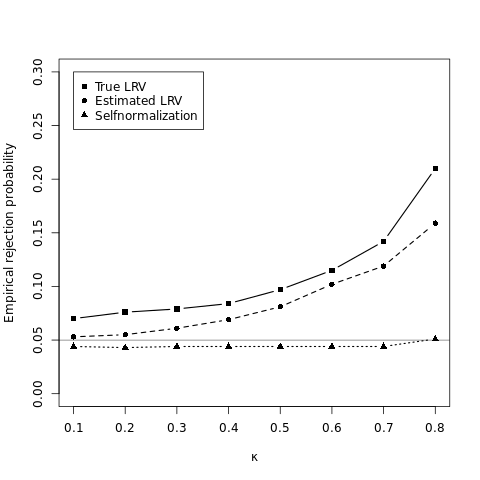}
    \includegraphics[width=5.65cm,height=5.25cm]{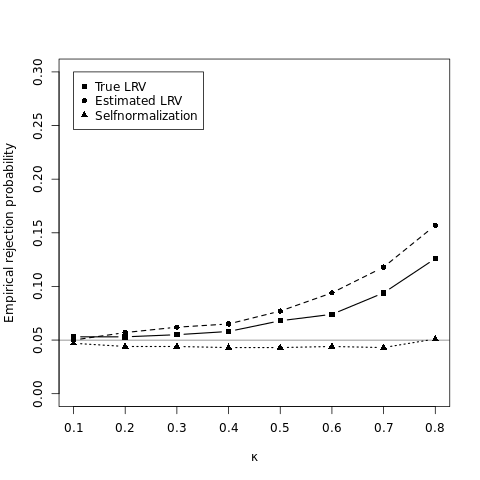}
    \caption{\it {Approximation of the test level for different values of $\kappa$. Errors are heavy-tailed fAR(1) processes defined by \eqref{eq:nonGaussianErrors}. 
    Top left: $\delta = 0.5, n = 100$;
    Top right: $\delta = 1.5, n = 100$;
    Bottom left: $\delta = 0.5, n = 200$;
    Bottom right: $\delta = 1.5, n = 200$
    }
    \label{fig:LRVlevelComparison_nongaussian}}
  }
\end{figure}

This section contains additional results comparing the performance of self-normalization and estimated long-run variance. First, we consider rejection probabilities under the null but for fAR(1) processes with heavy-tailed errors. More precisely we replace the errors $\eta_j$ in \eqref{eq:fAR1} by 
\begin{align} \label{eq:nonGaussianErrors}
  \eta_j^\prime = \sum_{i=1}^D \sqrt{3/(5 \, i^2)} \, t_{i,j} \, b_i 
  ~~~~ (j=1,\dots , n)
\end{align}
where \(t_{1,1},t_{2,1}, \dots, t_{D,n} \) are independent $t_5$-distributed random variables. The results are shown in Figure~\ref{fig:LRVlevelComparison_nongaussian}
and we  observe that the self-normalized test yields a much better approximation of the nominal level.

\newpage

\section{Proofs} \label{sec6}
\def\theequation{B.\arabic{equation}}
\setcounter{equation}{0}

We begin with some technical preliminaries, the notation introduced in this section will be used throughout all the proofs. Define the set of functions
\[
\mathcal{G} := \Big\{f: T\times[0,1] \to \R: \sup_{\lambda \in [0,1]} \int_T f^2(t,\lambda) dt < \infty \Big\}.
\]
Equip this set with the norm
\[
\|f-g\|_{\mathcal{G}} := \sup_{\lambda \in [0,1]} \Big\{\int_T [f(t,\lambda) - g(t,\lambda)]^2 dt \Big\}^{1/2}
\]
to obtain a normed vector space. We will frequently work with random elements with values in this space. Here, random elements need not be measurable and we will make use of the general theory of outer probabilities (see Chapter 1 in \cite{wellner1996}) where appropriate without explicitly mentioning this.

\subsection{Proofs of the results in Section \ref{sec2} }

First, we state some preliminary results that will be useful throughout the proofs. Define
\begin{equation}\label{eq:tilSntlam}
\tilde S_n (t, \lambda) =  \frac{1}{n} \sum_{j=1}^{\lfloor n\lambda \rfloor}   ( X_j(t)  - \mu (t) )~,~~\lambda \in [0,1] \, .
\end{equation}
Then it follows from Theorem 1.1 in \cite{berhorric2013}, that there exists a sequence of measurable random elements in $\mathcal{G}$, say $\{\Gamma_n (t,\lambda) \}_{\lambda,t \in [0,1]} $, such that
 \begin{eqnarray}
 \label{approx1}
 && \sup_{\lambda \in [0,1]} \int_T  \big ( \sqrt{n} \, \tilde S_n (t, \lambda)  - \Gamma_n (t,\lambda) \big )^2 dt  = o_\Prob (1) \\
  \label{approx2}
 && \{\Gamma_n (t,\lambda)  \}_{\lambda,t \in [0,1]}  \Dequal \{\Gamma (t,\lambda)  \}_{\lambda,t \in [0,1]} \, ,
\end{eqnarray}
where $\Gamma$ is defined by
 \begin{eqnarray}
 \label{approx3}
\Gamma (t,\lambda) &=& \sum_{i=1}^\infty \sqrt{\lambda_i} \phi_i(t) W_i(\lambda) \, .
\end{eqnarray}
$\{W_i\}_{i\in \N} $ is a sequence of independent Brownian motions and $\lambda_i$, $\phi_i$ are the eigenvalues and
(orthonormal) eigenfunctions
of the integral operator corresponding to the covariance kernel
 \begin{eqnarray}
 \label{approx4a}
C(s,t) &=&
 \sum_{i=1}^\infty {\lambda_i} \phi_i(s)\phi_i(t)
\end{eqnarray}
defined in \eqref{approx4},
that is
 \begin{eqnarray}
 \label{approx5}
\lambda_i \phi_i (s) = \int_T C(t,s) \phi_i (t) dt  ~~~~~  (i\in \N) \, .
\end{eqnarray}
Note that \cite{berhorric2013} also prove that $\sum_k \lambda_k < \infty$ (see their Lemma~2.2) and that
\begin{equation}\label{eq:intG2linf}
\sup_{0\leq \lambda \leq 1} \int_T \Gamma^2(t,\lambda) dt < \infty \quad a.s.
\end{equation}
The latter implies that for any square integrable function $\zeta: [0,1] \to \R$ the process
$$\bigg\{\int_T \zeta(t) \Gamma(t,\lambda) dt\bigg\}_{\lambda \in [0,1]}$$
can be viewed as an element of $\ell^\infty([0,1])$ and that the same is true for the process $\{\int_T \Gamma^2(t,\lambda) dt\}_{\lambda \in [0,1]}$. Moreover, summability of the sequence $(\lambda_k)_{k\in\N}$ together with properties of the modulus of continuity of Brownian motions implies that for any positive sequence $(\kappa_k)_{k\in\N}$ such that $\kappa_n \to 0$, it follows
\begin{align*}
& \sup_{\substack{\nu,\lambda \in[0,1]: \\ |\nu-\lambda|\leq \kappa_n}} \int_T \{\Gamma(t,\lambda)-\Gamma(t,\nu)\}^2 dt
\\
= &\sup_{\substack{\nu,\lambda \in[0,1]: \\ |\nu-\lambda|\leq \kappa_n}} \int_T \sum_{i=1}^\infty\sum_{j=1}^\infty\sqrt{\lambda_i\lambda_j} \{W_i(\lambda) - W_i(\nu)\}\{W_j(\lambda) - W_j(\nu)\} \phi_i(t)\phi_j(y) dt
\\
= &\sup_{\substack{\nu,\lambda \in[0,1]: \\ |\nu-\lambda|\leq \kappa_n}} \sum_{i=1}^\infty\sum_{j=1}^\infty\sqrt{\lambda_i\lambda_j} \{W_i(\lambda) - W_i(\nu)\}\{W_j(\lambda) - W_j(\nu)\} \int_T\phi_i(t)\phi_j(y) dt
\\
= &\sup_{\substack{\nu,\lambda \in[0,1]: \\ |\nu-\lambda|\leq \kappa_n}} \sum_{i=1}^\infty\lambda_i\{W_i(\lambda) - W_i(\nu)\}^2
\\
\leq &  \sum_{i=1}^\infty\lambda_i \sup_{\substack{\nu,\lambda \in[0,1]: \\ |\nu-\lambda|\leq \kappa_n}}\{W_i(\lambda) - W_i(\nu)\}^2 = o_{\mathbb{P}}(1)
\end{align*}
where the last line follows since by Fubini's Theorem
\[
\E\Big[\sum_{i=1}^\infty\lambda_i \sup_{\substack{\nu,\lambda \in[0,1]: \\ |\nu-\lambda|\leq \kappa_n}}\{W_i(\lambda) - W_i(\nu)\}^2\Big] =  \E\Big[\sup_{\substack{\nu,\lambda \in[0,1]: \\ |\nu-\lambda|\leq \kappa_n}}\{W_1(\lambda) - W_1(\nu)\}^2\Big]\sum_{i=1}^\infty\lambda_i = o(1).
\]
This implies
\begin{equation}
\label{eq:equicont2}
\sup_{\substack{\nu,\lambda \in[0,1]: \\ |\nu-\lambda|\leq \kappa_n}} \int_T \{\Gamma^2(t,\lambda)-\Gamma^2(t,\nu)\}^2 dt = o_{\mathbb{P}}(1) \quad \kappa_n \to 0.
\end{equation}

\subsubsection{Proof of Theorem \ref{thm_relevantfunc}} \label{sec61}

The main ingredients of the proof are the convergence result stated in \eqref{haupt} {when $\int_T \mu^2(t) dt > 0$ and the bounds {$\hat{\mathbb{T}}_{n} = o_{\Prob}(1), \hat{\mathbb{V}}_{n}  = o_\Prob(1)$} when $\int_T \mu^2(t) dt = 0$. We begin by considering the case $\int_T \mu^2(t) dt = 0$. In that case we have $\tilde S_n(t,\lambda) \equiv S_n(t,\lambda)$ for and hence by~\eqref{approx1} $\sup_{\lambda \in [0,1]}\int_T S_n^2(t,\lambda) dt = o_\Prob(1)$ which implies {$\hat{\mathbb{T}}_{n} = o_{\Prob}(1), \hat{\mathbb{V}}_{n}  = o_\Prob(1)$}.}


For the case  $ \int_T \mu^2(t) dt > 0$ note that a straightforward calculation shows
 \begin{eqnarray*}
\sqrt{n} \int_T \big ( S_n^2(t,\lambda) - \lambda^2 \mu^2(t)\big)  dt &=& \sqrt{n}  \int_T   \big (S_n(t,\lambda) - \lambda \mu (t) \big)^2  dt +  2 \sqrt{n}  \int_T
\lambda \mu (t)  \big (  S_n(t,\lambda)   - \lambda \mu (t) \big) dt  \\
&=& \sqrt{n}  \int_T \tilde  S_n^2(t,\lambda) dt +  2 \sqrt{n}  \int_T
\lambda \mu (t)   \tilde S_n(t,\lambda) dt  +  o_\Prob (1)  \\
&=&  \frac{1}{ \sqrt{n}} \int_T  \big ( \sqrt{n} \tilde S_n (t, \lambda)  - \Gamma_n (t,\lambda) \big )^2 dt + \frac{1}{ \sqrt{n}} \int_T \Gamma_n^2  (t,\lambda) dt   \\
&&  - \frac{2}{ \sqrt{n}} \int_T  \big ( \sqrt{n} \tilde S_n (t, \lambda)  - \Gamma_n(t,\lambda)  \big ) \Gamma_n (t,\lambda) dt \\
&& +  2 \int_T  \lambda \mu (t)  \big ( \sqrt{n} \tilde S_n (t, \lambda)  - \Gamma_n (t,\lambda)  \big ) dt +  2 \int_T  \lambda \mu (t)   \Gamma_n (t,\lambda) dt  +  o_\Prob (1)  \\
& = &  2 \int_T  \lambda \mu (t)   \Gamma_n (t,\lambda)  dt  +  o_\Prob (1)
\end{eqnarray*}
uniformly with respect to $\lambda \in [0,1]$, where we repeatedly used \eqref{approx1} and the Cauchy-Schwarz inequality. Therefore we obtain from \eqref{approx2} and Slutsky's Lemma that
 \begin{eqnarray}
 \label{approx6}
\Big\{\sqrt{n}\Big( \int_T S_n^2(t,\lambda) dt - \lambda^2 \int_T \mu^2(t) dt\Big)\Big\}_{\lambda\in [0,1]} \weak
 \Big\{  2  \lambda \int_T  \mu (t)   \Gamma(t,\lambda)  dt   \Big\}_{\lambda\in [0,1]}
 \end{eqnarray}
in $\ell^\infty([0,1])$ (recall that  by the discussion at the beginning of this  section
 the process on the right hand side is an element of $\ell^\infty([0,1])$), and observing \eqref{approx3}, \eqref{approx4a} and \eqref{approx5}  it follows by  a straightforward calculation that
 \begin{eqnarray*}
\mbox{Cov} \Big (
\int_T  \mu (t)   \Gamma (t,\lambda)  dt, \int_T  \mu (t)   \Gamma (t,\lambda^\prime)  dt \Big) &=& \big ( \lambda \wedge \lambda^\prime \big )
 \int_T  \int_T  \mu (t) \mu (s)  C(s,t)  ds  dt .
 \end{eqnarray*}
 Therefore
$$
 \Big\{  2  \lambda \int_T  \mu (t)   \Gamma (t,\lambda)   dt   \Big\}_{\lambda\in [0,1]} \Dequal
  \Big\{ \lambda \tau \Bb(\lambda) \Big\}_{\lambda\in [0,1]} \, ,
 $$
 where $\Bb$ denotes a standard Brownian motion on the interval $[0,1]$ and
  $\tau^2$ is defined in \eqref{tau}.
Consequently  the statement \eqref{haupt} in Section \ref{sec2} follows from \eqref{approx6}
and an application of the continuous mapping theorem observing that the mapping
 $$
 \Z \mapsto \frac{\Z (1)}{
 \big (\int_0^1 ( \Z (\lambda )  - \lambda^2  \Z (1  )  )^2 \nu (d\lambda ) \big )^{1/2} }
 $$
from the measurable functions in $\ell^\infty([0,1])$ onto $\mathbb{R}$ is continuous at points $f \in \ell^\infty([0,1])$ with $\int_0^1 (f(\lambda) - \lambda^2 f(1))^2 \nu(d\lambda) \neq 0$. This yields
\begin{align} \label{wkonv}
  \frac{\hat \T_n - d}{\hat{\mathbb{V}}_n}
  \Dkonv \frac{\Bb(1)}{\big( \int_0^1 \lambda^2 (\Bb(\lambda)
  - \lambda\Bb(1))^2 \nu (d\lambda ) \big)^{1/2}} = \mathbb{W} \, .
\end{align}

In the discussion following Theorem \ref{thm_relevantfunc} we already argued that
\begin{align*}
  \lim_{n \to \infty} \mathbb{P}
  \Big( \hat \T_n > \Delta + q_{1-\alpha}(\mathbb{W}) {\hat{\mathbb{V}}_n}  \Big)
  = 0 \, ,
\end{align*}
whenever $\int_T \mu^2(t) dt =0$. When $0 < \int_T \mu^2(t) dt < \Delta$, we have

\begin{align*}
  \lim_{n \to \infty} \mathbb{P}
  \Big( \hat \T_n > \Delta + q_{1-\alpha}(\mathbb{W}) {\hat{\mathbb{V}}_n}  \Big)
  = \lim_{n \to \infty} \mathbb{P}
  \Big( \frac{\hat \T_n - d}{ \hat{\mathbb{V}}_n}
  > \frac{\sqrt{n} (\Delta - d)}{\sqrt{n} \, \hat{\mathbb{V}}_n} + q_{1-\alpha}(\mathbb{W})  \Big)
  = 0
\end{align*}
since $\sqrt{n} \, \hat{\mathbb{V}}_n = O_{\mathbb{P}}(1), \frac{\hat \T_n - d}{ \hat{\mathbb{V}}_n} = O_{\mathbb{P}}(1)$, $\hat{\mathbb{V}}_n \geq 0$ a.s. and $\sqrt{n} (\Delta - d) \to +\infty$. In the case $ d = \int_T \mu^2(t) dt = \Delta$ we conclude
\begin{align*}
  \lim_{n \to \infty} \mathbb{P}
  \Big( \hat \T_n > \Delta + q_{1-\alpha}(\mathbb{W}) {\hat{\mathbb{V}}_n}  \Big)
  = \lim_{n \to \infty} \mathbb{P}
  \Big( \frac{\hat \T_n - d}{\hat{\mathbb{V}}_n} > q_{1-\alpha}(\mathbb{W}) \Big)
  = \alpha
\end{align*}
and, if $ d > \Delta$, we have
\begin{align*}
  \lim_{n \to \infty} \mathbb{P}
  \Big( \hat \T_n > \Delta + q_{1-\alpha}(\mathbb{W}) {\hat{\mathbb{V}}_n}  \Big)
= \lim_{n \to \infty} \mathbb{P}
 \Big( \hat \T_n - d
 > \Delta - d \,  + q_{1-\alpha}(\mathbb{W})\hat{\mathbb{V}}_n  \Big)
  = 1
\end{align*}
since $\hat{\mathbb{V}}_n = o_\mathbb{P}(1)$, $\hat \T_n - d = = o_\mathbb{P}(1)$ and $(\Delta - d) < 0$.
\hfill $\Box$

\subsubsection{Proof of Theorem \ref{thm_2s}} \label{sec62}

The processes $\{X_m \}_{m\in \Z} $ and $\{Y_n \}_{n\in \Z}$ satisfy assumptions
(B2) and thus admit the representation $X_j = \mu_1 + \eta^X_j, Y_j = \mu_2 + \eta^Y_j$ where
$(\eta^X_j)_{j\in\Z}$ and $(\eta^Y_j)_{j\in\Z}$ denote centered error processes that both satisfy (A2)-(A4). Define
\[
\tilde D_{m,n} (t,\lambda) := S_m^X (t,\lambda) - S_n^Y (t,\lambda) \, ,
\]
where the processes $S_m^X$ and $S_n^Y$ are given  by
\begin{eqnarray*}
S_m^X (t,\lambda)&=& \frac{1}{m} \sum_{j=1}^{\lfloor m\lambda \rfloor} \big ( X_j(t) - \mu_1 (t) \big )
= \frac{1}{m} \sum_{j=1}^{\lfloor m \lambda \rfloor} \eta^X_j(t) \, , \\
S_n^Y (t,\lambda)  &=& \frac{1}{n} \sum_{j=1}^{\lfloor n\lambda \rfloor} \big ( Y_j(t) - \mu_2 (t)  \big )
= \frac{1}{n} \sum_{j=1}^{\lfloor n \lambda \rfloor} \eta^Y_j(t) \, ,
 \end{eqnarray*}
 respectively.
A similar calculation as given in Section \ref{sec61} shows that
\begin{eqnarray*} 
\Z_{m,n}(\lambda)
&:=& \sqrt{n+m} \int_T \big ( D_{m,n}^2(t,\lambda) - \lambda^2 D^2(t)\big)  dt
\\
\nonumber &=&2 \sqrt{n+m} \int_T  \lambda D (t) \big ( D_{m,n}(t,\lambda) - \lambda D(t)\big)  dt  +  o_\Prob (1)
\\
& = &
\nonumber
2 \sqrt{n+m} \int_T  \lambda D (t)  \tilde  D_{m,n}(t,\lambda) dt  +  o_\Prob (1)
= \Z_{m}^X (\lambda) -\Z_{n}^Y (\lambda) + o_\Prob (1) \, ,
\end{eqnarray*}
where we use the fact that $\lambda \mu_1 = \frac{1}{m}\sum_{i=1}^{\lfloor m\lambda \rfloor} \mu_1 + o(1)$ uniformly in $\lambda$ in the third equality and the processes $\{ \Z_{m}^X (\lambda) \}_{\lambda \in [0,1]}$  and $\{ \Z_{n}^Y (\lambda) \}_{\lambda \in [0,1]}$
are given by
\begin{eqnarray} \label{ZmZn}
 \Z_{m}^X (\lambda) &:=& 2 \sqrt{n+m} \int_T  \lambda D (t) S_m^X (t,\lambda)  dt \, , \\
 \Z_{n}^Y (\lambda) &:=& 2 \sqrt{n+m} \int_T  \lambda D (t) S_n^Y (t,\lambda) dt \, ,
\end{eqnarray}
respectively.
As the times series  $\{X_n \}_{n\in \Z} $ and $\{Y_n \}_{n\in \Z} $ satisfy assumptions
(A1) - (A4) it follows from the proof of Theorem \ref{thm_relevantfunc}
that the processes $\{ \Z_{m}^X (\lambda) \}_{\lambda \in [0,1]}$  and
$\{ \Z_{n}^Y (\lambda) \}_{\lambda \in [0,1]}$ converge weakly in $\ell^\infty ( [0,1]) $ with both limits corresponding to scaled Brownian motions. 
Furthermore, both processes are independent and therefore
\[
  \{ \Z_{m,n} (\lambda) \}_{\lambda \in [0,1]}
  \weak \big\{ \lambda \tau_{D} \Bb(\lambda) \big\}_{\lambda\in [0,1]}
\]
in $\ell^\infty([0,1])$, where $\{ \Bb(\lambda) \}_{\lambda\in [0,1]}$ is a Brownian motion and $\tau_D$ is a real number depending on the auto-covariance structures of $\{X_m \}_{m\in \Z}$ and $\{Y_n \}_{n\in \Z}$.

The assertion now follows exactly in the same way as in the proof of Theorem~\ref{thm_relevantfunc} and the details are omitted for the sake of brevity. \hfill $\Box$

\subsection{Proofs of the results in  Section~\ref{sec3}} \label{sec63}
 \subsubsection{A technical result} \label{sec:jointpr}

In this section we prove a useful intermediate result. Now consider the situation which is described in model~\eqref{modcp} (see also Remark \ref{knownCP}(a)). Set
\begin{align*}
  \eta^{(1)}_i &= f_1(\eps_i,\eps_{i-1,...}),\quad i = 1,\dots, N
	\\
  \eta^{(2)}_i &= f_2(\eps_i,\eps_{i-1,...}),\quad i = 1,\dots, N
\end{align*}
where $f_1,f_2, (\eps_j)_{j\in\Z}$ satisfy the conditions in (A2), $\E[\eta^{(1)}_1] = \E[\eta^{(2)}_1] = 0 $, and $\eta^{(1)}_i,\eta^{(2)}_i$ satisfy assumptions (A3), (A4).

\begin{lemma} \label{jointlemma}
In the setting above consider a fixed (but arbitrary) function $\zeta$ in $L^2(T)$.
For $\lambda \in [0,1]$ define the processes
\[
\tilde Z_N^{(k)}(\lambda) := \frac{1}{\sqrt{N}}\sum_{i=1}^{\floor{N\lambda}} \int_T \eta^{(k)}_i(t) \zeta(t) dt, \quad k=1,2 \, ,
\]
then
\[
(\tilde Z_N^{(1)},\tilde Z_N^{(2)})^\top \weak \Sigma^{1/2} (\Bb_1,\Bb_2)^\top \quad \mbox{in  } \ell^{\infty}([0,1])^2  \, ,
\]
where $\Bb_1,\Bb_2$ are two independent standard Brownian motions on the interval $[0,1]$ and $\Sigma$ is a symmetric $2\times 2$ matrix with finite entries given by
\[
\Sigma_{ij} = \sum_{h \in \Z} \int_T\int_T \text{Cov}(\eta^{(i)}_0(s),\eta^{(j)}_h(t)) \zeta(s)\zeta(t) ds dt \, .
\]
\end{lemma}

\textbf{Proof}
It suffices to prove tightness of the processes $\tilde Z_N^{(1)}, \tilde Z_N^{(2)}$ individually and joint finite-dimensional convergence. Tightness of $\tilde Z_N^{(k)}$ follows from weak convergence of $\tilde Z_N^{(k)}$, which can be established by an application of Theorem 2.1 in \cite{berhorric2013} and similar arguments as given in Section~\ref{sec61}. Thus it remains to prove that for arbitrary $s_1,...,s_L \in [0,1]$ we have
\[
(\tilde Z_N^{(1)}(s_1),\tilde Z_N^{(2)}(s_1),...,\tilde Z_N^{(1)}(s_L),\tilde Z_N^{(2)}(s_L)) \weak \mathcal{N}(0,\Sigma(s_1,...,s_L))
\]
where $\Sigma(s_1,...,s_L)$ denotes the covariance matrix of the vector $(\Gb_1(s_1), \Gb_2(s_1),...,\Gb_1(s_L),\Gb_2(s_L))$ and $(\Gb_1,\Gb_2)^\top := \Sigma (\Bb_1,\Bb_2)^\top$. Following \cite{berhorric2013} we define the random variables
\[
\eta^{(k)}_{j,m} := f_k(\varepsilon_j, \varepsilon_{j-1}, \ldots \varepsilon_{j-m + 1},\boldsymbol{\varepsilon}_{j,m}^*), \quad k=1,2, m \in \N \, ,
\]
where $\boldsymbol{\varepsilon}_{j,m}^* = (\varepsilon_{j,m,j-m }^*,\varepsilon_{j,m,j-m -1}^*, \ldots )$ is given in Assumption (A4). Let $\Sigma_m$ denote matrices with entries (below we shall prove that all entries are finite for any $m \geq 1$)
\[
(\Sigma_{m})_{ij} := \sum_{|h|\leq m} \int_T\int_T \text{Cov}(\eta^{(i)}_{0,m}(s),\eta^{(j)}_{h,m}(t)) \zeta(s)\zeta(t) ds dt \, ,
\]
define $(\Gb_{1,m},\Gb_{2,m})^\top := \Sigma_m^{1/2} (\Bb_1,\Bb_2)^\top$ and
\[
\tilde Z_{N,m}^{(k)}(\lambda) := \frac{1}{\sqrt{N}}\sum_{i=1}^{\floor{N\lambda}} \int_T \eta^{(k)}_{i,m}(t) \zeta(t) dt, \quad k=1,2 \, .
\]
By an application of Example 11 in Chapter IV of \cite{pollard1984} it suffices to prove that
\begin{enumerate}
\item[(i)] $\Sigma_m \to \Sigma$ as $m \to \infty$.
\item[(ii)] For any $\delta >0, k\in\{1,2\}, \ell \in \{1,...,L\} $ we have
\[
\lim_{m\to\infty}\limsup_{N\to\infty} \mathbb{P}\Big(|\tilde Z_{N,m}^{(k)}(s_\ell) - \tilde Z_{N}^{(k)}(s_\ell)| > \delta\Big) = 0 \, .
\]
\item[(iii)] For any fixed $m$ we have
\[
(\tilde Z_{N,m}^{(1)}(s_1),\tilde Z_{N,m}^{(2)}(s_1),...,\tilde Z_{N,m}^{(1)}(s_L),\tilde Z_{N,m}^{(2)}(s_L)) \weak \mathcal{N}(0,\Sigma_m(s_1,...,s_L)) \, .
\]
\end{enumerate}
{
In order to show the claim in (i), we prove $(\Sigma_{m})_{ij} \to \Sigma_{ij}$ as $m \to \infty$ for $i,j = 1,2$. For $i = j$ this assertion directly follows from Lemma 2.2 in \cite{berhorric2013}. For $i\neq j$, one can use similar arguments as in the proof of the latter Lemma. More precisely, assume without loss of generality that $i=1, j=2$ and show that
\begin{align}
|\Sigma_{12}| &< \infty \, , \label{Sigma-finite}
\\
|(\Sigma_{m})_{12}| &< \infty \, , \quad m\geq 1 \, ,  \label{Sigma_m-finite}
\\
(\Sigma_{m})_{12} &\to \Sigma_{12}, \quad \text{as $m \to \infty$} \, . \label{Sigma_m-to-Sigma}
\end{align}
From the estimate
\begin{align} \label{Sigma-0}
  |\Sigma_{12}| \leq& \, \bigg|\int_T\int_T \E[\eta^{(1)}_0(s)\eta^{(2)}_0(t)] \zeta(s)\zeta(t) ds dt \bigg|
  + 2\bigg|\sum_{h=1}^\infty \int_T\int_T \E[\eta^{(1)}_0(s) \eta^{(2)}_h(t)] \zeta(s)\zeta(t) ds dt \bigg| \, ,
\end{align}
\eqref{Sigma-finite} follows if each of the terms above is finite. For the first term, we apply the Cauchy-Schwarz inequality to obtain by (A3)
\begin{align*}
  \bigg|\int_T\int_T \E[\eta^{(1)}_0(s)\eta^{(2)}_0(t)] \zeta(s)\zeta(t) ds dt \bigg|
  &\leq \|\zeta\|^2 \bigg(\int_T\int_T \E[\eta^{(1)}_0(s)\eta^{(2)}_0(t)]^2 ds dt \bigg)^{1/2} \\
  &\leq \|\zeta\|^2 \bigg(\int_T\int_T \E[\eta^{(1)}_0(s)^2] \E[\eta^{(2)}_0(t)^2] ds dt \bigg)^{1/2} \\
  &= \|\zeta\|^2 \big(\E\|\eta^{(1)}_0\|^2 \E\|\eta^{(2)}_0\|^2 \big)^{1/2} < \infty \, .
\end{align*}
We proceed with the second term in \eqref{Sigma-0} and drop the constant 2. For any $i \geq 1$,
$\E[\eta^{(1)}_0\eta^{(2)}_{i,i}] = \E[\eta^{(1)}_0] \E[\eta^{(2)}_{i,i}]= 0$ since
$\eta^{(1)}_0$ and $\eta^{(2)}_{i,i}$ are independent. Applying the triangle inequality and the Cauchy-Schwarz
inequality yields
\begin{align} \label{Sigma-h}
\begin{split}
  &\bigg|\sum_{h=1}^\infty \int_T\int_T \E[\eta^{(1)}_0(s) \eta^{(2)}_h(t)] \zeta(s)\zeta(t) ds dt \bigg| \\
  &\leq \sum_{h=1}^\infty \|\zeta\|^2 \bigg( \int_T\int_T \E[\eta^{(1)}_0(s) \eta^{(2)}_h(t)]^2 ds dt \bigg)^{1/2} \\
  &= \sum_{h=1}^\infty \|\zeta\|^2 \bigg( \int_T\int_T \E[\eta^{(1)}_0(s)
  (\eta^{(2)}_h(t) -  \eta^{(2)}_{h,h}(t))]^2 ds dt \bigg)^{1/2} \\
  &\leq \sum_{h=1}^\infty \|\zeta\|^2 \bigg( \int_T\int_T \E[\eta^{(1)}_0(s)^2]
  \E[(\eta^{(2)}_h(t) -  \eta^{(2)}_{h,h}(t))^2] ds dt \bigg)^{1/2} \\
  &= \|\zeta\|^2 \E\big[\|\eta^{(1)}_0\|^2\big]^{1/2} \sum_{h=1}^\infty\E\big[\|\eta^{(2)}_0 -  \eta^{(2)}_{0,h}\|^2\big]^{1/2} \, .
\end{split}
\end{align}
Due to condition (A3) we have $\E\big[\|\eta^{(1)}_0(s)\|^2\big]^{1/2}<\infty$ and by an application of the H{\"o}lder inequality we get
\begin{align*}
  \sum_{h=1}^\infty \E\big[\|\eta^{(2)}_h -  \eta^{(2)}_{h,h}\|^2\big]^{1/2}
  \leq \sum_{h=1}^\infty \E\big[\|\eta^{(2)}_0 - \eta^{(2)}_{0,h}\|^{2+\psi}\big]^{1/(2+\psi)}
\end{align*}
which is finite by (A4). This completes the proof of \eqref{Sigma-finite}.}

For $(\Sigma_m)_{12}$, we proceed similarly. We have
$\E[\eta^{(1)}_0(s)\eta^{(2)}_0(t)] = \E[\eta^{(1)}_{0,m}(s)\eta^{(2)}_{0,m}(t)]$
and therefore
\begin{align*}
  \bigg|\int_T\int_T \E[\eta^{(1)}_{0,m}(s)\eta^{(2)}_{0,m}(t)] \zeta(s)\zeta(t) ds dt \bigg| < \infty \, .
\end{align*}
Note that the vectors $(\eta^{(1)}_{0,m}, \eta^{(2)}_{h,m})$ and $(\eta^{(1)}_{0}, \eta^{(2)}_{h,m})$ have the same distribution for all $h=1,\dots,m$. Furthermore, the vectors $(\eta^{(1)}_{h,m}, \eta^{(2)}_{h,h})$ and
$(\eta^{(1)}_{0}, \eta^{(2)}_{0,h})$ have the same distribution for all $h=1,\dots,m$ (this follows from the definition of all quantities involved and the i.i.d. structure of the $\eps_{i,j,m}^*$).
Thus, using similar arguments as in \eqref{Sigma-h}, we obtain
\begin{align*}
 &\bigg|\sum_{h=1}^m \int_T\int_T \E[\eta^{(1)}_{0,m}(s) \eta^{(2)}_{h,m}(t)] \zeta(s)\zeta(t) ds dt \bigg| \leq \|\zeta\|^2 \sum_{h=1}^m \bigg( \int_T\int_T \E[\eta^{(1)}_{0,m}(s) \eta^{(2)}_{h,m}(t)]^2 ds dt \bigg)^{1/2}
\\
 =& \|\zeta\|^2 \sum_{h=1}^m \bigg( \int_T\int_T \E[\eta^{(1)}_{0}(s) \eta^{(2)}_{h,m}(t)]^2 ds dt \bigg)^{1/2}
  \leq \|\zeta\|^2\E\big[\|\eta^{(1)}_0\|^2\big]^{1/2} \sum_{h=1}^m
  \E\big[\|\eta^{(2)}_{h,m} -  \eta^{(2)}_{h,h}\|^2\big]^{1/2}
\\
  \leq& \|\zeta\|^2 \E\big[\|\eta^{(1)}_0\|^2\big]^{1/2} \sum_{h=1}^\infty   \E\big[\|\eta^{(2)}_{0} -  \eta^{(2)}_{0,h}\|^2\big]^{1/2}
  < \infty
\end{align*}
which proves \eqref{Sigma_m-finite}.

In order to establish \eqref{Sigma_m-to-Sigma}, we begin by observing that

\begin{align*}
&\bigg|\sum_{h=0}^\infty \int_T\int_T \E[\eta^{(1)}_0(s) \eta^{(2)}_h(t)] \zeta(s)\zeta(t) ds dt - \sum_{h=0}^m \int_T\int_T \E[\eta^{(1)}_{0,m}(s) \eta^{(2)}_{h,m}(t)] \zeta(s)\zeta(t) ds dt \bigg|
\\
\leq&~ \bigg|\sum_{h=1}^m \int_T\int_T \{\E[\eta^{(1)}_0(s) \eta^{(2)}_{h}(t)] - \E[\eta^{(1)}_{0,m}(s) \eta^{(2)}_{h,m}(t)]\} \zeta(s)\zeta(t) ds dt \bigg|
\\
& + \sum_{h=m+1}^\infty \bigg|\int_T\int_T \E[\eta^{(1)}_0(s) \eta^{(2)}_h(t)] \zeta(s)\zeta(t) ds dt \bigg|
\\
\leq&~ \sum_{h=1}^m \bigg|\int_T\int_T \E\Big[\eta^{(1)}_0(s) \{\eta^{(2)}_{h}(t) -  \eta^{(2)}_{h,m}(t)\}\Big] \zeta(s)\zeta(t) ds dt \bigg|
\\
& + \|\zeta\|^2 \E\big[\|\eta^{(1)}_0\|^2\big]^{1/2} \sum_{h=m+1}^\infty\E\big[\|\eta^{(2)}_0 -  \eta^{(2)}_{0,h}\|^2\big]^{1/2} \, ,
\end{align*}
where the last inequality follows by similar arguments as~\eqref{Sigma-h}.
Now the second term converges to zero as $m \to \infty$ and for the first term we obtain
\begin{align*}
& \sum_{h=1}^m \bigg|\int_T\int_T \E\Big[\eta^{(1)}_0(s) \{\eta^{(2)}_{h}(t) -  \eta^{(2)}_{h,m}(t)\}\Big] \zeta(s)\zeta(t) ds dt \bigg|
\\
\leq~& \|\zeta\|^2 \sum_{h=1}^m \Big(\int_T\int_T \E\Big[\eta^{(1)}_0(s) \{\eta^{(2)}_{h}(t) -  \eta^{(2)}_{h,m}(t)\}\Big]^2 ds dt\Big)^{1/2}.
\end{align*}
Now by (A4) we have for any fixed $h$
\begin{align*}
\Big(\int_T\int_T \E\Big[\eta^{(1)}_0(s) \{\eta^{(2)}_{h}(t) -  \eta^{(2)}_{h,m}(t)\}\Big]^2 ds dt\Big)^{1/2}
&\leq \E\big[\|\eta^{(1)}_0\|^2\big]^{1/2} \E\big[\|\eta^{(2)}_h -  \eta^{(2)}_{h,m}\|^2\big]^{1/2}
\\
&= \E\big[\|\eta^{(1)}_0\|^2\big]^{1/2} \E\big[\|\eta^{(2)}_0 -  \eta^{(2)}_{0,m}\|^2\big]^{1/2} \to 0 \, .
\end{align*}
Since also by similar arguments as in~\eqref{Sigma-h} and in the proof of~\eqref{Sigma_m-finite}
\begin{align*}
& \Big(\int_T\int_T \E\Big[\eta^{(1)}_0(s) \{\eta^{(2)}_{h}(t) -  \eta^{(2)}_{h,m}(t)\}\Big]^2 ds dt\Big)^{1/2}
\\
\leq & \, 2 \, \Big\{ \Big(\int_T\int_T \E\Big[\eta^{(1)}_0(s)\eta^{(2)}_{h}(t)\Big]^2 ds dt\Big)^{1/2} + \Big(\int_T\int_T \E\Big[\eta^{(1)}_0(s)\eta^{(2)}_{h,m}(t)\Big]^2 ds dt\Big)^{1/2}\Big\}
\\
\leq & \, 4 \, \E\big[\|\eta^{(1)}_0\|^2\big]^{1/2} \E\big[\|\eta^{(2)}_{0} -  \eta^{(2)}_{0,h}\|^2\big]^{1/2}
\end{align*}
and since the right-hand side is summable over $h \geq 1$ it follows that
\[
\sum_{h=1}^m \bigg|\int_T\int_T \E\Big[\eta^{(1)}_0(s) \{\eta^{(2)}_{h}(t) -  \eta^{(2)}_{h,m}(t)\}\Big] \zeta(s)\zeta(t) ds dt \bigg| \to 0 \, , \quad m\to \infty
\]
by the dominated convergence theorem for series.
%
%

The assertion in \eqref{Sigma_m-to-Sigma} follows and this also completes the proof of (i).

The claim in (ii) follows by a direct application of Lemma 2.1 in \cite{berhorric2013}.

For a proof of claim (iii) note that for each fixed $m$ the sequence
$$\bigg(\int_T \eta^{(1)}_{i,m}(t) \zeta(t) dt \, , \int_T \eta^{(2)}_{i,m}(t) \zeta(t) dt \bigg)_{i \in \Z}$$
form a collection of stationary, $m$-dependent random vectors with finite variance. Now (iii) follows by a straightforward application of the Cramer-Wold device and the CLT for m-dependent random variables, see for instance Theorem 9.1 in \cite{dasguptaasymptotic}.  \hfill $\Box$

\subsubsection{Proof of Proposition~\ref{propcpest}} \label{sec:ratecp}

\textbf{Step 1:}
Recall the definition of $\hat f$ in \eqref{hol0}. We begin by proving the following preliminary result
\begin{align} \label{G convergence}
  \big\{\mathbb{G}_N(\theta)\big\}_{\theta \in [0,1]}
  := \big\{\sqrt{N}(\hat{f}(\lfloor N\theta \rfloor) - d(\theta))\big\}_{\theta \in [0,1]}
  \rightsquigarrow \big\{\mathbb{G}(\theta)\big\}_{\theta \in [0,1]}
\end{align}
in \(\ell^\infty([0,1])\) as \(N\to\infty\), where
\begin{align} \label{d definition}
  d(\theta) := \tilde{d}(\theta) \int_T \delta(t)^2 dt \, ,
  \quad
  \tilde{d}(\theta) = \theta(1-\theta)
  \begin{cases}
    ( \theta_0 / \theta )^2, \quad & 1 > \theta > \theta_0 \\
    ( (1 - \theta_0) / (1 - \theta) )^2, \quad & 0 < \theta \leq \theta_0
  \end{cases}
\end{align}
$\tilde{d}(0) = \tilde d(1) = 0$, and the process \(\mathbb{G}\) is a random element in \(\ell^\infty([0,1])\) with a.s. continuous sample paths. To this end define for $k=1,...,N-1$
\begin{align*}
A_N(t, k) &:= \frac{1}{k} \sum_{j=1}^{k} (X_j(t) - \E[X_j(t)])
- \frac{1}{N - k} \sum_{j = k + 1}^{N} (X_j(t) - \E[X_j(t)])
\\
B_N(t, k) &:= \frac{1}{k} \sum_{j=1}^{k} \E[X_j(t)] - \frac{1}{N - k} \sum_{j = k + 1}^{N}
  \E[X_j(t)]
\end{align*}
and let $A_N(t,N) = A_N(t,0) = B_N(t,N) = B_N(t,0) \equiv 0$. With those definitions we can write
\begin{align} \label{hat f decomposition}
\begin{split}
\hat{f}(k) &= \int_T (A_N(t, k) + B_N(t, k))^2 dt \ \frac{k}{N}\Big(1-\frac{k}{N}\Big)
\\
&= \Big\{ \int_T A_N(t, k)^2 dt + 2 \int_T A_N(t, k) B_N(t, k) dt
+ \int_T B_N(t, k)^2 dt \Big\} \ \frac{k}{N}\Big(1-\frac{k}{N}\Big).
\end{split}
\end{align}
From Theorem 1.1 in \cite{berhorric2013} it follows that
\begin{equation} \label{hol1}
\frac{k}{N}\Big(1-\frac{k}{N}\Big) \int_T A_N(t, k)^2 dt = o_\mathbb{P}(N^{-1/2}) \, ,
\end{equation}
uniformly with respect to $k$. For \(1 \leq k\leq k_0 := \lfloor N\theta_0 \rfloor\), straightforward calculations yield
\begin{align*}
  B_N(t, k) &= \frac{1}{k} \sum_{j=1}^{k} \mu(t)
  - \frac{1}{N - k} \sum_{j = k + 1}^{k_0}
  \mu(t)
  - \frac{1}{N - k} \sum_{j = k_0 + 1}^{N} (\mu(t) + \delta(t))
	\\
  &= \mu (t) \bigg( 1 - \frac{k_0 - k}{N - k} \bigg)
  - (\mu(t) + \delta(t)) \frac{N - k_0}{N - k}
	\\
  &= \mu (t) \bigg( 1 - \frac{\theta_0 - k/N}{1 - k/N} \bigg)
  - (\mu(t) + \delta(t)) \frac{1 - \theta_0 }{1 - k/N} + O(N^{-1})
	\\
  &= - \frac{1 - \theta_0 }{1 - k/N} \delta(t) + O(N^{-1})
\end{align*}
and in the case \(N > k > k_0\) we have (again uniformly in $k$)
\begin{align*}
  B_N(t, k) &= \frac{1}{k} \sum_{j=1}^{k_0} \mu(t)
  + \frac{1}{k} \sum_{j = k_0 + 1}^{k} (\mu(t) + \delta(t))
  - \frac{1}{N - k} \sum_{j = k + 1}^{N} (\mu(t) + \delta(t))
\\
  &= \frac{\theta_0}{k/N} \mu(t) + \frac{k/N- \theta_0}{k/N} (\mu(t) + \delta(t))
  - \frac{1 - k/N}{1 - k/N} (\mu(t) + \delta(t)) + O(N^{-1})
\\
  &= - \frac{\theta_0}{k/N} \delta(t) + O(N^{-1}) \, .
\end{align*}
Hence we obtain
\begin{align} \label{hol2}
  \frac{k}{N}\Big(1-\frac{k}{N}\Big)  \int_T B_N(t, k)^2 dt = d(k/N) + O(N^{-1}) \, ,
\end{align}
uniformly with respect to \(k\) and
\begin{align*}
\frac{k}{N}\Big(1-\frac{k}{N}\Big) \int_T A_N(t, k) B_N(t, k) dt
= \int_T A_N(t, k) \delta(t) dt ~ \tilde d(k/N) + o_{\mathbb{P}}(N^{-1/2}) \, .
\end{align*}
{
Therefore we obtain from \eqref{hat f decomposition}, \eqref{hol1}, \eqref{hol2}, Lipschitz continuity of $\theta \mapsto d(\theta), \theta \mapsto \tilde d(\theta)$ and the line above
\begin{equation}\label{hol3}
\mathbb{G}_N(\theta) =  2 \sqrt{N} \bigg\{ \int_T A_N(t, \lfloor N\theta \rfloor) \delta(t) dt \ \tilde{d}(\theta) \bigg\} + o_{\mathbb{P} }(1)
\end{equation}
uniformly with respect to $\theta \in [0,1]$. In order to investigate the leading term on the right hand side observe for  any $\theta \in [1/N, 1)$ the representation
\begin{align*}
\begin{split}
  A_N(t, \lfloor \theta N \rfloor) &= \frac{1}{\lfloor \theta N \rfloor}
  \bigg( \sum_{j=1}^{\lfloor (\theta \wedge \theta_0) N \rfloor} \eta_j^{(1)}(t)
  + \mathbbm{1}\{\theta_0 < \theta\} \sum_{j=\lfloor \theta_0 N \rfloor + 1}
  ^{\lfloor \theta N \rfloor} \eta_j^{(2)}(t) \bigg) \\
  &\hspace{10pt}- \frac{1}{N - \lfloor \theta N \rfloor} \bigg(
  \mathbbm{1}\{\theta_0 \geq \theta\} \sum_{j=\lfloor \theta N \rfloor + 1}^{\lfloor \theta_0 N \rfloor} \eta_j^{(1)}(t)
  + \sum_{j = \lfloor (\theta \vee \theta_0) N \rfloor + 1}^N \eta_j^{(2)}(t) \bigg) \\
  &= \frac{1}{\lfloor \theta N \rfloor}
  \sum_{j=1}^{\lfloor (\theta \wedge \theta_0) N \rfloor} \eta_j^{(1)}(t)
  - \frac{\mathbbm{1}\{\theta_0 \geq \theta\}}{N - \lfloor \theta N \rfloor}
  \sum_{j=\lfloor \theta N \rfloor + 1}^{\lfloor \theta_0 N \rfloor} \eta_j^{(1)}(t) \\
  &\hspace{10pt}+ \frac{\mathbbm{1}\{\theta_0 < \theta\} }{\lfloor \theta N \rfloor}
  \sum_{j=\lfloor \theta_0 N \rfloor + 1}^{\lfloor \theta N \rfloor} \eta_j^{(2)}(t)
  - \frac{1}{N - \lfloor \theta N \rfloor} \sum_{j = \lfloor (\theta \vee \theta_0) N \rfloor + 1}^N \eta_j^{(2)}(t), 
  \end{split}
\end{align*}
which yields
\begin{align*}
  & 2 \sqrt{N} \int_T A_N(t, \lfloor N\theta \rfloor) \delta(t) dt
	\\
  &= \frac{N}{\lfloor \theta N \rfloor} \tilde \Z^{(1)}_N(\theta\wedge\theta_0)
  - \mathbbm{1}\{\theta_0 \geq \theta\} \frac{N}{N - \lfloor \theta N \rfloor}
  (\tilde \Z^{(1)}_N(\theta_0) - \tilde \Z^{(1)}_N(\theta)) \\
  &\hspace{10pt}+ \mathbbm{1}\{\theta_0 < \theta\} \frac{N}{\lfloor \theta N \rfloor}
  (\tilde \Z^{(2)}_N(\theta) - \tilde \Z^{(2)}_N(\theta_0))
  - \frac{N}{N - \lfloor \theta N \rfloor} (\tilde \Z^{(2)}_N(1) - \tilde \Z^{(2)}_N(\theta_0\vee \theta))~,
\end{align*}
where
\begin{align} \label{tilde Z_N}
  \tilde \Z^{(i)}_N(\lambda) = \frac{1}{\sqrt{N}} \sum_{j=1}^{\lfloor \lambda N \rfloor} \int_T \eta_j^{(i)}(t) \delta(t) dt \, .
\end{align}
Finally, note that we have
\[
\sup_{\theta \in [1/N,1)} \Big|\tilde{d}(\theta)\frac{N}{\lfloor \theta N \rfloor} - \frac{\tilde d(\theta)}{\theta}\Big| = o(1),
\quad
\sup_{\theta \in [1/N,1)} \Big|\tilde{d}(\theta)\frac{N}{N-\lfloor \theta N \rfloor} - \frac{\tilde d(\theta)}{1-\theta}\Big| = o(1) \, .
\]
Hence Lemma \ref{jointlemma}, Slutskys Lemma and the continuous mapping theorem yield
 \begin{align} \label{A_N convergence}
  2 \sqrt{N} \bigg\{ \int_T A_N(t, \lfloor N\theta \rfloor) \delta(t) dt \ \tilde{d}(\theta) \bigg\}_{\theta \in [0,1]}
  \rightsquigarrow \big\{\mathbb{G}(\theta)\big\}_{\theta \in [0,1]} \, .
\end{align}
Combing \eqref{A_N convergence} with \eqref{hol3} gives us the weak convergence in \eqref{G convergence}. }

\bigskip

\textbf{Step 2:}
Given the weak convergence in~\eqref{G convergence} we are ready to prove~\eqref{eq:ratecp}. The proof will proceed in three steps. First, we show that $\hat \theta = \theta_0 + o_{\mathbb{P}}(1)$. In the second step we show that
\begin{equation}\label{eq:helpratecp}
\hat \theta = \theta_0 + o_{\mathbb{P}}(N^{-1/4}) \, .
\end{equation}
In the final step we derive~\eqref{eq:ratecp}.

Observe that the function \(\theta \mapsto d(\theta)\), defined in \eqref{d definition}, is strictly increasing in
\([0, \theta_0]\) and strictly decreasing in \((\theta_0, 1]\). Therefore, for any
\(\tilde{\delta} > 0\), there is an \(\varepsilon > 0\) such that \(|\theta - \theta_0| > \tilde{\delta}\) implies \(d(\theta_0) - d(\theta) > \varepsilon\). Now let \(\tilde{\delta} > 0\) be arbitrary and assume \(|\hat{\theta} - \theta| > \tilde{\delta}\). Using that \(\hat k := N\hat{\theta}\) is the maximizer of the function \(k \mapsto f(k)\), the result from Step 1 and the previously mentioned monotonicity property, we obtain
\begin{align} \label{CP consistency}
\begin{split}
  0 &\geq \hat{f}(k_0) - \hat{f}(\hat{k})
  = \hat{f}(k_0) - d(\theta_0) - (\hat{f}(\hat{k}) - d(\hat{\theta}))
  + d(\theta_0) - d(\hat{\theta}) \\
  &= O_\mathbb{P}(N^{-1/2}) + d(\theta_0) - d(\hat{\theta})
  > O_\mathbb{P}(N^{-1/2}) + \varepsilon
\end{split}
\end{align}
for some \(\varepsilon > 0\), where $k_0 := \lfloor N \theta_0 \rfloor$. This means that
\begin{align*}
  \mathbb{P}\big(|\hat{\theta} - \theta_0| > \tilde{\delta} \, \big)
  \leq \mathbb{P}(O_\mathbb{P}(N^{-1/2}) < -\varepsilon)
  \to 0
\end{align*}
as \(N \to \infty\) and therefore, \(\hat{\theta}\) converges to the true change point \(\theta_0\)
in probability.

Next we show that \(|\hat{\theta} - \theta_0| = O_\mathbb{P}(N^{-1/2})\). Making a Taylor expansion of \(d\)
at the point \(\theta_0\), we obtain, as \(\theta \to \theta_0\),
\begin{align*}
  d(\theta) = d(\theta_0) + c\big(-(\theta - \theta_0) \mathds{1}\{ \theta > \theta_0 \}
  + (\theta - \theta_0) \mathds{1}\{ \theta \leq \theta_0 \} \big)
  + O\big( (\theta - \theta_0)^2 \big)
\end{align*}
for some constant \(c > 0\). Therefore, as \(\theta \to \theta_0\), we can find a constant
\(\tilde{\delta} > 0\) such that
\begin{align} \label{taylor property}
  d(\theta_0) - d(\theta) \geq \tilde{\delta} \, |\theta - \theta_0| + O( (\theta - \theta_0)^2 ) \, .
\end{align}
Since \(\hat{\theta}\) is a consistent estimator of \(\theta_0\) (by the discussion in the previous paragraph), we can use this property
and similar arguments as in \eqref{CP consistency} to obtain
\begin{align*}
  0 \leq \hat{f}(\hat{k}) - \hat{f}(k_0)
  = O_\mathbb{P}(N^{-1/2}) + d(\hat{\theta}) - d(\theta_0)
  \leq O_\mathbb{P}(N^{-1/2}) - \tilde{\delta} \, |\hat{\theta} - \theta_0|
\end{align*}
which means that \(|\hat{\theta} - \theta_0| = O_\mathbb{P}(N^{-1/2})\).

Thus, with probability converging to \(1\), we have
$\hat{\theta} \in \argmax_{\theta : |\theta - \theta_0| \leq N^{-1/4}} \hat{f}(\lfloor N \theta \rfloor)$.
Since the process \(\mathbb{G}_N\) in \eqref{G convergence} is stochastically equicontinuous, we get
\begin{align*}
  | \hat{f}(\hat k) - \hat{f}(k_0)
  - (d(\hat{\theta}) - d(\theta_0)) |
  &\leq \sup_{\theta : |\theta - \theta_0| \leq N^{-1/4}}
  | \hat{f}( \lfloor N\theta \rfloor) - \hat{f}(k_0)
  - (d(\theta) - d(\theta_0)) | \\
  &\leq \sup_{\theta, \theta^\prime : |\theta - \theta^\prime| \leq N^{-1/4}}
  N^{-1/2} |\mathbb{G}_N(\theta) - \mathbb{G}_N(\theta^\prime)| \\
  &= o_\mathbb{P}(N^{-1/2}) \, .
\end{align*}
Using this rate and the bound in \eqref{taylor property} yields
\begin{align*}
  0 \leq \hat{f}(\hat k) - \hat{f}(k_0)
  &\leq d(\hat{\theta}) - d(\theta_0) + | \hat{f}(\hat k) - \hat{f}(k_0)
  - (d(\hat{\theta}) - d(\theta_0)) | \\
  &\leq d(\hat{\theta}) - d(\theta_0) + o_\mathbb{P}(N^{-1/2}) \\
  &\leq -\tilde{\delta} \, |\theta_0 - \hat{\theta}| + o_\mathbb{P}(N^{-1/2})
\end{align*}
which finally implies \(|\hat{\theta} - \theta_0| = o_\mathbb{P}(N^{-1/2})\). \hfill $\Box$

\subsubsection{Proof of Theorem \ref{thmcp}}

We begin by stating some useful technical results and notations. Define
\begin{align} \label{S tilde}
  \tilde S_N^{(k)}(t, \lambda)
  = \frac{1}{N} \sum_{j=1}^{\lfloor \lambda N \rfloor} \eta^{(k)}_j(t), \quad k=1,2
\end{align}
where $\eta^{(1)}_j := f_1(\eps_j,\eps_{j-1,...}), \eta^{(2)}_j := f_2(\eps_j,\eps_{j-1,...})$ for $j\in \Z$. Since $f_1,f_2$ satisfy assumptions (A3), (A4), it follows from Theorem~1.1 in~\cite{berhorric2013} that there exist random elements in $\mathcal{G}$ (recall the beginning of Section~\ref{sec6}), say $\Gamma_N^{(i)}$, with
\begin{equation}\label{eq:tSnGauss}
\sup_{\lambda \in [0,1]} \int_T (\sqrt{N}\tilde S_N^{(i)}(t, \lambda) - \Gamma_N^{(i)}(t,\lambda))^2 dt = o_P(1) \, , \quad i=1,2 \, ,
\end{equation}
where each $\Gamma_N^{(i)}$ satisfies the analogue of~\eqref{approx2}-\eqref{eq:equicont2} with covariance kernels corresponding to $\eta_i^{(1)}$ and $\eta_i^{(2)}$, respectively. 

\bigskip
%


{First consider the case $\int \delta^2(t) dt \neq 0$.} Recalling that $\hat{\theta} = \theta_0 + o_\mathbb{P}(N^{-1/2})$ by Proposition~\ref{propcpest} we proceed in several steps.
First, we show that for the process
\begin{align*}
  \Z_N(\lambda,\theta) = \sqrt{N} \int_T \big( D_N^{cp}(t,\lambda,\theta)^2
  - \lambda^2 \delta(t)^2 \big) dt \, ,
\end{align*}
we have
\begin{align} \label{Z_N knownCP}
  \big\{ \Z_N(\lambda,\theta_0) \big\}_{\lambda \in [0,1] }
  \weak \big\{ \lambda \tau_{\delta,\theta_0} \Bb(\lambda) \big\}_{\lambda\in [0,1]}
\end{align}
in $\ell^\infty([0,1])$,
where $\{ \Bb(\lambda) \}_{\lambda\in [0,1]}$ is a Brownian motion and $\tau_{\delta,\theta_0}$ is a parameter depending on the covariance structure of $\{(\eta_{j}^{(1)}\},\{\eta_{j}^{(2)})\}_{j\in \Z}$ and the true change point location $\theta_0$. Second we prove
\begin{align} \label{Z_N assertion}
  \sup_{\lambda \in [0,1]} |\Z_N(\lambda,\theta_0) - \Z_N(\lambda,\hat{\theta})|
  = o_{\mathbb{P}}(1) \, ,
\end{align}
where \(\hat{\theta}\) is the estimator of \(\theta_0\) defined in \eqref{eq:hattheta}.
Finally we can again use the same arguments as in the proof of Theorem \ref{thm_relevantfunc}
to obtain the assertion.

\bigskip

{Next, consider the case $\int_T \delta^2(t)dt = 0$. It suffices to show that $\hat{\mathbb{D}}_{N}^{cp} = o_{\Prob}(1), \hat{\mathbb{V}}_{N}^{cp}  = o_\Prob(1)$. To this end define the partial sum process
\[
W_N(t,\lambda) := \frac{1}{N} \Big\{ \sum_{i=1}^{\floor{N(\lambda\wedge\theta_0)}} \eta_i^{(1)}(t) + \sum_{i=\floor{N\theta_0} + 1}^{\floor{N(\lambda\vee\theta_0)}} \eta_i^{(2)}(t)\Big\}
\]
and observe that by~\eqref{eq:tSnGauss} and some elementary computations we have
\[
\sup_{\lambda \in [0,1]} \int_T W_N^2(t,\lambda) dt = o_\Prob(1) \, .
\]
Next, observing that
\begin{multline*}
D^{cp}_N(t,\lambda,\theta)
\\
= \frac{N}{\floor{N\theta}} W_N\Big(t,\frac{\floor{\lambda\floor{N\theta}}}{N}\Big) - \frac{N}{N-\floor{N\theta}}\Big\{W_N\Big(t,\frac{\floor{N\theta}+\floor{\lambda(N-\floor{N\theta})}}{N}\Big) - W_N\Big(t,\frac{\floor{N\theta}}{N}\Big)\Big\} \, ,
\end{multline*}
some elementary calculations taking into account that by definition $\hat \theta \in [\eps,1-\eps]$ show that
\begin{align*}
\hat{\mathbb{V}}_N^{cp} &\leq 4 \sup_{\lambda \in [0,1], \theta \in [\eps,1-\eps]} \int_T \{D_N^{cp}(t,\lambda,\theta) \}^2 dt \lesssim \frac{1}{\eps^2 } \sup_{\lambda \in [0,1]} \int_T W_N^2(t,\lambda) dt = o_\Prob(1) \, .
\end{align*}
Similar but simpler arguments show that $\hat{\mathbb{D}}_N^{cp} = o_\Prob(1)$ and this completes the proof in the case $\int_T \delta^2(t)dt = 0$.
}

%

\bigskip

\textbf{Proof of~\eqref{Z_N knownCP}}. Define the processes
\begin{align*}
  S_N^{(1)} (t,\lambda,\theta) &= \frac{1}{\lfloor \theta N \rfloor}
  \sum_{j=1}^{\lfloor \lambda \lfloor \theta N \rfloor \rfloor}
  (X_j(t) - \mu(t)) \\
  S_N^{(2)} (t,\lambda,\theta) &= \frac{1}{N - \lfloor \theta N \rfloor}
  \sum_{j = \lfloor \theta N \rfloor + 1}^{\lfloor \theta N \rfloor + \lfloor \lambda (N - \lfloor \theta N \rfloor ) \rfloor} (X_j(t) - \mu(t) - \delta(t))
\end{align*}
and similar to the calculations in Section~\ref{sec61} we can write
\begin{align} \label{Z_N rewrite1}
\begin{split}
  \Z_N(\lambda, \theta)
  &= \sqrt{N} \int_T \big( D_N^{cp}(t,\lambda,\theta)
  - \lambda \delta(t) \big)^2 dt
  + 2 \sqrt{N} \int_T \lambda \delta(t)
  \big( D_N^{cp}(t,\lambda,\theta) - \lambda \delta(t) \big) dt \\
  &= \sqrt{N} \int_T \big( S_N^{(1)}(t,\lambda,\theta)
  - S_N^{(2)}(t,\lambda,\theta ) \big)^2 dt \\
  &\hspace{10pt}+ 2 \sqrt{N} \int_T \lambda \delta(t)
  \big( S_N^{(1)}(t,\lambda,\theta)
  - S_N^{(2)}(t,\lambda,\theta) \big) dt
  + o_{\mathbb{P}}(1)
 \end{split}
\end{align}
uniformly in $\lambda\in[0,1], \theta \in (\eps,1-\eps)$ for any $\eps > 0$. For $\theta = \theta_0$, the first term at the end of the calculation above converges to zero (as in the two sample case)
and the second term can be rewritten such that Lemma \ref{jointlemma} can be applied
\begin{align} \label{Z_N rewrite2}
\begin{split}
  \Z_N(\lambda, \theta_0)
  &= 2 \sqrt{N} \int_T \lambda \delta(t)
  \big( S_N^{(1)}(t,\lambda,\theta_0)
  - S_N^{(2)}(t,\lambda,\theta_0) \big) dt
  + o_{\mathbb{P}}(1) \\
  &= \Z_N^{(1)}(\lambda, \theta_0) - \Z_N^{(2)}(\lambda, \theta_0)
  + o_{\mathbb{P}}(1) \, ,
\end{split}
\end{align}
where
\begin{align} \label{Z_N^{(i)} def}
  \Z_N^{(i)}(\lambda,\theta) = 2 \sqrt{N} \int_T \lambda \delta(t)
  S_N^{(i)}(t,\lambda,\theta) dt \, , ~ i=1,2 \, .
\end{align}
We can rewrite the processes in \eqref{Z_N rewrite2} as
\begin{align*}
  \Z_N^{(1)}(\lambda, \theta_0) - \Z_N^{(2)}(\lambda, \theta_0)
  &= 2 \frac{N}{\lfloor \theta_0 N \rfloor} \lambda \, \frac{1}{\sqrt{N}}
  \sum_{j=1}^{\lfloor \lambda \lfloor \theta_0 N \rfloor \rfloor} \int_T \eta_j^{(1)}(t) \delta(t) dt \\
  &\quad - 2 \frac{N}{N - \lfloor \theta_0 N \rfloor} \lambda \, \frac{1}{\sqrt{N}}
  \sum_{j=\lfloor \theta_0 N \rfloor + 1}^{\lfloor \theta_0 N \rfloor + \lfloor \lambda (N - \lfloor \theta_0 N \rfloor) \rfloor}
  \int_T \eta_j^{(2)}(t) \delta(t) dt
	\\
  &= \frac{2\lambda}{\theta_0} \, \tilde \Z^{(1)}_N(\lfloor \lambda \lfloor \theta_0 N \rfloor \rfloor / N) \\
  &\quad - \frac{2\lambda}{1-\theta_0} \Big\{ \tilde \Z^{(2)}_N( (\lfloor \theta_0 N \rfloor + \lfloor \lambda (N - \lfloor \theta_0 N \rfloor) \rfloor)/N)
  - \tilde \Z^{(2)}_N(\theta_0) \Big\} + o_{\mathbb{P}}(1) \, ,
\end{align*}
where the remainder is uniform in $\lambda \in [0,1]$ and the processes $\tilde \Z^{(1)}_N$ and $\tilde \Z^{(2)}_N$ are defined in \eqref{tilde Z_N}.
An application of Lemma \ref{jointlemma} with $\zeta = \delta$, asymptotic equicontinuity of the sample paths of $\tilde \Z^{(i)}_N$, and the continuous mapping theorem yield the assertion in \eqref{Z_N knownCP}. To see that the limit has the right structure, observe that
\begin{align} \label{Z_N limit}
  \Z_N^{(1)}(\lambda, \theta_0) - \Z_N^{(2)}(\lambda, \theta_0)
  &\weak \frac{2\lambda}{\theta_0} \, \tilde \Z^{(1)}(\lambda  \theta_0 )
  - \frac{2\lambda}{1-\theta_0} \Big\{ \tilde \Z^{(2)}( \theta_0
  +  \lambda (1 - \theta_0 )) - \tilde \Z^{(2)}(\theta_0) \Big\}
\end{align}
where
\begin{align*}
\tilde \Z^{(1)} = \tilde \Sigma_{11} \Bb_1 + \tilde \Sigma_{12} \Bb_2 ;
\qquad \tilde \Z^{(2)} = \tilde \Sigma_{21} \Bb_1 + \tilde \Sigma_{22} \Bb_2 \, ,
\end{align*}
and $\tilde \Sigma_{ij}$ denotes the $ij$-th entry of $\Sigma^{1/2}$.
By straightforward calculations one obtains
\begin{align*}
\text{Cov} \big( \tilde \Z^{(1)}(\lambda_1  \theta_0 ), \tilde \Z^{(2)}( \theta_0
  +  \lambda_2 (1 - \theta_0 )) - \tilde \Z^{(2)}(\theta_0) \big) &= 0 \, , \\
\text{Cov}(\tilde \Z^{(1)}(\lambda_1  \theta_0 ), \tilde \Z^{(1)}(\lambda_2  \theta_0 )) &= (\lambda_1 \wedge \lambda_2) \theta_0 (\tilde \Sigma_{11}^2 + \tilde \Sigma_{12}^2) {= (\lambda_1 \wedge \lambda_2) \theta_0\Sigma_{11}}
\end{align*}
{where $\Sigma$ is defined in the statement of Lemma~\ref{jointlemma} and the last equation follows from the fact that $\tilde\Sigma$ is symmetric and that $\tilde\Sigma \tilde\Sigma = \Sigma$.}
Furthermore, $ \tilde \Z^{(2)}(\theta_0 + \lambda (1 - \theta_0)) - \tilde \Z^{(2)}(\theta_0)$ has the same
distribution as $\tilde \Z^{(2)}(\lambda (1 - \theta_0))$ and
\begin{align*}
\text{Cov}(\tilde \Z^{(2)}(\lambda_1 (1 - \theta_0)), \tilde \Z^{(2)}(\lambda_2 (1 - \theta_0)))
= (\lambda_1 \wedge \lambda_2) (1-\theta_0) (\tilde \Sigma_{21}^2 + \tilde \Sigma_{22}^2) {= (\lambda_1 \wedge \lambda_2) (1-\theta_0)\Sigma_{22} \, .}
\end{align*}
Combining the calculations above, we can conclude that the limit process in \eqref{Z_N limit} is of the form as claimed in \eqref{Z_N knownCP}. 


\bigskip

\textbf{Proof of~\eqref{Z_N assertion}}. For $\theta = \hat \theta$, we
show that the first term at the end of the calculation in \eqref{Z_N rewrite1}

vanishes by proving
\begin{align} \label{Sni}
  N^{1/4} \sup_{\lambda \in [0,1]}
  \| S_N^{(i)}(\ \cdot \ ,\lambda,\hat{\theta}) \|
  = o_{\mathbb{P}}(1)
\end{align}
for \(i=1,2\). Since both cases are similar we only consider the case \(i=1\). Write
\begin{align} \label{S1=Q1+Q2}
  N^{1/4} S_N^{(1)}(t,\lambda,\hat{\theta})
  = \frac{1}{N^{1/4}}
  \frac{N}{\lfloor \hat{\theta} N \rfloor}
  \big(Q_N^{(1)}(t,\lambda,\hat{\theta}) + Q_N^{(2)}(t,\lambda,\hat{\theta}) \big)
\end{align}
where
\begin{align} \label{Q1Q2}
\begin{split}
  Q_N^{(1)}(t,\lambda,\hat{\theta}) &=  \frac{1}{\sqrt{N}}
  \sum_{j=1}^{\lfloor \lambda \lfloor \hat{\theta} N \rfloor \rfloor \wedge \lfloor \theta_0 N \rfloor}
  (X_j(t) - \mu(t))
  = \frac{1}{\sqrt{N}}
  \sum_{j=1}^{\lfloor \lambda \lfloor \hat{\theta} N \rfloor \rfloor \wedge \lfloor \theta_0 N \rfloor}
  \eta^{(1)}_j(t) \\
  Q_N^{(2)}(t,\lambda,\hat{\theta}) &= \frac{1}{\sqrt{N}}
  \sum_{j=\lfloor \lambda \lfloor \hat{\theta} N \rfloor \rfloor \wedge \lfloor \theta_0 N \rfloor+1}
  ^{\lfloor \lambda \lfloor \hat \theta N \rfloor \rfloor}
  (X_j(t) - \mu(t)) \, .
  \end{split}
\end{align}
We have by~\eqref{eq:tSnGauss} and the properties of $\Gamma_N^{(i)}$
\begin{align*}
  &\lim_{p\to\infty} \limsup_{N\to\infty} \mathbb{P} \Big(\sup_{\lambda\in[0,1]}
  \| Q_N^{(1)}(\ \cdot \ ,\lambda,\hat{\theta}) \| > p\Big)
	\\
  \leq& \lim_{p\to\infty} \limsup_{N\to\infty} \mathbb{P} \Big(
  \sup_{\lambda\in[0,1]} \sqrt{N}\| \tilde S_N^{(1)}(\cdot, \lambda) \| > p\Big)
\\
=& \lim_{p\to\infty} \limsup_{N\to\infty} \mathbb{P} \Big(\sup_{\lambda\in[0,1]}
  \int_T \{\Gamma_N^{(1)}(t,\lambda) \}^2 dt > p^2 \Big) + o(1)  = 0
\end{align*}
where $\tilde S_N^{(1)}$ is defined in~\eqref{S tilde}. Therefore \( \sup_{\lambda\in[0,1]} \| Q_N^{(1)}(\ \cdot \ ,\lambda,\hat{\theta}) \| = O_{\mathbb{P}}(1) \). The second term in \eqref{Q1Q2} is zero if $\floor{\lambda\floor{\hat \theta N}} \leq \floor{\theta_0 N}$ or, if $\floor{\lambda\floor{\hat \theta N}}> \floor{\theta_0 N}$, we can write
\begin{align*}
  \sup_{\lambda\in[0,1]} \| Q_N^{(2)}(\ \cdot \ ,\lambda,\hat{\theta}) \|
  = \sup_{\lambda\in[0,1]} \Big\| \frac{1}{\sqrt{N}}
  \sum_{j= \lfloor \theta_0 N \rfloor+1}
  ^{\lfloor \lambda \lfloor \hat \theta N \rfloor \rfloor}
  (\eta_j^{(2)} + \delta) \Big\| \, .
\end{align*}
The number of terms in the sum above is bounded by the distance between \(\theta_0\) and \(\hat{\theta}\) in the sense that
\begin{align} \label{Q2bound}
\begin{split}
&\sup_{\lambda\in[0,1]} \Big\| \frac{1}{\sqrt{N}}
  \sum_{j=\lfloor \theta_0 N \rfloor+1}
  ^{\lfloor \lambda \lfloor \hat \theta N \rfloor \rfloor}
  (\eta_j^{(2)} + \delta) \Big\|
\\
\lesssim&
  \sup_{\substack{\nu,\lambda \in[0,1]: \\ |\nu-\lambda|\leq |\theta_0-\hat{\theta}|}}
  \sqrt{N} \big\| \tilde S_N^{(2)}( \ \cdot \ ,\nu) - \tilde S_N^{(2)}( \ \cdot \ ,\lambda) \big\|
  + \frac{\lfloor \hat \theta N \rfloor - \lfloor \theta_0 N \rfloor }
  {\sqrt{N}} \|\delta\|
\\
=& \sup_{\substack{\nu,\lambda \in[0,1]: \\ |\nu-\lambda|\leq |\theta_0-\hat{\theta}|}} \sqrt{\int_T \{\Gamma_N^{(2)}( t,\nu) - \Gamma_N^{(2)}( t,\lambda)\}^2 dt }
+ \frac{\lfloor \hat \theta N \rfloor - \lfloor \theta_0 N \rfloor }
  {\sqrt{N}} \|\delta\| + o_{\mathbb{P}}(1) \, ,
\end{split}
\end{align}
where the last equality follows by~\eqref{eq:tSnGauss} and the definition of $\|\cdot\|$. The first term in \eqref{Q2bound} converges to zero in probability since~\eqref{eq:equicont2} holds with $\Gamma_N^{(2)}$ instead of $\Gamma$ and since by~\eqref{eq:ratecp}
\(|\theta_0-\hat{\theta}| = o_\mathbb{P}(1/\sqrt{N})\). The latter also implies that the second term converges to zero in probability. Therefore we have \( \sup_{\lambda\in[0,1]} \| Q_N^{(2)}(\ \cdot \ ,\lambda,\hat{\theta}) \|= o_{\mathbb{P}}(1) \). Recalling \eqref{S1=Q1+Q2}, we conclude that \eqref{Sni} holds in the case $i=1$ and similar arguments yield the statement
for $i=2$. This means that we can continue the calculations in \eqref{Z_N rewrite1} for
$\theta = \hat \theta$ and obtain
\begin{align*}
  &\sqrt{N} \int_T \big( S_N^{(1)}(t,\lambda,\hat{\theta})
  - S_N^{(2)}(t,\lambda,\hat{\theta} ) \big)^2 dt + 2 \sqrt{N} \int_T \lambda \delta(t)
  \big( S_N^{(1)}(t,\lambda,\hat{\theta})
  - S_N^{(2)}(t,\lambda,\hat{\theta}) \big) dt \\
  &=  2 \sqrt{N} \int_T \lambda D^{cp}(t)
  \big( S_N^{(1)}(t,\lambda,\hat{\theta})
  - S_N^{(2)}(t,\lambda,\hat{\theta}) \big) dt + o_{\mathbb{P}}(1) \\
  &= \Z_N^{(1)}(\lambda,\hat{\theta}) - \Z_N^{(2)}(\lambda,\hat{\theta})
  + o_{\mathbb{P}}(1) \, ,
\end{align*}
where $\Z_N^{(i)}$, for $i=1,2$, is defined by \eqref{Z_N^{(i)} def}.
In order to prove \eqref{Z_N assertion} it consequently remains to show
\begin{align*}
  \sup_{\lambda\in[0,1]} | \Z_N^{(i)}(\lambda,\hat{\theta})
  - \Z_N^{(i)}(\lambda,\theta_0) | = o_\mathbb{P}(1)
\end{align*}
for \(i=1,2\).
For that purpose we write
\begin{align*}
& \sup_{\lambda\in[0,1]} | \Z_N^{(1)}(\lambda,\hat{\theta}) - \Z_N^{(1)}(\lambda,\theta_0) |
\\
&=\sup_{\lambda\in[0,1]} \bigg|2\sqrt{N} \int_T \lambda \delta(t) \bigg\{
  \bigg( \frac{1}{\lfloor \hat{\theta} N \rfloor} - \frac{1}{\lfloor \theta_0 N \rfloor}
  \bigg) \sum_{j=1}^{\lfloor\lambda\lfloor (\hat{\theta}\wedge\theta_0)N\rfloor \rfloor}
  \big( X_j(t) - \mu(t) \big)
	\\
  &\hspace{10pt} + \big(\mathds{1}\{\hat{\theta}\geq\theta_0\}
  - \mathds{1}\{\hat{\theta}<\theta_0\}\big)
  \frac{1}{\lfloor (\hat{\theta}\vee\theta_0)N \rfloor}
  \sum_{j = \lfloor \lambda \lfloor (\hat{\theta}\wedge\theta_0)N \rfloor \rfloor + 1}
  ^{\lfloor \lambda \lfloor (\hat{\theta}\vee\theta_0)N \rfloor \rfloor}
  \big( X_j(t) - \mu(t) \big) \bigg\} dt \bigg| \\
  &\leq 2\|\delta \| \lambda \bigg\{ \frac{N}{\lfloor \theta_0 N \rfloor} \
  \frac{| \lfloor \theta_0 N \rfloor - \lfloor \hat{\theta} N \rfloor |}
  {\lfloor \hat{\theta} N \rfloor}
  \sup_{\lambda\in [0,1]} \bigg\| \frac{1}{\sqrt{N}}
  \sum_{j=1}^{\lfloor\lambda\lfloor (\hat{\theta}\wedge\theta_0)N\rfloor \rfloor}
  ( X_j - \mu) \bigg\|
	\\
  &\hspace{10pt} + \frac{N}
  {\lfloor (\hat{\theta}\vee\theta_0)N\rfloor}
  \sup_{\lambda\in [0,1]} \bigg\|
  \frac{1}{\sqrt{N}}
  \sum_{j = \lfloor \lambda \lfloor (\hat{\theta}\wedge\theta_0)N \rfloor \rfloor + 1}
  ^{\lfloor \lambda \lfloor (\hat{\theta}\vee\theta_0)N \rfloor \rfloor}
  ( X_j - \E[X_j]) \bigg\|
\\
  &\hspace{10pt} + \frac{N}
  {\lfloor (\hat{\theta}\vee\theta_0)N\rfloor}
  \sup_{\lambda\in [0,1]}
  \frac{1}{\sqrt{N}}
  \sum_{j = \lfloor \lambda \lfloor (\hat{\theta}\wedge\theta_0)N \rfloor \rfloor + 1}
  ^{\lfloor \lambda \lfloor (\hat{\theta}\vee\theta_0)N \rfloor \rfloor}
  \|\delta\| \bigg\}
\\
&\lesssim \frac{|\theta_0 - \hat{\theta}|}{\theta_0\hat{\theta}}
\sup_{\lambda\in [0,1]} \sqrt{N} \| \tilde S_N^{(1)}( \ \cdot \ , \lambda)\|
\\
 &\hspace{10pt} + \frac{1}{\hat{\theta}\vee\theta_0} \bigg( \sum_{i=1}^2
\sup_{\substack{\nu,\lambda \in[0,1]: \\ |\nu-\lambda|\leq |\theta_0-\hat{\theta}|}}
\sqrt{N} \big\| \tilde S_N^{(i)}( \ \cdot \ ,\nu)
- \tilde S_N^{(i)}( \ \cdot \ ,\lambda) \big\|
+ \frac{| \lfloor \hat \theta N \rfloor - \lfloor \theta_0 N \rfloor | }
{\sqrt{N}} \|\delta\|  \bigg)
+ o(1)
\\
&= o_\mathbb{P}(1) \, .
\end{align*}
The last equality holds since
\(|\theta_0-\hat{\theta}| = o_\mathbb{P}(1/\sqrt{N})\),
\( \sup_{\lambda\in[0,1]} \sqrt{N} \| \tilde S_N^{(1)}(\ \cdot \ ,\lambda) \| = O_{\mathbb{P}}(1) \) and since
\[
\sup_{\substack{\nu,\lambda \in[0,1]: \\ |\nu-\lambda|\leq |\theta_0-\hat{\theta}|}} \sqrt{N} \big\| \tilde S_N^{(i)}( \ \cdot \ ,\nu)
- \tilde S_N^{(i)}( \ \cdot \ ,\lambda) \big\|
= \sup_{\substack{\nu,\lambda \in[0,1]: \\ |\nu-\lambda|\leq |\theta_0-\hat{\theta}|}} \sqrt{\int_T \{\Gamma_N^{(i)}( t,\nu) - \Gamma_N^{(i)}( t,\lambda)\}^2 dt } + o_{\mathbb{P}}(1)
\]
which is $o_{\mathbb{P}}(1)$ by similar arguments as given right after~\eqref{Q2bound}. Similar arguments prove
\[
\sup_{\lambda\in[0,1]} | \Z_N^{(2)}(\lambda,\hat{\theta})
  - \Z_N^{(2)}(\lambda,\theta_0) | = o_\mathbb{P}(1)
\]
which finally implies \eqref{Z_N assertion}. \hfill $\Box$


\subsection{Proof of Remark \ref{rem7}} \label{proofrem}
{
The proof is based on the following generalization of Lemma~\ref{jointlemma}. Assume model~\eqref{modmult} and (m1). For an arbitrary finite collection of functions $\zeta_1,...,\zeta_M$ in $L^2(T)$ consider the processes
\[
Z_N^{(j,m)}(\lambda) := \frac{1}{\sqrt{N_j}} \int_T \zeta_m(t) \tilde  S_{N,j}(t, \lambda) dt, \qquad j=1,...,K, ~ m=1,...,M
\] 
where
\[
\tilde  S_{N,j}(t, \lambda) = \frac {1}{N_j}
\sum^{ \lfloor \lambda  N_j \rfloor}_{i= 1  }
\big (  X_{\lfloor N \theta_{j-1}\rfloor+i}(t) - \delta_{j}(t) \big)
\]
with $N_j := \lfloor N\theta_j \rfloor - \lfloor N\theta_{j-1} \rfloor$. Then 
\begin{equation}\label{eq:genLe}
\{Z_N^{(j,m)}(\cdot) \}_{j=1,...,K, m=1,...,M} \rightsquigarrow \Big\{ \sum^\infty_{i=1} \sqrt{\lambda_i} \int_T \phi_i (t) \zeta_m(t) dt~ W^j_i (\cdot) \Big\}_{j=1,...,K, m=1,...,M}
\end{equation}
in $\ell^\infty([0,1])^{MK}$; here $\phi_i$ and $\lambda_i$ are the eigenfunctions and eigenvalues of the integral operator corresponding to the error process $\eta_j := f(\eps_j,\eps_{j-1},...)$ and $\{W^j_i\}_{i,j \in \mathbb{N}}$ is an array of independent Brownian motions. The proof is similar to the proof of Lemma~\ref{jointlemma} and details are omitted for the sake of brevity. \\
\\
We now sketch the proof of Remark \ref{rem7}. Let $\hat {\mathbb{D}}^{cp}_{N,j}(\lambda,\hat \theta), \hat {\mathbb{D}}^{L^2}_{N}(\lambda,\hat \theta)$ and $\hat {\mathbb{V}}^{L^2}_{N}(\hat \theta)$ be defined as in \eqref{remDj}, \eqref{remD} and \eqref{remV}, respectively, and let ${\mathbb{D}}^{cp}_{N,j}(\lambda), {\mathbb{D}}^{L^2}_{N}(\lambda)$ and ${\mathbb{V}}^{L^2}_{N}$ denote the corresponding quantities with $\hat \theta$ replaced by the vector of true change points. Now similar arguments as given in the proof of Theorem \ref{thmcp} show that
\begin{equation}\label{prorem1}
\frac{\hat {\mathbb{D}}^{L^2}_{N}(1,\hat \theta) - \sum_{j=1}^K \Psi_j}{\hat {\mathbb{V}}^{L^2}_{N}(\hat \theta)} = \frac{{\mathbb{D}}^{L^2}_{N}(1)- \sum_{j=1}^K \Psi_j}{ {\mathbb{V}}^{L^2}_{N}} + o_P(1) \, .
\end{equation}
Consequently, it is sufficient to establish the weak convergence of the right-hand side of \eqref{prorem1} to $\mathbb{W}$. To this end consider the partial sum processes 
\[
S_{N,j}(t, \lambda) = \frac {1}{N_j}
 \sum^{ \lfloor \lambda  N_j \rfloor}_{i= 1  }  X_{\lfloor N \theta_{j-1}\rfloor+i} (t) \qquad ( j = 1,\ldots, K+1) \, . 
\]
Then 
\[
{\mathbb{D}}^{cp}_{N,j}(\lambda) = \int_T \{S_{N,j+1}(t, \lambda) - S_{N,j}(t, \lambda)\}^2 dt \qquad ( j = 1,\ldots, K) \, . 
\]
Now a similar calculation as in the proof of Theorem \ref{thmcp} shows
\[
{\mathbb{D}}^{cp}_{N,j}(\lambda) - \lambda^2 \Psi_j = - 2 \lambda \int_T \{ \delta_j (t) -  \delta_{j-1} (t) \} \{\tilde S_{N,j+1} (t, \lambda) - \tilde S_{N,j} (t, \lambda) \} dt + o_{\mathbb{P}}(N^{-1/2})
\]
where $\tilde S_{N,j}$ are centered versions of $S_{N,j}$ defined in the beginning of this proof.\\ 
Applying~\eqref{eq:genLe} for the collection $\zeta_m = \delta_m - \delta_{m-1}, m= 1 ,...,K+1$ in combination with the continuous mapping theorem shows that
\begin{multline*}
\Big\{ \sqrt{N}\Big( {\mathbb{D}}^{cp}_{N}(\lambda) - \lambda^2 \sum_{j=1}^K \Psi_j\Big) \Big\}_{\lambda \in [0,1]}
\\ 
\rightsquigarrow \Big\{ 2 \lambda  \sum_{j=1}^K \sum_{i=1}^\infty
\sqrt{\lambda_i}  \int \{\delta_j (t) -  \delta_{j-1} (t)\}\phi_i(t) dt \Big( \frac { W_{i}^j(\lambda)}{\sqrt{\theta_{j+1}-\theta_j}} - \frac{W_i^{j+1}(\lambda)}{\sqrt{\theta_{j+2}-\theta_{j+1}}}  \Big) \Big\}_{\lambda \in [0,1]}
\end{multline*}
After some tedious but straightforward covariance manipulations it follows that the process on the right-hand side above has the same distribution as 
$\{\tau^2 \lambda \mathbb{B}(\lambda)\}_{\lambda\in[0,1]}$ where $\mathbb{B}$ denotes a standard Brownian motion and $\tau^2$ is a constant depending on various quantities in model~\eqref{modmult}. Now exactly the same arguments as given in the second part of Section~\ref{sec61} show the weak convergence of the right-hand side of~\eqref{prorem1} to the random variable $\mathbb{W}$ which completes the proof. \hfill $\Box$ 
}

\subsection{Outline of proofs for Section~\ref{sec4}}

We begin by some preliminary observations. Assume that $\{X_j\}_{j\in \Z}$ is a functional time series with values in $L^2([0,1])$ satisfying (A1) with mean $\mu$ and errors $\eta_j$ and that those errors satisfy (A2), (A3'), (A4'). Note that the functions $(X\otimes X)(s,t) = X(s)X(t)$ can be interpreted as random elements in $L^2([0,1]^2)$ since
\[
\int_T \int_T \{X_1(s) X_1(t)\}^2 dsdt= \|X_1\|_2^4 < \infty \, .
\] 
Moreover, by some elementary computations
\begin{equation} \label{eq:lappr-tensor}
\sum_{\ell=1}^\infty \Big(\E\Big\|\eta_0\otimes\eta_0 - \E[\eta_0\otimes\eta_0]-  \eta_{0,\ell}\otimes\eta_{0,\ell} + \E[\eta_{0,\ell}\otimes\eta_{0,\ell}]\Big\|^{2+\psi/2}\Big)^{2/\kappa} < \infty
\end{equation}
where $\|\cdot\|$ now denotes the $L^2$ norm on $L^2([0,1]^2)$, see also the proof of Lemma A.3 in~\cite{AueRiceSonmez2018} for similar arguments. 
Now going through the proofs in~\cite{berhorric2013} we find that most of their results hold for spaces of square integrable functions on general subsets of $\R^d$ and that there is no special structure of $T = [0,1]$ that they use. In particular, we obtain the following generalizations of the results in~\cite{berhorric2013}. Let  
\[
\tilde S_n((s,t),\lambda) := \frac{1}{n} \sum_{j=1}^{\lfloor n \lambda \rfloor} \eta_j(t)\eta_j(s)
\]
and equip
\[
\mathcal{G}_2 := \Big\{f: T^2\times[0,1] \to \R: \sup_{\lambda \in [0,1]} \int_T\int_T f^2((s,t),\lambda) dsdt < \infty \Big\}
\]
with the norm 
\[
\|f\|_{\mathcal{G}_2} := \sup_{\lambda \in [0,1]} \int_T\int_T f^2((s,t),\lambda) dsdt \, .
\]
A generalization of Theorem 1.1 in \cite{berhorric2013} implies that there exists a sequence of measurable random elements in $\mathcal{G}_2$,
say $\{\Gamma_n (s,t,\lambda) \}_{\lambda,s,t \in [0,1]} $, such that
\begin{eqnarray}
\label{approx1t}
 && \sup_{\lambda \in [0,1]} \Big\| \sqrt{n} \, \tilde S_n (\cdot, \lambda)  - \Gamma_n (\cdot,\lambda) \Big\|  = o_\Prob (1) \\
\label{approx2t}
 && \{\Gamma_n ((s,t),\lambda)  \}_{\lambda,t \in [0,1]}  \Dequal \{\Gamma ((s,t),\lambda)  \}_{\lambda,s,t \in [0,1]} \, ,
\end{eqnarray}
where $\Gamma$ is defined by
\begin{eqnarray}
\label{approx3t}
\Gamma ((s,t),\lambda) &=& \sum_{i=1}^\infty \sqrt{\lambda_i} \phi_i(s,t) W_i(\lambda) \, .
\end{eqnarray}
$\{W_i\}_{i\in \N} $ is a sequence of independent Brownian motions and $\lambda_i$, $\phi_i$ are the eigenvalues and (orthonormal) eigenfunctions
of the integral operator corresponding to the covariance kernel
\begin{align*}
C((s,t),(s',t')) := &\cov(X_0\otimes X_0(s,t), X_0\otimes X_0(s',t')) + \sum_{\ell=1}^\infty \cov(X_0\otimes X_0(s,t),X_\ell\otimes X_\ell(s',t'))
\\
&+ \sum_{\ell=1}^\infty Cov(X_0\otimes X_0(s,t),X_{-\ell}\otimes X_{-\ell}(s',t')) \, ,
\end{align*}
i.e. 
\begin{eqnarray}
\label{approx4at}
C((s,t),(s',t')) &=&
 \sum_{i=1}^\infty {\lambda_i} \phi_i(s,t)\phi_i(s',t') \, .
\end{eqnarray}
with
\begin{eqnarray}
\label{approx5t}
\lambda_i \phi_i (s,t) = \int_T \int_T C((s',t'),(s,t)) \phi_i (s',t') ds'dt'  ~~~~~  (i\in \N) \, .
\end{eqnarray}
A generalization of Lemma~2.2 in \cite{berhorric2013} further shows that $\sum_k \lambda_k < \infty$ and that
\begin{equation} 
\sup_{0\leq \lambda \leq 1} \int_T \Gamma^2((s,t),\lambda) dt < \infty \quad a.s.
\end{equation}
The latter implies that for any square integrable function $\zeta: [0,1]^2 \to \R$ the process
\[
\bigg\{\int_T\int_T \zeta(s,t) \Gamma((s,t),\lambda) dsdt\bigg\}_{\lambda \in [0,1]}
\]
can be viewed as an element of $\ell^\infty([0,1])$; the same is true for the process $\{\int_T \Gamma^2((s,t),\lambda) dt\}_{\lambda \in [0,1]}$. Moreover, summability of the sequence $(\lambda_k)_{k\in\N}$ together with properties of the modulus of continuity of Brownian motions and similar arguments as give in the derivation of~\eqref{eq:equicont2} imply that for any positive sequence $(\kappa_k)_{k\in\N}$ such that $\kappa_n \to 0$
\begin{equation}
\label{eq:equicont2t}
\sup_{\substack{\nu,\lambda \in[0,1]: \\ |\nu-\lambda|\leq \kappa_n}} \int_T \{\Gamma^2((s,t),\lambda)-\Gamma^2((s,t),\nu)\}^2 dt = o_{\mathbb{P}}(1) \quad (\kappa_n \to 0) \, .
\end{equation}

\subsubsection{Outline of proof for Section~\ref{sec:2sCov}}

Observe that under the assumptions made we have 
\begin{equation}\label{eq:supnorm}
\sup_{\lambda\in[0,1]} \Big\|\frac{1}{m} \sum_{i=1}^{\lfloor m\lambda \rfloor} \eta_i^X \Big\|_2^2 = O_P(m^{-1})
\end{equation}
and the same is true with $\eta_i^Y$ and $n$ replacing $m$. Next note that we have for any $\lambda \geq \eps>0$ for all $m > 1/\eps$
\begin{align*}
& \frac{1}{m-1} \sum_{j=1}^{\lfloor m \lambda \rfloor}
\Big\{\bigg(X_j(s)-\frac{1}{\lfloor m \lambda \rfloor \vee 1}\sum_{i=1}^{\lfloor m \lambda \rfloor} X_i(s)\bigg)
\bigg(X_j(t)-\frac{1}{\lfloor m \lambda \rfloor \vee 1}\sum_{i=1}^{\lfloor m \lambda \rfloor} X_i(t)\bigg) -  \eta_j^X(t)\eta_j^X(s) \Big\}
\\
= & \frac{1}{m-1} \sum_{j=1}^{\lfloor m \lambda \rfloor} \Big\{
\bigg(\eta_j^X(s)-\frac{1}{\lfloor m \lambda \rfloor \vee 1}\sum_{i=1}^{\lfloor m \lambda \rfloor} \eta_i^X(s)\bigg)
\bigg(\eta_j^X(t)-\frac{1}{\lfloor m \lambda \rfloor \vee 1}\sum_{i=1}^{\lfloor m \lambda \rfloor} \eta_i^X(t)\bigg) -  \eta_j^X(t)\eta_j^X(s) \Big\}
\\
= & - \frac{\lfloor m \lambda \rfloor \vee 1}{m-1} \bigg(\frac{1}{\lfloor m \lambda \rfloor \vee 1}\sum_{i=1}^{\lfloor m \lambda \rfloor} \eta^X_i(s)\bigg)\bigg(\frac{1}{\lfloor m \lambda \rfloor \vee 1}\sum_{i=1}^{\lfloor m \lambda \rfloor} \eta^X_i(t)\bigg).
\end{align*}
Define 
\[
\widetilde D_{n,m}(s,t,\lambda) := \frac{1}{m} \sum_{j=1}^{\lfloor m \lambda \rfloor} \eta_j^X(t)\eta_j^X(s) - \frac{1}{n} \sum_{j=1}^{\lfloor n \lambda \rfloor} \eta_j^Y(t)\eta_j^Y(s).
\]
The above calculation combined with~\eqref{eq:supnorm} shows that
\begin{equation}\label{eq:Dequiv}
\sup_{1 \geq \lambda \geq \eps}\int_T \int_T \{D_{n,m}(s,t,\lambda) - \tilde D_{n,m}(s,t,\lambda)\}^2 dsdt = o_P((n+m)^{-1}) \, .
\end{equation}
Indeed, observe that
\begin{align*}
D_{n,m}(s,t,\lambda) - \tilde D_{n,m}(s,t,\lambda) = & - \frac{\lfloor m \lambda \rfloor \vee 1}{m-1} \bigg(\frac{1}{\lfloor m \lambda \rfloor \vee 1}\sum_{i=1}^{\lfloor m \lambda \rfloor} \eta^X_i(s)\bigg)\bigg(\frac{1}{\lfloor m \lambda \rfloor \vee 1}\sum_{i=1}^{\lfloor m \lambda \rfloor} \eta^X_i(t)\bigg)
\\
&+ \frac{\lfloor n \lambda \rfloor \vee 1}{n-1} \bigg(\frac{1}{\lfloor n \lambda \rfloor \vee 1}\sum_{i=1}^{\lfloor n \lambda \rfloor} \eta^Y_i(s)\bigg)\bigg(\frac{1}{\lfloor n \lambda \rfloor \vee 1}\sum_{i=1}^{\lfloor n \lambda \rfloor} \eta^Y_i(t)\bigg)
\end{align*}
It suffices to bound both terms on the right individually. Since $\lambda \geq \eps$ we have $m/\floor{m\lambda} = O(1)$. Moreover
\begin{align*}
& \int_t\int_T \Big\{\bigg(\frac{1}{\lfloor m \lambda \rfloor \vee 1}\sum_{i=1}^{\lfloor m \lambda \rfloor} \eta^X_i(s)\bigg)\bigg(\frac{1}{\lfloor m \lambda \rfloor \vee 1}\sum_{i=1}^{\lfloor m \lambda \rfloor} \eta^X_i(t)\bigg)\Big\}^2 dsdt
\\
= & \Big\|\frac{1}{\lfloor m \lambda \rfloor \vee 1}\sum_{i=1}^{\lfloor m \lambda \rfloor} \eta^X_i \Big\|^2 \Big\|\frac{1}{\lfloor m \lambda \rfloor \vee 1}\sum_{i=1}^{\lfloor m \lambda \rfloor} \eta^X_i\Big\|^2
\\
= & O(1) \Big\|\frac{1}{m}\sum_{i=1}^{\lfloor m \lambda \rfloor} \eta^X_i \Big\|^4 = O_P(m^{-2})
\end{align*} 
where the last inequality follows from~\eqref{eq:supnorm}. The other term can be bounded similarly.
 
Hence by a simple calculation involving the Cauchy-Schwarz inequality we find that 
\begin{align*}
\sqrt{n+m}~\hat{\mathbb{D}}_{n,m}^C &= \int_T\int_T \widetilde D_{n,m}(s,t,\lambda)^2 ds dt + o_P(1)
\\
\sqrt{n+m}~\hat{\mathbb{V}}_{m,n}^C &=
\Big( \int_0^1  \Big[ \int_T\int_T \widetilde D_{m,n}(s,t,\lambda )^2 dsdt
- \lambda^2 \int_T\int_T \widetilde D_{m,n}(s,t,1)^2 dsdt\Big]^2 \nu (d\lambda) \Big)^{1/2} + o_P(1) 
\end{align*}
where we used the fact that $\nu$ does not place any mass in a neighbourhood of zero in the second identity. Hence it suffices to analyze the test based on $\widetilde D_{m,n}$ instead of $D_{m,n}$. This can now be done by exactly the same arguments as in the proof of Theorem~\ref{thm_2s} after replacing~\eqref{approx1t}-\eqref{approx3t} in that proof by~\eqref{approx1}-\eqref{approx3}. Details are omitted for the sake of brevity. \hfill $\Box$

\subsubsection{Outline of proof for Section~\ref{sec:cpCov}}  

Similarly as in the two sample case, the arguments are very similar to those given in Section~\ref{sec63} and hence we only outline the main steps. We begin by stating an extension of Lemma~\ref{jointlemma} in Section~\ref{sec:jointpr}. Recall the model described in~\eqref{modcp_cov} and define $\eta_i := X_i - \mu$. By assumption, $\eta_i^{(k)}$ satisfy (A4'). 

\begin{lemma} \label{jointlemma_2}
In the setting above consider a fixed (but arbitrary) function $\zeta$ in $L^2(T^2)$. For $\lambda \in [0,1]$ define the processes
\[
\tilde Z_N^{(k)}(\lambda) := \frac{1}{\sqrt{N}}\sum_{i=1}^{\floor{N\lambda}} \int_T\int_T \eta^{(k)}_i(s)\eta^{(k)}_i(t) \zeta(s,t) dsdt, \quad k=1,2 \, ,
\]
then
\[
(\tilde Z_N^{(1)},\tilde Z_N^{(2)})^\top \weak \Sigma^{1/2} (\Bb_1,\Bb_2)^\top \quad \mbox{in  } \ell^{\infty}([0,1])^2  \, ,
\]
where $\Bb_1,\Bb_2$ are two independent standard Brownian motions on the interval $[0,1]$ and $\Sigma$ is a symmetric $2\times 2$ matrix with finite entries given by
\[
\Sigma_{ij} = \sum_{h \in \Z} \int_T\int_T\int_T\int_T \text{Cov}(\eta^{(i)}_0\otimes \eta^{(i)}_0(s,t),\eta^{(j)}_h\otimes \eta^{(j)}_h(s',t'))~\zeta(s,t)\zeta(s',t') ds dt ds' dt' \, .
\]
\end{lemma}
The proof of this Lemma follows by very similar arguments as the proof of Lemma~\ref{jointlemma} upon observing that all results from~\cite{berhorric2013} used in that proof continue to hold with $\eta_i$ replaced by $\eta_i\otimes\eta_i$, $L^2(T)$ replaced by $L^2(T^2)$ upon noting that under (A4') the $\eta_i\otimes\eta_i$ satisfy~\eqref{eq:lappr-tensor}.

The next key step is to prove that $\hat\theta^{Cov} = \theta_0 + o_P(N^{-1/2})$. Begin by observing that 
\begin{align*}
\hat C_{1:k}(s,t) =& \frac{1}{k-1}\sum_{i=1}^k \eta_i(s)\eta_i(t) - \frac{k}{k-1}\Big(\frac{1}{k}\sum_{i=1}^k \eta_i(s) \Big)\Big(\frac{1}{k}\sum_{i=1}^k \eta_i(t) \Big),
\\
\hat C_{k+1:N}(s,t) =& \frac{1}{N-k-1}\sum_{i=k+1}^N \eta_i(s)\eta_i(t) - \frac{N-k}{N-k-1}\Big(\frac{1}{N-k}\sum_{i=k+1}^N \eta_i(s) \Big)\Big(\frac{1}{N-k}\sum_{i=k+1}^N \eta_i(t) \Big) \, .
\end{align*}
Noting that~\eqref{eq:supnorm} holds with $N, \eta_i^{(j)}$ $(j=1,2)$ instead of $m, \eta_i^X$ shows after some computations that 
\begin{align*}
& {\sup_{\floor{N\eps}+1 \leq k \leq N-\floor{N\eps}}}\int_T\int_T\Big\{ \frac{k}{k-1}\Big(\frac{1}{k}\sum_{i=1}^k \eta_i(s) \Big)\Big(\frac{1}{k}\sum_{i=1}^k \eta_i(t) \Big) \Big\}^2dsdt = O_P(N^{-2})
\\
& {\sup_{\floor{N\eps}+1 \leq k \leq N-\floor{N\eps}}}\int_T\int_T\Big\{\frac{N-k}{N-k-1}\Big(\frac{1}{N-k}\sum_{i=k+1}^N \eta_i(s) \Big)\Big(\frac{1}{N-k}\sum_{i=k+1}^N \eta_i(t) \Big)\Big\}^2dsdt = O_P(N^{-2}) \, .
\end{align*}
Define 
\[
\tilde C_{1:k} := \frac{1}{k-1} \sum_{i=1}^k \eta_i\otimes\eta_i, \quad \tilde C_{k+1:k} := \frac{1}{N-k-1} \sum_{i=k+1}^N \eta_i\otimes\eta_i \, .
\]
Next observe that 
\[
{\sup_{\floor{N\eps}+1 \leq k \leq N-\floor{N\eps}}}\Big\| \frac{1}{k} \sum_{i=1}^k \eta_i\otimes\eta_i - \E[\eta_i\otimes\eta_i] \Big\|_{L^2(T^2)} + \Big\|\frac{1}{N-k}\sum_{i=k+1}^N \eta_i\otimes\eta_i - \E[\eta_i\otimes\eta_i] \Big\|_{L^2(T^2)} = O_P(N^{-1/2}) \, .
\]
This in particular implies that 
\[
{\sup_{\floor{N\eps}+1 \leq k \leq N-\floor{N\eps}}} \|\tilde C_{1:k}\| + \|\tilde C_{k+1:N}\| = O_P(1) \, .
\]
Hence by the reverse triangle inequality and the usual triangle inequality we have
\begin{align*}
&{\sup_{\floor{N\eps}+1 \leq k \leq N-\floor{N\eps}}} \Big|\hat f^{Cov}(k) - \Big\| \tilde C_{1:k} - \tilde C_{k+1:N}\Big\|_{L^2(T^2)}^2 \Big|
\\
= & {\sup_{\floor{N\eps}+1 \leq k \leq N-\floor{N\eps}}} \Big|~\Big\| \hat C_{1:k} - \hat C_{k+1:N}\Big\|_{L^2(T^2)}^2 - \Big\| \tilde C_{1:k} - \tilde C_{k+1:N}\Big\|_{L^2(T^2)}^2\Big|
\\
\leq & {\sup_{\floor{N\eps}+1 \leq k \leq N-\floor{N\eps}}} \Big\{\|\tilde C_{1:k} - \hat C_{1:k}\| + \|\tilde C_{k+1:N} - \hat C_{k+1:N}\|\Big\}\Big\{\|\tilde C_{1:k}\| + \| \hat C_{1:k}\| + \|\tilde C_{k+1:N}\| + \|\hat C_{k+1:N}\|\Big\}
\\
=&~ o_P(N^{-1/2}) \, .
\end{align*}
Hence, applying the same arguments as in the proof of Proposition~\ref{propcpest} but replacing all instances $X_i,  \E[X_i]$ there by $\eta_i\otimes\eta_i, \E[\eta_i\otimes\eta_i]$ and making corresponding adjustments to integrals and norms we find that $\hat\theta^{Cov} = \theta_0 + o_P(N^{-1/2})$. 

The remaining proof consists in observing that by similar arguments as above we have uniformly in $\theta \in [\eps,1-\eps], \lambda \in [\eps,1]$
\begin{multline*}
\Big\| D_{N}^{cp,Cov}(\cdot,\lambda, \theta) - \frac{1}{\lfloor N \theta \rfloor-1} \sum_{j=1}^{\lfloor \lfloor N \theta \rfloor \lambda \rfloor}
\eta_j\otimes\eta_j - 
\\
\frac{1}{N - \lfloor N \theta \rfloor-1} \sum_{j=\lfloor \theta N \rfloor + 1}^{\lfloor \theta N \rfloor + \lfloor \lambda (N - \lfloor \theta N \rfloor ) \rfloor} 
\eta_j\otimes\eta_j \Big\|_{L^2(T^2)} = o_P(N^{-1/2})
\end{multline*}
and following similar arguments as in the proof of Theorem~\ref{thmcp}. 
 \hfill $\Box$

\section{Extensions beyond functional time series}\label{sec:concalt}
\def\theequation{C.\arabic{equation}}
\setcounter{equation}{0}

In this section we briefly discuss how the ideas presented in this paper  can be extended beyond the context of functional time series. We begin by introducing a general setup which will be used throughout this section. Let $X_1,...,X_n$ denote a sample of (potentially  dependent) random elements in some measurable space $\mathcal{S}$. Assume that we are interested in inference on a parameter $\mu = \mu_\mathbb{P}$ that can be assigned to distributions $\mathbb{P}$ on $\mathcal{S}$. We will assume that $\mu$ takes values in $\mathcal{M}$, a subset of a real Hilbert space $\mathcal{H}$ equipped with  an inner product $\langle \cdot,\cdot\rangle$ and  induced norm by $\|\cdot\|_\mathcal{H}$. Further, assume that for each $m \in \N$ there exists a mapping  $f_m: \mathcal{S}^m \to \mathcal{M}$, where $f_m(X_1,...,X_m)$ is interpreted as estimator for $\mu$ based on the observations $X_1,...,X_m$. The situation considered in Sections~\ref{sec1} - \ref{sec3}  corresponds to the choice $\mathcal{S} = L^2([0,1])$,  $\Hc = L^2([0,1])$ and  $\mu_\mathbb{P}$ is the mean (function) of $X_i$ (see the discussion in Example~\ref{ex1} below for more details). Finally, define for $\Lambda\subset [0,1]$ the space of functions
\[
\Bc(\Lambda,\Hc) := \{f: \Lambda \to \Hc: \sup_{ \lambda \in \Lambda} \|f(\lambda)\|_\Hc < \infty\}
\]
 equipped with the norm
\[
\|g\|_{\Bc} := \sup_{\lambda\in \Lambda} \|g( \lambda)\|_\Hc \, .
\]
This  space  will be used to characterize the joint behaviour of estimators of $\mu$ computed from several sub-samples (with sub-sample proportion corresponding to the index $\lambda$). Note that if the set $\Lambda$ contains only finitely many elements, say $|\Lambda|$, the normed space $(\Bc(\Lambda,\Hc),\|\cdot\|_\Bc ) $ can be identified with the $|\Lambda|$-fold
Cartesian product of $\Hc$ (viewed as a normed space).

\subsection{The one sample and two sample case}

We begin by considering the one sample case since in this setting the relevant conditions are particularly simple and transparent. The self-normalized statistic is based on a probability measure $\nu$ with support $\Lambda_\nu \subset (0,1)$. Using the notation $\hat \mu_{1:k} := f_k(X_1,...,X_k)$  we  define random elements in $\Bc(\Lambda,\Hc)$ through $g_n(\lambda) := \lambda \sqrt{n} (\hat \mu_{1:\floor{\lambda n}} - \mu)$. Assume that
\begin{equation} \label{eq:genmu2weak}
g_n \weak \mathbb{H} \quad \mbox{in } \Bc(\Lambda_\nu\cup 1,\Hc) \, .
\end{equation}
Note that we do not require measurability of $g_n$ and weak convergence is defined in the sense of Hoffman-Jorgensen, see Section 1.3 in \cite{wellner1996}. Further, assume that the limit $\mathbb{H}$ has the following additional properties:
\begin{eqnarray} \label{eq:genmuweak}
&&\Big\{ \langle \mathbb{H}(\lambda), \mu \rangle \Big\}_{\lambda \in \Lambda_\nu\cup 1} \stackrel{\mathcal{D}}{=} \sigma^2 \Big\{ \lambda \mathbb{B}(\lambda) \Big\}_{\lambda \in \Lambda_\nu\cup 1} \, ,
\\
 \label{eq:Hnonnull}
&&\Big(\int_{\Lambda_\nu} \Big[ \|\mathbb{H}(\lambda)\|_\Hc^2 - \lambda^2 \|\mathbb{H}(1)\|_\Hc^2\Big]^2 \nu(d\lambda) \Big)^{-1} = O_P(1) \, ,
\end{eqnarray}
where $\sigma^2 $ is a (nonnegative)  real-valued parameter that can depend on the distribution  $ \mathbb{P} $ of $X_i$ and $\mathbb{B}$ is a standard Brownian motion on the interval $[0,1]$. Moreover, we assume that  $\sigma^2 \neq 0$ if $\mu \neq 0$.

If \eqref{eq:genmu2weak} - \eqref{eq:Hnonnull} are satisfied, an asymptotic level $\alpha$ and consistent test for the (relevant)  hypotheses
\[
H_0: \| \mu \|^2_\Hc \leq  \Delta  \quad\mbox{ versus  }\quad H_1: \| \mu \|^2_\Hc > \Delta
\]
is given  by rejecting $H_0$  whenever the inequality  \eqref{testrelone}  holds,
where the statistic  $ \hat{\mathbb{T}}_n$ and  $ \hat{\mathbb{V}}_n$ are  now defined by
$\hat{\mathbb{T}}_n := \| \hat \mu_{1:n} \|^2_\Hc$  and
\begin{align*}
\hat { \mathbb{V}}_n := \Big( \int_{\Lambda_\nu} \Big[ \lambda^2\| \hat \mu_{1:\floor{n\lambda}}\|^2_\Hc - \lambda^4 \| \hat \mu_{1:n}\|^2_\Hc \Big]^2 \nu(d\lambda) \Big)^{1/2},
\end{align*}
respectively,
and  $q_{1-\alpha}(\mathbb{W})$ is the $(1-\alpha)$-quantile of the distribution of the random variable $\mathbb{W}$ defined in \eqref{Wdef}.

\begin{example} \label{ex1} ~~
{\rm
\begin{itemize}
\item[(a)]
In the setting of Section~\ref{sec21} the random variables
$X_1,...,X_n$ with distribution $ \mathbb{P} $ take values  in $\mathcal{S} = L^2([0,1])$, $\mu_\mathbb{P}$ is the mean (function) of $X_i$
which is an element of $\Hc = L^2([0,1])$,  and $f_m(X_1,...,X_m) = m^{-1}(X_1+...+X_m)$.
 The estimators $\hat \mu_{1:\floor{n\lambda}}$ then take the form $\hat \mu_{1:\floor{n\lambda}}(t) = \frac{1}{\floor{n\lambda}} \sum_{i=1}^{\floor{n\lambda}} X_i(t)$. The limit $\mathbb{H}$ in~\eqref{eq:genmu2weak} is given by $\mathbb{H}(\lambda) = \Gamma(\cdot,\lambda)$ where the process  $\Gamma$ is defined in equation~\eqref{approx3}.
\item[(b)]
When $\Hc$ is $\R$ with $\langle x,y\rangle = xy$, the space $\Bc(\Lambda,\Hc)$ can be identified with $\ell^\infty(\Lambda)$. Conditions~\eqref{eq:genmu2weak} and~\eqref{eq:genmuweak} follow from the functional CLT
\begin{equation}\label{eq:funclt}
\Big\{\lambda \sqrt{n} (\hat \mu_{1:\floor{\lambda n}} - \mu) \Big\}_{\lambda \in \Lambda_\nu\ \cup 1}  \weak \tau^2 \Big\{ \mathbb{B}(\lambda) \Big\}_{\lambda \in \Lambda_\nu \cup 1} \, .
\end{equation}
In this case $\mathbb{H} = \tau^2 \Bb$ and $\sigma^2 = \mu^2 \tau^2$. Moreover, condition~\eqref{eq:Hnonnull} follows from elementary properties of the multivariate normal distribution and the assumption that $\nu$ is a probability measure on the interval $(0,1)$.
\\
When $\mu$ is the mean of $\mathbb{P}$ and $\hat \mu_{1:\floor{\lambda n}}$ is the sample mean of the first $\floor{\lambda n}$ observations,~\eqref{eq:funclt} holds under a wide variety of assumptions on the serial dependence of time series. When $\Sc = \R^d$ and $\mu$ is a smooth function of the (multivariate) cdf of $X$, this condition can be verified
using the general framework developed in \cite{vosh2014}. This  includes quantities such as quantiles or Kendall's $\tau$ and other  dependence measures.
\item[(c)]
If  $\Hc= \R^d$  (with the usual inner product), the space  $\Bc(\Lambda,\Hc)$ can be identified with $[\ell^\infty(\Lambda)]^d$. The norm $\|f\|_{\Bc} = \sup_{\lambda \in \Lambda} \|f(\lambda)\|_2$ $(f \in \Bc(\Lambda,\R^d))$ is equivalent to the usual norm $\|(g_1,...,g_d)\| := \max_{j=1,...,d} \sup_{\lambda \in \Lambda} |g_j(\lambda)|$ where $(g_1,...,g_d) \in [\ell^\infty(\Lambda)]^d$ and the weak convergence in \eqref{eq:genmu2weak} is interpreted as weak convergence in the product space $[\ell^\infty(\Lambda)]^d$ as discussed in Section 1.4 of \cite{wellner1996}.
Let, $\hat \mu_{1:\floor{\lambda n},j}$ and   $\mu_j$ denote the $j$'th components of the vectors $\hat \mu_{1:\floor{\lambda n}}$ and $ \mu$, respectively,   and let $g_{n,j}(\lambda) := \lambda \sqrt{n} (\hat \mu_{1:\floor{\lambda n},j} - \mu_j)$. Conditions ~\eqref{eq:genmu2weak} and~\eqref{eq:genmuweak} now follow from a multivariate version of the functional CLT, i.e.
\begin{equation}\label{eq:funcltm}
(g_{n,1},...,g_{n,d})^\top \weak \Sigma (\mathbb{B}_1,...,\mathbb{B}_d )^\top \quad
\end{equation}
in $[\ell^\infty(\Lambda)]^d$,  where $\mathbb{B}_1,...,\mathbb{B}_d$ are independent Brownian motions on the interval
$[0,1]$ and $\Sigma$ denotes a matrix which can depend on $\mathbb{P}$ and is {non-singular}
when $\mu \neq 0$. The functional weak convergence in~\eqref{eq:funcltm} can be verified in a similar fashion as discussed in (b) and details are omitted for the sake of brevity.
\end{itemize}
}
\end{example}

Next we briefly discuss the two sample case, where  $X_1,...,X_m$ and $Y_1,...,Y_n$
are random variables  with $X_i \sim \mathbb{P}, Y_i \sim \mathbb{Q}$. For simplicity we shall further assume that the samples $X_1,...,X_m$ and $Y_1,...,Y_n$ are independent, while dependence within the samples is explicitly allowed. Let $\mu_\mathbb{P}, \mu_\mathbb{Q}$ denote the parameters of interest corresponding to $\mathbb{P},\mathbb{Q}$, respectively. Introduce the notation $\hat\mu_{1:k}^X := f_k(X_1,...,X_k), \hat\mu_{1:k}^Y := f_k(Y_1,...,Y_k)$ and let $g_m^X(\lambda) := \lambda \sqrt{m} (\hat \mu_{1:\floor{\lambda m}}^X - \mu_\mathbb{P})$, $g_n^Y(\lambda) := \lambda \sqrt{n} (\hat \mu_{1:\floor{\lambda n}}^Y - \mu_\mathbb{Q})$. Provided that $m/(m+n) \to \rho \in (0,1)$ and $g_m^X \weak \mathbb{H}^X$ and $g_n^Y \weak \mathbb{H}^Y$ in $\Bc(\Lambda_\nu\cup 1,\Hc)$ with limiting processes $\mathbb{H}^X, \mathbb{H}^Y$ satisfying~\eqref{eq:genmuweak} and~\eqref{eq:Hnonnull}, an asymptotic level $\alpha$ and consistent test for the (relevant) hypotheses
\[
H_0: \| \mu_\mathbb{P} - \mu_\mathbb{Q} \|^2_\Hc \leq  \Delta  \quad\mbox{ versus  }\quad H_1: \| \mu_\mathbb{P} - \mu_\mathbb{Q} \|^2_\Hc > \Delta
\]
is given  by rejecting $H_0$ whenever $\hat{\mathbb{T}}_{m,n} > \Delta + q_{1-\alpha}(\mathbb{W}) {\hat{\mathbb{V}}_{m,n}}$. Here the statistics $\hat{\mathbb{T}}_{m,n}$ and  $\hat{\mathbb{V}}_{m,n}$ are defined by $\hat{\mathbb{T}}_{m,n} := \|\hat \mu_{1:m}^X - \hat \mu_{1:n}^Y\|_\Hc^2$ and
\begin{eqnarray*}
\hat{\mathbb{V}}_{m,n}  &=& \Big( \int_{\Lambda_\nu}
\Big\{ \lambda^2\Big\|\hat \mu_{1:\floor{m \lambda}}^X - \hat \mu_{1:\floor{n \lambda}}^Y\Big\|_\Hc^2  - \lambda^4 \Big\|\hat \mu_{1:m}^X - \hat \mu_{1:n}^Y\Big\|_\Hc^2 \Big\}^2
\nu(d\lambda) \Big)^{1/2},
\end{eqnarray*}
respectively,  and  $q_{1-\alpha}(\mathbb{W})$ is the $(1-\alpha)$-quantile of the distribution of the random variable $\mathbb{W}$ in~\eqref{Wdef}.
The discussion in Example~\ref{ex1} also applies to the two sample case with obvious modifications.

\bigskip
\subsection{Testing for relevant change points in a general setting}
Next we discuss the problem of testing for relevant change points in the general context introduced in the beginning of Section~\ref{sec:concalt}. Assume that $X_i \sim \mathbb{P}$ for $1 \leq i \leq n\theta_0$ and $X_i \sim \mathbb{Q}$ for  $n\theta_0 < i \leq n$ where  $\theta_0 \in (0,1)$. Technically this is a triangular array model, but we will not stress this in the notation.
We explicitly allow $\mathbb{P}=\mathbb{Q}$ which corresponds to the case of no change point
in the sequence $X_1, \ldots , X_n$. Let $\mu_\mathbb{P}, \mu_\mathbb{Q}$ denote the parameters of interest corresponding to $\mathbb{P},\mathbb{Q}$, respectively. Our aim is to test the hypotheses of a relevant change in the parameter $\mu$ of the sequence $X_1 , \ldots , X_n$, that is
\begin{equation} \label{nullcpgen}
H_0: \| \mu_\mathbb{P} - \mu_\mathbb{Q} \|^2_\Hc \leq  \Delta  \quad\mbox{ versus  }\quad H_1: \| \mu_\mathbb{P} - \mu_\mathbb{Q} \|^2_\Hc > \Delta \, .
\end{equation}
{
Following the developments in Section~\ref{sec3},  we assume that   $\hat \theta$ is a consistent estimator for $\theta_0$ and
introduce the notation $\hat \mu_{j:k} := f_{k-j+1}(X_j,...,X_k)$. The {null hypothesis of no relevant difference} in \eqref{nullcpgen} is rejected
if $\hat{\mathbb{T}}^{cp} > \Delta + q_{1-\alpha}(\mathbb{W}) {\hat{\mathbb{V}}^{cp}}$, where
the statstics $\hat{\mathbb{T}}^{cp} $ and  $\hat{\mathbb{V}}^{cp} $ are defined by
\begin{eqnarray*}
\hat{\mathbb{T}}^{cp} &=& \|\hat \mu_{1:\floor{n \hat \theta}} - \hat \mu_{\floor{n \hat \theta}+1:n}\|_\Hc^2  ~,\\
\hat{\mathbb{V}}^{cp}  &=& \Big( \int_{\Lambda_\nu}
\Big\{ \lambda^2\Big\|\hat \mu_{1:\floor{n \lambda \hat \theta}} - \hat \mu_{\floor{n \hat \theta}+1:\floor{n \hat \theta} + \floor{n\lambda(1-\hat\theta)}}\Big\|_\Hc^2  - \lambda^4 \Big\|\hat \mu_{1:\floor{n \hat \theta}} - \hat \mu_{\floor{n \hat \theta}+1:n}\Big\|_\Hc^2 \Big\}^2
\nu(d\lambda) \Big)^{1/2},
\end{eqnarray*}
respectively,  and  $q_{1-\alpha}(\mathbb{W})$ is the $(1-\alpha)$-quantile of the distribution of the random variable $\mathbb{W}$ in~\eqref{Wdef}.
}
The asymptotic validity of this  test  can be verified provided that
\begin{equation}\label{wDth}
\Big\{\lambda \sqrt{n} (\hat \mu_{1:\floor{\lambda \hat\theta n}} - \hat \mu_{1+\floor{\hat\theta n}+\floor{\lambda (1-\hat\theta) n}:n} + \mu_\mathbb{Q} - \mu_\mathbb{P}) \Big\}_{\lambda \in \Lambda_\nu\cup 1}
 \weak \mathbb{H} \quad \mbox{in } \Bc(\Lambda_\nu \cup 1,\Hc) \, ,
\end{equation}
where the limiting process $\mathbb{H}$ satisfies~\eqref{eq:genmuweak} and~\eqref{eq:Hnonnull}.
This is a non-trivial high-level assumption since the estimator $\hat\theta$ appears on the left-hand side.

In concrete situations, for example in the situation considered in Section~\ref{sec3}, it is possible to replace~\eqref{wDth} by more tractable and elementary assumptions.
As a further illustration we will discuss the setting of a real-valued parameter $\mu$ that can be represented as smooth functional of the distribution function $F$ of the random variable $X$ with values in  $\Sc = \R^d$.
First, we note that there exists a general and well-developed machinery for
establishing convergence rates of change point estimators  [see, for example  \cite{carlstein1988}, \cite{dumbgen1991} or \cite{hariz2007},  among others],
and we assume
\begin{equation}\label{eq:cprate}
\hat \theta = \theta_0 + o_\mathbb{P}(n^{-1/2}) \, ,
\end{equation}
throughout this section, which is satisfied in many cases of practical interest.
The following discussion is a little informal and we refer the interested reader to the work of
 \cite{vosh2014} for some of the technical details that are omitted here for the sake of brevity. Assume that $\mu_\mathbb{P} = \Phi(F_\mathbb{P} ), \mu_\mathbb{Q}  = \Phi(F_\mathbb{Q} )$ for some smooth map $\Phi$,where $F_\mathbb{P} , F_\mathbb{Q} $ denote the cdf  of $\mathbb{P} ,\mathbb{Q} $, respectively. Let $\hat \mu_{i:j} = \Phi(\hat F_{i:j})$ where $\hat F_{i:j}$ is the empirical CDF of $X_i,...,X_j$. Under suitable assumptions on the temporal dependence structure of $X_1,...,X_n$ it is possible to prove the weak convergence

\begin{multline*} 
\Big(\Big\{ \lambda \sqrt{n} (\hat F_{1:\floor{n\theta_0\lambda}}(u) - F_\mathbb{P}(u)) \Big\}_{\lambda\in [0,1], u \in \R^d},  \Big\{\lambda \sqrt{n} (\hat F_{\floor{n \theta_0}+1:\floor{n \theta_0} + \floor{n\lambda(1-\theta_0)}}(u) - F_\mathbb{Q}(u))\Big\}_{\lambda\in [0,1], u \in \R^d}\Big)
\\
\weak \Big(\Big\{\Gb_1(\lambda,u) \Big\}_{\lambda\in [0,1], u \in \R^d}, \Big\{\Gb_2(\lambda,u) \Big\}_{\lambda\in [0,1], u \in \R^d} \Big) \quad
\end{multline*}
in $ \ell^\infty([0,1]\times \R^d)^2$, where $\Gb_1,\Gb_2$ denote two independent centered Gaussian processes with covariance structure of the form $\E[\Gb_j(s,u)\Gb_j(t,v)] = (s \wedge t)K_j(u,v)$ ($j=1,2$). By elementary calculations we have
\[
\sup_{\lambda \geq \eps} \sup_u |\hat F_{1:\floor{n\theta_0\lambda}}(u) - \hat F_{1:\floor{n\hat \theta\lambda}}(u)| = O_\mathbb{P}(|\theta_0 - \hat \theta|) = o_\mathbb{P}(n^{-1/2}) \, ,
\]
(note that two empirical cdf which are based on $k$ and $k+l$ observations with an overlap of $k$ observations differ by at most $l/k$). Similarly, it can be shown that
for any  $\eps > 0$
\[
\sup_{\lambda \geq \eps} \sup_u |\hat F_{\floor{n \theta_0}+1:\floor{n \theta_0} + \floor{n\lambda(1-\theta_0)}}(u) - \hat F_{\floor{n \hat\theta}+1:\floor{n \hat\theta} + \floor{n\lambda(1-\hat\theta)}}(u)| = O_\mathbb{P}(|\theta_0 - \hat \theta|) = o_\mathbb{P}(n^{-1/2}) \, ,
\]
and the two displays above imply the weak convergence 
\begin{multline*} 
\Big(\Big\{ \lambda \sqrt{n} (\hat F_{1:\floor{n\hat \theta\lambda}}(u) - F_\mathbb{P}(u)) \Big\}_{\lambda\in [\eps,1], u \in \R^d},  \Big\{\lambda \sqrt{n} (\hat F_{\floor{n \hat \theta}+1:\floor{n \hat \theta} + \floor{n\lambda(1-\hat \theta)}}(u) - F_\mathbb{Q}(u))\Big\}_{\lambda\in [\eps,1], u \in \R^d}\Big)
\\
\weak \Big(\Big\{\Gb_1(\lambda,u) \Big\}_{\lambda\in [\eps,1], u \in \R^d}, \Big\{\Gb_2(\lambda,u) \Big\}_{\lambda\in [\eps,1], u \in \R^d} \Big) \quad
\end{multline*}
in $ \ell^\infty([\eps,1]\times \R^d)^2$. Assume that the mapping  $\Phi$ is Hadamard differentiable at the points $F_\mathbb{P}, F_\mathbb{Q}$, tangentially to a vector space $V \subset \ell^\infty(\R^d)$, with linear derivatives $d\Phi_{F_\mathbb{P}}, d\Phi_{G_\mathbb{P}}$ and that $\Gb_1(\lambda,\cdot), \Gb_2(\lambda,\cdot) \in V$ a.s. Then the results in Section 4 of \cite{vosh2014} combined with linearity of $d\Phi_{F_\mathbb{P}}, d\Phi_{F_\mathbb{Q}}$ and some calculations show that the weak convergence in \eqref{wDth} holds for the functional $\mu = \Phi (F)$ with $\hat \mu_{i:j} = \Phi(\hat F_{i:j})$ provided that $\Lambda_\nu \subset [\eps,1]$. In particular we have have $\mathbb{H} \stackrel{\mathcal{D}}{=} \sigma^2 \Bb$ for some $\sigma^2 > 0$.

\end{document}